\documentclass[smallextended]{svjour3}       
\smartqed  
\usepackage{graphicx}
\usepackage{natbib}  
\usepackage{amsmath,amssymb,amsfonts}
\usepackage[colorlinks=true,linkcolor=blue,citecolor=blue,urlcolor=blue]{hyperref}
 \newcommand{\lae}{\mathrel{<\kern-1.0em\lower0.9ex\hbox{$\sim$}}}
 \newcommand{\gae}{\mathrel{>\kern-1.0em\lower0.9ex\hbox{$\sim$}}}

 \newcommand{\arcsec}{^{\prime\prime}}
\newcommand{\gsim}{\gtrsim}
\newcommand{\lsim}{\lesssim}

\journalname{The Astronomy and Astrophysics Review}
\begin{document}

\title{The nature of compact radio sources: \\ the case of FR\,0 radio galaxies}



\author{Ranieri D. Baldi}  


\institute{R.~D. Baldi \at
              INAF - Istituto di Radioastronomia, Via Piero Gobetti 101, 40129 Bologna, Italy \\
              \email{ranieri.baldi@inaf.it}           
}


\maketitle

\begin{abstract}

Radio-loud compact radio sources (CRSs) are characterised by morphological {\sl compactness} of the jet structure centred on the active nucleus of the galaxy. Most of the local elliptical galaxies are found to host a CRS  with nuclear luminosities lower than those of typical quasars, $\lsim$10$^{42}\, {\rm erg\, s}^{-1}$. Recently, low-luminosity CRSs with a LINER-like optical spectrum have been named Fanaroff--Riley (FR) type~0 to highlight their lack of substantially extended radio emission at kpc scales, in contrast with the other Fanaroff--Riley classes, full-fledged FR~Is and FR~II radio galaxies. FR~0s are the most abundant class of radio galaxies in the local Universe,  and characterised by a higher core dominance, poorer Mpc-scale environment and smaller (sub-kpc scale, if resolved) jets than FR~Is. However, FR~0s share similar host and nuclear properties with FR~Is. A different accretion-ejection paradigm from that in place in FR~Is is invoked to account for the parsec-scale FR~0 jets. This review revises the state-of-the-art knowledge about FR~0s, their nature, and which open issues the next generation of radio telescopes can solve in this context.

\end{abstract}

\setcounter{tocdepth}{3}
\tableofcontents

\section{Introduction}
\label{sec:intro}

A minority of the Active Galactic Nuclei (AGN), named radio-loud AGN (RLAGN)\footnote{Radio loudness, $R$, is defined  as ratio between the flux densities in the radio  (6 cm, 5 GHz) and in the optical band ($\sim$4400 \AA) \citep{kellermann89}: RLAGN have $R \geq 10$, whereas radio-quiet AGN have $R < 10$. See also \citet{terashima03}  for a radio loudness definition based on radio and X-ray (2\,--\,10 keV) luminosity ratio.} ($\sim$10\,--\,20\%, e.g. \citealt{urry95,kratzer15,macfarlane21}), is known to launch relativistic collimated jets from parsec (pc) to Mpc scales, which connect the active supermassive black hole (BH) with the interstellar medium (ISM)  to the furthest intra-cluster medium (ICM) (e.g., \citealt{blandford19,hardcastle20,saikia22}). A significant excess of radio emission over that expected from star formation (SF) processes is attributed to non-thermal jet emission in RLAGN (e.g., \citealt{condon92,bonzini13,padovani15}). However, a fraction of radio-quiet (RQ) AGN can also produce non-thermal emission from jets, which are, though, uncollimated and sub-relativistic (e.g. \citealt{padovani16,panessa19}). Jets, whether relativistic or not, are found to play a key role in galaxy evolution and in the maintenance of massive galaxies in the present-day Universe (e.g. \citealt{croton06,kharb23}), by releasing a large
amount of energy to their surrounding environment (the so-called radio-mode feedback) \citep{fabian12}. In fact, RLAGN are typically associated with the most evolved systems, i.e. massive early-type galaxies (ETGs, and mostly ellipticals) and the most massive BHs (e.g. \citealt{heckman14}), although a small number of exceptions have been found (e.g. \citealt{hota11,singh15,kaviraj15,kotilaninen16,mao18,webster21a,davis22}).

One of the most historically important classifications of RLAGN was introduced by \citet{fanaroff74}, based on the extended radio structure: edge-darkened sources were classified as Fanaroff--Riley type I (FR~Is), while edge-brightened as type II (FR~IIs): the latter generally more radio luminous ($> 2 \times 10^{25}\, {\rm W\, Hz}^{-1}$ at 178 MHz) than the former.  Extended plumes, lobes, and tails account for typically 90\% of their total radio emission \citep{miley80}. Other than the radio linear size and morphology, the jet orientation with respect to the line of sight is another important variable to characterise RLAGN: extended sources inclined at small angles may appear compact due to projection effects and their radio emission can be boosted due to relativistic beaming of the nuclear jets moving at relativistic velocities. A further crucial aspect of RLAGN is the radio spectrum (flux density $S_{\nu}$, varies with frequency $\nu$ as  $\nu^{-\alpha}$ and $\alpha$ is the spectral index): steep-spectrum sources ($\alpha > 0.5$) are typically associated with optically-thin emission from extended jets, while flat/inverted-spectrum sources ($\alpha < 0.5$) are largely due to (synchrotron-self or free-free) absorption process involved in compact cores and jet knots. While misaligned (type-2) RLAGN (with respect to the line of sight, e.g. FR~I/II), commonly named as radio galaxies (RGs), are typically dominated by their extended emission with steep spectra, aligned (type-1) RLAGN, named blazars\footnote{The blazar population consists in two sub-classes, the flat-spectrum radio quasars and BL~Lacs, generally at high and low luminosities, respectively.)}, show flat, inverted or complex radio spectra of the dominant cores. 

Traditionally, compact radio sources (CRSs) associated with misaligned RLAGN are believed to represent the early stages of evolution of full-fledged RLAGN (FR~I/IIs, \citealt{odea98}) and are characterised by a peaked radio spectrum (see Sect.~\ref{sec:compact}). Recently,  \citet{baldi15} introduced a new class of {\it low-power CRSs} ($\lsim$ 10$^{24}\, {\rm W\, Hz}^{-1}$), named {\it FR~0} RGs, whose compact radio emission is dominated by the core and pc-scale jets and not related to a juvenile radio activity. The characteristic property of such an abundant class of RGs is the substantial lack of kpc-scale jet emission, that has changed  the classical view on the kpc/Mpc-scale RLAGN phenomenology, particularly at low luminosities, where jets are scarcely studied. An increasingly-incomplete list  of RLAGN classes is given in Table~\ref{tab:agn_zoo}.

\begin{table*}
\caption{Radio-optical AGN taxonomy}
\begin{tabular}{lp{0.68\textwidth}r}
\hline 
Acronyms & Main properties  & Reference \\
\hline 
Quasar & Quasi-stellar radio source   & 1 \\
RLAGN & radio-loud AGN (relativistic collimated jets) & 2 \\
RQAGN & {\scriptsize radio-quiet AGN (thermal/non-thermal emission, uncollimated sub-relativistic jet)} & 2,3  \\
CRS & RL or RQ radio compact source & 4 \\
FR~I &Fanaroff--Riley class I radio source; radio core-brightened &  5 \\ 
FR~II &Fanaroff--Riley class II radio source; radio edge-brightened & 5 \\ 
FR~0 & {\scriptsize Fanaroff--Riley class 0 radio source; RL CRS lacking kpc-scale extended emission} & 6 \\
RG & Radio galaxy, misaligned RL AGN & 7 \\ 
Blazar & aligned RL AGN  & 7 \\
REAF & radiatively-efficient accretion disc & 8 \\
RIAF & radiatively-inefficient accretion disc & 9 \\ 
LERG  & Low-Excitation Radio galaxies & 10\\
HERG & High-Excitation Radio galaxies & 10\\
LLAGN & Low-luminosity AGN & 11  \\
CSS & Compact steep spectrum radio source; young RG  & 12 \\
GPS & Gigahertz-peaked radio source; young RG & 12 \\
HPF & High frequency peakers; young RG & 12 \\
CSO &  Compact symmetric object; young RG & 12 \\
MSO & Medium-sized symmetric object; young RG & 12\\
Seyfert & High-ionisation nuclear emission-line regions, RQ AGN & 13 \\
LINER & Low-ionisation nuclear emission-line regions, RQ or RL AGN & 13,14 \\
CoreG & Core Galaxies, nearby low-luminosity FR~0-like RGs & 15 \\
\hline 
\end{tabular}

The complex radio-optical AGN taxonomy includes several acronyms. Here a partial but helpful list of labels for AGN, their properties and references (first/key papers or recent papers, which give up-to-date details). References: 1. \citet{schmidt63}, 2.  \citet{padovani16}, 3. \citet{panessa19}, 4. \citet{kellermann81}, 5. \citet{fanaroff74}, 6. \citet{baldi15}, 7. \citet{urry95}, 8. \citet{shakura73} 9. \citet{yuan14}, 10. \citet{heckman14}, 11. \citet{ho08}, 12. \citet{odea21}, 13. \citet{kewley06},  14. \citet{heckman80}, 15. \citet{balmaverde06core}.
\label{tab:agn_zoo}
\end{table*}

The complex radio taxonomy of RLAGN needs to find a correspondence with the optical classification schemes to modes of accretion onto the
BHs  \citep[e.g.,][]{jackson97,heckman14,hardcastle00}. Based on their optical spectra, RGs have been classified into Low Excitation Radio Galaxies (LERGs) and High Excitation Radio Galaxies (HERGs) (Tab.~\ref{tab:agn_zoo}), which basically reflect two BH accretion states \citep{buttiglione10,tadhunter16a} with distinct distributions of Eddington ratios\footnote{The Eddington ratio  $\dot{L_{E}}$ gauges the BH accretion properties and is defined by the ratio between the bolometric AGN luminosity and the Eddington luminosity given its BH mass.}. LERGs are typically accreting below 1\% of their Eddington accretion rate limit, while HERGs have typical accretion rates between 1 and 10\% (or even higher) \citep{heckman14}. HERGs, accretion-dominated RLAGN, are characterized by radiatively efficient accretion flows (REAF),  i.e. standard optically thick, geometrically thin discs  \citep{shakura73}. LERGs, jet-dominated RLAGN, are powered by radiatively inefficient accretion flows (RIAF), which  include the disc solutions of geometrically thick advection-dominated accretion flows  (ADAF) (e.g., \citealt{narayan91,narayan95,Narayan00,yuan14}). LERGs  prefer redder, gas-poorer, more massive ETGs with lower star-formation rates, which inhabit richer and more dynamically relaxed environment and feed more massive BHs than HERGs \citep{baldi08,heckman14}.  LERGs are generally thought to be fuelled by the cooling of hot gas from haloes present in their massive host galaxies, whereas the HERGs generally tend to accrete cold gas efficiently, from  processes external or internal to the galaxy \citep{hardcastle07}. Several radio-optical studies of RGs concluded that the two FR radio morphologies are not representative of two distinct  accretion states, but can co-exist in the same optical class (e.g., \citealt{gendre13,mingo19,mingo22}). In fact, local LERGs are associated with a FR~I or FR~II morphology, whereas HERGs,  which are on average of higher luminosity, are generally FR~IIs. The new low-power FR~0 class has thus further entangled, although already complex, the radio-optical classification scheme (Tab.~\ref{tab:agn_zoo}).

\begin{figure}
\centerline{\includegraphics[width=0.72\textwidth,angle=-90]{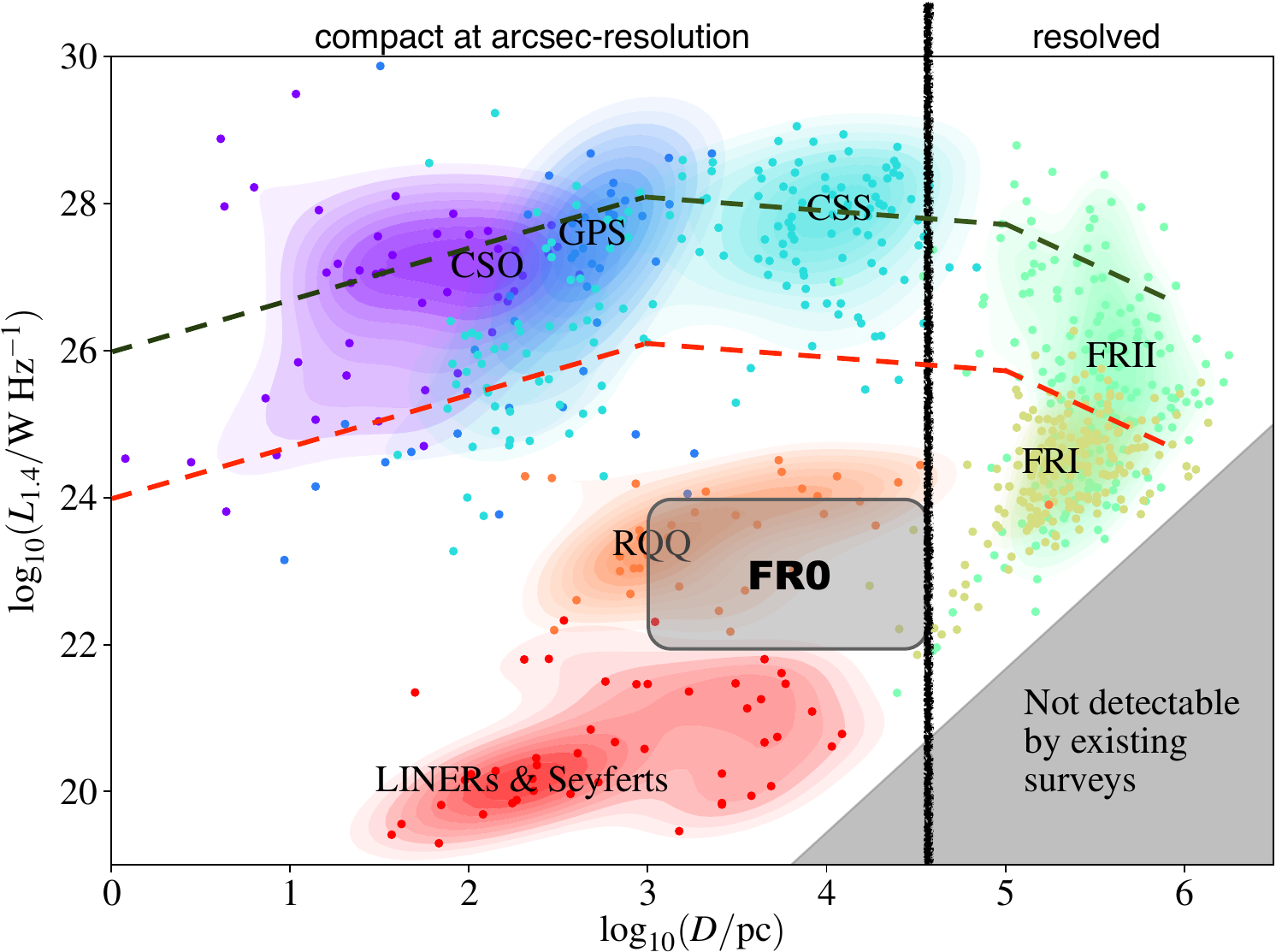}}
\caption{Radio power/linear-size plot ($P$-–$D$ diagram) for different types of RL and RQ AGN, adapted from plots presented by  \citet{an12,jarvis19,hardcastle20}. Points show individual objects and coloured contours represent a smoothed estimator of source density. The
different categories of source shown are: CSO, GPS, CSS, FR~I, FR~II, RQ quasars, Seyferts and LINERS, and FR~0s (see Sect.~\ref{sec:intro} and \ref{sec:compact} for the definition of the classes). Red and dark-green dashed lines represents the classical evolutionary tracks of FR~Is and FR~IIs (e.g., \citealt{an12}). The shaded bottom-right corner shows the effect of surface-brightness limitations by existing radio surveys: very recently, deep LOFAR and MeerKAT surveys are starting to explore this region of the $P$-–$D$ diagram \citep{whittam22,best23}. The vertical line roughly represents the separation between resolved and unresolved/compact sources based on arcsec angular resolution, generally provided by the VLA array. The black box depicts the VLA detected FR~0s and represents an upper limit on their actual radio physical size. This figure is a modified version of Fig.~2 from \citet{hardcastle20}.}
\label{pdhardcastle}
\end{figure}

Other than radio and optical bands, decades of observations of accreting BHs at different wavelengths have shed new light on specific aspects of the accretion-jet phenomena (e.g. X-ray, broad/narrow optical lines, IR excess), collecting evidence for an anisotropic AGN emission. The attempt to unify all the AGN classes in one single picture concluded with the Unification Model (UM,  e.g. \citealt{barthel89,antonucci93,urry95}), which states that, despite their differences, RLAGN have the same basic structure (attested for powerful sources): optically-thick circumnuclear matter (torus) obscuring the accretion disc in an edge-on view, perpendicular to a relativistic jet, Doppler boosted  when seen at small angles to the line of sight. This orientation-based scheme represents the most courageous way to characterise the fact that the nuclear continuum and emission-line radiation from all types of AGN are simply a function of wavelength, inclination to the line of sight and source luminosity. However, the advent of modern sensitive and survey-mode telescopes has unveiled new regions in the space parameters of RLAGN phenomenology (see e.g., in time domain astronomy,  radio/optical/X-ray spectroscopy and polarimetry, jet/wind structure, disc and dust properties, \citealt{padovani16,padovani17b,spinoglio21}), which have defined specific accretion-ejection states of AGN and relative transitions, which the simplistic UM cannot explain. Although the UM is still generally valid, the  most logical way to relieve the tension is the inclusion of the time variable in the UM, i.e. the parameters can evolve across time. An evolutionary scheme of RLAGN offers a more adaptable method to fine-tune the AGN parameters observed in distinct and transitioning states of accretion and ejections \citep{antonucci12,netzer15,tadhunter16a}.

The dynamic evolution of the accretion-ejection coupling in RLAGN is traditionally explained as a progression of the radio power with the linear size of the radio structure (see \citealt{an12} and references therein). Figure~\ref{pdhardcastle} shows the radio power $P$ versus the total extent of the source, $D$ (the so-called ``$P$-–$D$'' diagram, \citealt{baldwin82}): different populations of radio-emitting AGN (quiet and loud)  span over a very wide range in radio luminosities (nearly ten orders of magnitude) and source sizes (six orders of magnitude) \citep{hardcastle19,hardcastle20}. For RLAGN, in Fig.~\ref{pdhardcastle}, two representative evolutionary tracks within the $P$-–$D$ diagram are shown and predict RL CRSs to evolve into traditional $\sim$100-kpc double RGs (FR~Is or FR~IIs, e.g. \citealt{kunert10,an12,kunert16}). However, there is an important caveat. All the evolutionary models and our current knowledge on RG populations have long been based on samples of powerful sources, mostly above $10^{24}\, {\rm W\, Hz}^{-1}$, selected from high-flux low-frequency radio surveys such as the Third Cambridge (3C) catalogue \citep{bennett62a}. In opposition to the past, recent large-area sensitive surveys have revealed that the local RG population is dominated by sources with radio power below $10^{24}\, {\rm W\, Hz}^{-1}$ \citep{best12}, which includes mostly compact FR~0-type RGs. The UM model is not able to successfully reproduce such an abundant population of 'low-luminosity' 
RLAGN.

A milestone in the comprehension of the RLAGN phenomenon is the work by \citet{best05a}, which selected  the largest complete sample of low-luminosity RGs ($\lsim 10^{41}\, {\rm erg\, s}^{-1}$) by cross-matching Sloan Digital Sky Survey (SDSS, \citealt{york00}), National Radio Astronomy Observatory (NRAO) Very Large Array
(VLA) Sky Survey (NVSS, \citealt{condon98}), and the Faint Images
of the Radio Sky at Twenty centimeters survey (FIRST, \citealt{becker95}) with flux densities $>$ 5 mJy a 1.4 GHz. This flux-density cut is much below than the selection of, e.g. the 3C catalogue (178 MHz flux density $>$ 9 Jy, \citealt{bennett62a}), on which most of our comprehension of the radio-AGN phenomenon is based.   The most interesting result from the radio-optical survey is that their radio morphology appears unresolved at the scale of the FIRST radio maps, i.e. 5$\arcsec$, which corresponds to 10--20 kpc with $z<0.3$. These compact RGs, named later {\it FR0s}, which belong to a heterogeneous population of LERG-type red massive ellipticals, represent the bulk of the RG population of the local Universe with a space density $>$ 100 times higher than 3C/RGs. The study of the FR~0 population has a relevant role in the modern astrophysics because: i) since they are the most common RLAGN in the local Universe, their comprehension provides an important insight on the accretion-ejection mechanism for ordinary RGs; ii) since their radio emission is on galactic scale, their jets can have a tremendous impact on the galaxy evolution in the context of radio-mode feedback. 

In this review, we provide an overview of the observational properties and theoretical understanding of this interesting class of compact RGs, FR~0s.  
We introduce the class of CRSs in Sect.~\ref{sec:compact} to then focus on the FR~0s (Sect.~\ref{sec:fr0}), by discussing their selection (radio and host properties). Then we derive the radio luminosity function of local RGs to demonstrate the abundance of FR~0s with respect to the FR~I/IIs (Sect.~\ref{sec:rlf}). Then we review their multi-band properties from radio (Sect.~\ref{sec:radio}), optical and IR (Sect.~\ref{sec:optIR}) to high energy bands (Sect.~\ref{sec:HE}) to picture their typical spectral energy distribution (SED). A discussion of the accretion-ejection coupling (Sect.~\ref{sec:a&e}), environmental properties (Sect.~\ref{sec:environment}) and their role of compact sources in AGN feedback (Sect.~\ref{sec:feedback}) lead to drawing static and dynamic scenarios to account for the multi-band properties of FR~0s in relation to the other FR classes (Sect.~\ref{sec:friilerg} and \ref{sec:models}). A final chapter on  future perspective is also included (Sect.~\ref{sec:fututre}).

We adopt in this work $H_0 = 70$~km~s$^{-1}$~Mpc$^{-1}$, $\Omega_{m_0} = 0.3$, $\Omega_{\Lambda_0} = 0.7$.

\section{Compact radio sources}
\label{sec:compact}

In this review we consider CRSs, those AGN which appear unresolved at small angular sizes ($\lsim$ a few arcseconds). CRSs can be RQ or RL, and here we will focus on the latter. In RL CRSs, the compact component is generally  ascribed to the radio core, which is interpreted as non-thermal self-absorbed synchrotron emission from the base of a relativistic jet, that extracts energy from the spinning BH  and/or the accretion disc \citep{blandford77,blandford82}. The connection between the compact core emission and the pc-kpc-Mpc scale extended jet emission, has been discussed in previous reviews (e.g.,  \citealt{condon78,odell78,kellermann80,kellermann81,odea98,falcke04a,lobanov06,tadhunter16b,odea21}), which all gather increasing evidence of a large population of CRSs in the local Universe.

{\it What does 'compact' mean and what defines the compactness?} The angular size of a compact source can vary from milli-arcsecond (mas) to a few arcseconds depending on the frequency, resolution and depth of observations. There is no a specific limit on the physical scale for a compact source. The morphological compactness can be defined as the 'unresolved' structure of a radio source, when the deconvolved size is smaller or equal to the radio-map beam width and when its visibility function is flat across the entire spatial-frequency plane.  Starting with the
Rayleigh–Jeans limit for brightness temperature $T_{b}$ [K] \citep{condon16}, the angular dimension $\theta$ of a compact source with its peak intensity  S$_{\nu}$ [mJy beam$^{-1}$]  at the frequency $\nu$ [GHz] is
\begin{equation}
    \theta \sim 35 \,\ S_{\nu}^{1/2} \,\ T_{b}^{-1/2} \,\ \nu^{-1} \,\ arcsec.
    \end{equation}
For  brightness temperatures above the threshold to discriminate between an AGN and stellar origin \citep{falcke00}, $T_{b} \gg 10^{7}$ K, for a spectral peak frequency between hundreds of MHz to GHz, the angular size varies between a few mas to a few arcseconds. This corresponds to a linear dimension range of several hundreds of pc to kpc in the nearby Universe ($z<0.3$). However, in the literature, a CRS is trivially classified as a morphologically point-like source based on the corresponding angular resolution.

In theory, a conventional definition of a compact source  predicts the presence of a characteristic self absorption in the radio spectrum at low frequencies (below GHz regime).  A varying opacity throughout the source entails a spectral characterisation: a flat, inverted or undulating spectrum over a wide range of frequencies due to the superposition of several radio-emitting partially opaque sources. Both the  generally flat-spectrum and the compactness of the source can lead to the interpretation of an unresolved radio-emitting nucleus.

A (flat-spectrum) compact radio core can be observed across all types of galaxies and AGN. Even normal spiral, star forming galaxies, RQAGN and late-type galaxies (LTG) in general can reveal a compact  nucleus, whose radio origin is thermal and non-thermal emission from several physical processes \citep{condon92,panessa19}. Since the most
radio-loud AGN are preferentially hosted by bulge-dominated evolved galaxies with masses larger than $10^{11}\,M_{\odot}$ and much less signs of morphological disturbance (spirals, bars) and SF than RQAGN (e.g. \citealt{best05b,ho08,best12,koziel17a,koziel17b,magliocchetti22}), a prior selection of the optical hosts, ETGs rather than LTGs, can help to exclude RQAGN from genuine RL CRS samples at the cost of completeness in the RLAGN population.

The presence of an arcsec-scale compact radio emission at the centre of ETGs has been affirmed since '70s as the main results of shallow VLA radio surveys up to now (e.g. \citealt{rogstad69,heeschen70,ekers73,kellermann81,sadler84,fanti86,fanti87,wrobel91b,slee94,giroletti05,capetti09,nyland16,hardcastle19,roy21,grossova22,wojtowicz22,capetti23}).  More massive galaxies and earlier in type appear to be more probably connected to the presence of a RLAGN (e.g., \citealt{smith86,best05b,floyd10,kim17,zheng20}), able to launch from the weakest to the most powerful jets in the Universe (large range of luminosities, morphologies, duty cycles and speeds, e.g. \citealt{heckman14,morganti17b,morganti21,saikia22}).

Hosted in ETGs, RL CRSs have been classified based on radio spectral and morphological properties. Other than blazars which have a large intrinsic radio size but appear compact because of projection effects and are affected by relativistic beaming (one-sidedness, superluminal
motions, and high brightness temperatures), misaligned RL CRSs \citep{readhead94b,odea98,orienti16,odea21}  have been studied mainly at high powers ($L_{\rm 1.4\, GHz} > 10^{25}\, {\rm W\, Hz}^{-1}$) and are characterised by a  convex synchrotron radio spectrum: the peak position around 100 MHz in the case of compact-steep spectrum (CSS) sources (well determined only by LOFAR and MWA observations in the recent years, e.g. \citealt{mahony16,callingham17,slob22}), and at about 1 GHz in the case of
GHz-peaked spectrum (GPS) sources, or even up to a few GHz in the sub-population
of high frequency peakers (HFP) \citep{fanti85,spencer89,stanghellini98,snellen98,dallacasa00,kunert02,orienti07,hancock10} (Fig.~\ref{pdhardcastle}). Morphologically, lobes and/or hot spots are typically resolved with very-long baseline interferometry (VLBI) observations and a weak component hosting the core is occasionally present (e.g. \citealt{wilkinson91,gugliucci05,an10,an12b,wu13}). Depending on their size, CSS/GPS may be termed as compact symmetric objects (CSO) if they are smaller than 1 kpc, or medium-sized symmetric objects (MSO) if they extend up to 10 - 15 kpc \citep{conway02,fanti01}. The existence of a relation between the rest-frame peak frequency and the projected linear size (e.g. \citealt{odea97}) indicates that the mechanism responsible for the curvature of the spectrum is the youth: these sources are small because they are
still in an early stage of their evolution, and will develop into FR~I/II sources (e.g.,  \citealt{phillips82,fanti90,snellen00,an12}). The alternative scenarios point to a  
dense medium which might limit and frustrate the jet growth \citep{vanbreugel84,carvalho94,carvalho98,ghisellini04,giroletti05}, or to a short or recurrent activity due to occasional BH accretion \citep{readhead94b,gugliucci05,kunert10,kunert11,an12,kiehlmann23}.

In conclusion, the CRS category can embrace a large population of radio-emitting sources: RQAGN, star-forming galaxies, blazars, young RGs and the FR~0s. In the next section, we will focus on the properties of this `new' class of compact RGs, FR~0s, in relation to the large-scale RLAGN population.

\section{Low-luminosity CRSs: the FR0s}
\label{sec:fr0}

 A significant fraction of nearby galaxies shows evidence of weak nuclear activity unrelated to normal stellar processes. Recent high-resolution, multi-wavelength observations indicate that this activity derives from BH accretion with a wide range of
accretion rates and is associated with a CRS (e.g., \citealt{nagar05,ho08,zuther12,saikia18,williams22,williams23}).  In fact, moving to lower luminosities generally corresponds to selecting AGN with smaller and weaker jet (compact) structures and flatter radio spectra (e.g., \citealt{nagar05,sadler14,baldi10b,gurkan18,sabater19,hardcastle19,debhade23}), but with an increasing contribution from spurious RQAGN \citep{mezcua14,bonzini13,baldi21b}. Current radio surveys of the local Universe have unearthed a large population of low-luminosity AGN (LLAGN, with bolometric luminosities $\lsim$10$^{40}\, {\rm erg\, s}^{-1}$), which were poorly explored in the past.  \citet{best12}, up-dating the sample of \citet{best05a}, select  18,286 RGs (the SDSS/NVSS sample, hereafter), with low powers ($L_{\rm 1.4\, GHz} < 10^{24}\, {\rm W\, Hz}^{-1}$) at low redshifts ($z<0.3$), whose the majority ($\sim$80\%) are LLAGN and radio compact (5$\arcsec$), with linear sizes $\lsim$10--20 kpc. 

The role of LLAGN  and their compact jet emission in galaxy-BH co-evolution \citep{ho08,kormendy13} is crucial for several aspects:

i) since LLAGN outnumber the quasar population by a few orders of magnitudes at $z<0.3$ \citep{nagar05,best05b,saikia18}, they provide the snapshot of the ordinary relation between an accreting BH and its host. The absence of an outshining AGN
at the galaxy centre allows us to better study the co-evolutionary link between host and BH;

ii) since LLAGN reside in less massive galaxies, the identification of LLAGN would help to constrain the occupation fraction of active BH in galaxies at low stellar masses $>10^{9-10}\,M_{\odot}$ \citep{greene12,gallo19}, and the BH mass density function at $M_{\rm BH} <10^8\,M_{\odot}$. These quantities are 
 fundamental to calibrate the prescriptions for BH-galaxy growth of semi-analytical and numerical models (e.g.,  \citealt{shankar09,barausse17});

iii) due to the lack of sensitive surveys in the past, the role of LLAGN in galaxy evolution has been always downgraded with respect to powerful
quasars, which by definition can offer a larger energetic budget
to the host. Yet, recently the advent of deep radio surveys is reversing our view on AGN activity: LLAGN are always switched on at
some level at low radio powers ($L_{\rm 150\,MHz} \gsim 10^{21}$ W
Hz$^{-1}$, \citealt{sabater19}) and have galactic-scale jets, that can have a tremendous impact on their hosts by continuously injecting energy into the host, a crucial aspect for the jet-mode (or radio-mode) feedback \citep{fabian12}.

While in the optical band the role of LLAGN in BH-galaxy co-evolution and their BH accretion properties have been largely studied \citep{ho08,fanidakis11}, their connection with the radio band has recently started to be explored. The past and current optical-radio studies of radio-emitting LLAGN collect observational evidence that three states of accretion-ejection exist: RQ Seyferts, RQ LINERs and RL LINERs (Low-Ionization Nuclear Emission line Regions, \citealt{hine79,heckman80,kewley06}), different from the accretion-ejection states at higher luminosities, LERGs and HERGs and RQ quasars\footnote{At low luminosities the presence of intermediate RL Seyferts is still unclear and a luminosity gap between Seyferts and HERGs seems to exist (e.g. \citealt{baldi10b,pierce19,pierce20}). At high luminosities the studies on LERG-type REAF-type quasars are still contradictory (see e.g. \citealt{younes12}).}. LINERs have lower accretion rates ($\dot{m}$), are usually more radio-loud and reside in earlier type galaxies than Seyferts \citep{ho08}. In fact, LINERs tend to host compact cores \citep{cohen69,falcke00,filho02,maoz07}, more radio luminous as the BH mass (or galaxy mass) increases (e.g. \citealt{laor00,best05a,mauch07}).  RQ LINERs and Seyferts exhibit sub-relativistic and not collimated jets (e.g.,  \citealt{ulvestad99,wrobel00,ulvestad01a,gallimore06,singh15b,baldi21b}). Conversely, RL LINERs have been generally interpreted as the scaled-down version of powerful RLAGN in terms of accretion and jet luminosities \citep{chiaberge05,balmaverde06core}.
The nuclei of RL LINERs can be described with 
a model of  synchrotron self-absorbed base of a low-power (mildly) relativistic jet coupled with an underluminous RIAF disc (typically an ADAF, \citealt{narayan94}),  analogous to FR~I/LERG disc-jet coupling (e.g. \citealt{balmaverde06b,hardcastle09}). The low-power CRS population selected from the SDSS/NVSS sample \citep{best12}
in the same luminosity range ($\lsim$10$^{41}\, {\rm erg\, s}^{-1}$) of classical 3C/FR~Is includes a heterogeneous population of mostly LINER/LERGs\footnote{LERG and RL LINER are equivalent classes at low luminosities.}  with a broad distribution of BH mass and host properties.

\citet{baldi10} analysed in detail the photometric and spectroscopic properties of the SDSS/NVSS sample to select the bona-fide RLAGN population (see Sect.~\ref{sec:selfr0}). They found that the majority of the SDSS/NVSS sample ($\sim$ 80\%) consists of compact LERGs, that are characterised by a total jet power up to a factor $\sim$1000  lower than what expected by RGs with bolometric AGN luminosity similar to those of the 3C/FR~Is ($\sim$ 10$^{40}\, {\rm erg\, s}^{-1}$). This remarkable result that the local Universe is dominated by low-luminosity CRSs lacking of substantial extended emission, expresses the need to include these sources in the taxonomy of RGs.  \citet{ghisellini11} for the first time introduced in the literature the name {\it FR~0} to characterise a population of weak RL CRSs hosted in ellipticals, named  Core Galaxies\footnote{The Core Galaxy nomenclature comes from the (core-type) optical flat surface brightness profile in innermost region of an ETG (e.g. \citealt{faber97}).} (CoreG), which exhibit  radio core and AGN bolometric luminosities similar to the weakest 3C/FR~Is (M87), but with an extended radio emission hundreds of times weaker \citep{baldi09}. CoreG  host genuine 'miniature' RGs with LINER-like nuclei, which extend the nuclear luminosity correlations reported for 3C/FR~Is by a factor of $\sim$1000 toward lower luminosities \citep{balmaverde06core,kharb12}: this has been interpreted as a sign of a common central engine (RIAF disc) \citep{balmaverde06core,kharb12}. CoreG are characterised by kpc-scale jets and a deficit of total radio emission in analogy to the SDSS/NVSS sample, but at lower radio luminosities.

In analogy to CoreG, the FR~0 classification (see Sect.~\ref{sec:selfr0}) does not correspond to a pure radio morphological selection of CRSs, but also includes an optical identification (host and AGN properties) to separate the genuine FR~0s which are all RLAGN, from spurious RQAGN and star forming galaxies (bluer LTGs with emission line ratios consistent with Seyfert or SF and steeper radio spectra,  \citealt{baldi16}).

\begin{figure}
\centering
\includegraphics[width=\textwidth]{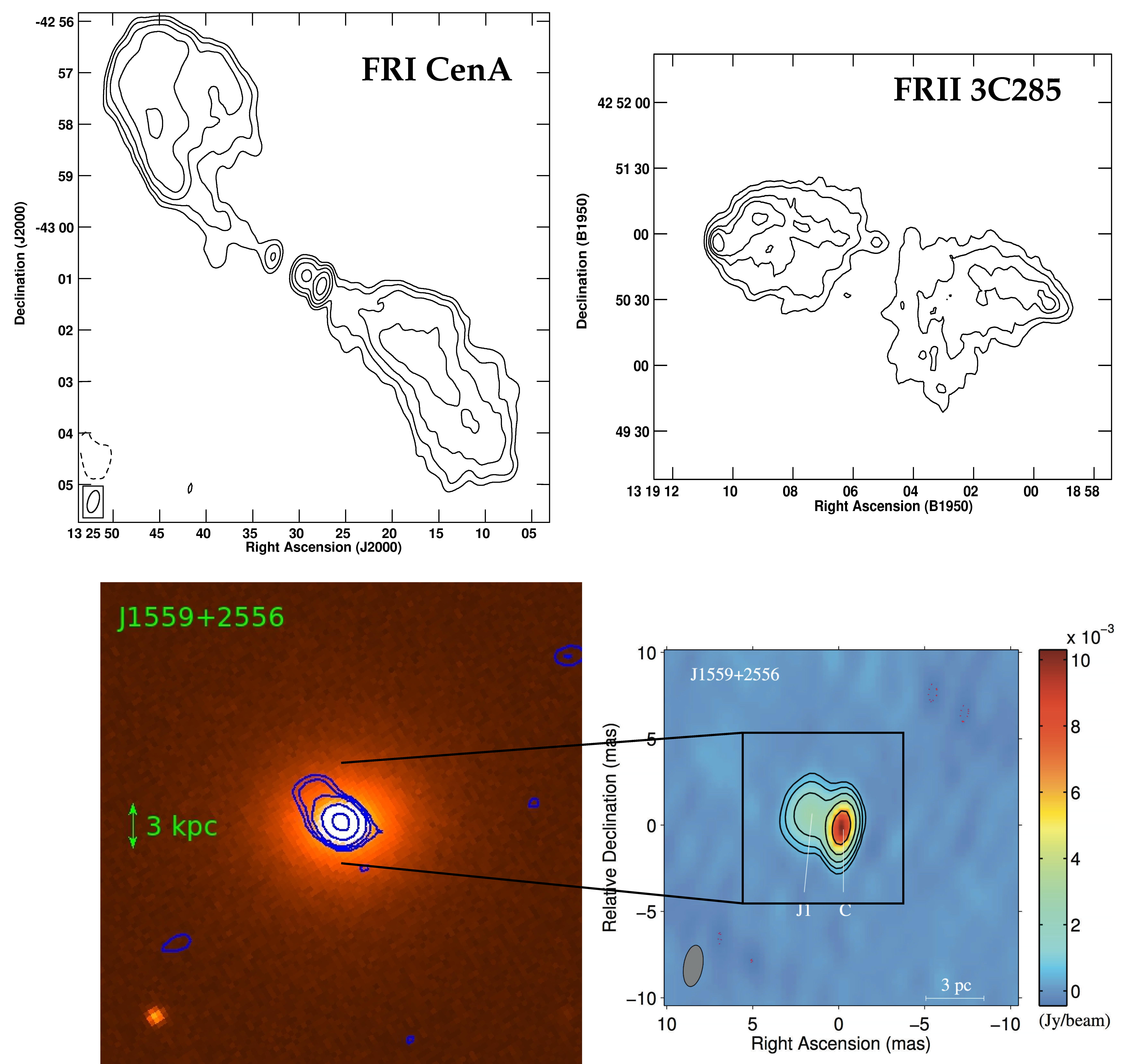}
\caption{Multi-band composite panel of RGs. On the top two examples of typical radio morphologies of a FR~I (Cen~A, \citealt{burns83} at 1.4 GHz) and a FR~II (3C~285, \citealt{alexander87} at 1.4 GHz). On the bottom, we show an example of FR~0. The left panel  displays the r-band SDSS image of the ETG which hosts the FR~0 with the blue VLA 4.5-GHz radio contours \citep{baldi19} (3 kpc scale set by the green arrow). The right panel represents the high-resolution zoom on the radio core (on the scale of 3 pc) provided by the VLBI image from \citet{cheng18}. Image reproduced with permission from \citet{baldi19rev}, copyright by the authors.}
\label{fr0panel}
\end{figure}

A closer look at the FR~0s at sub-arcsec/mas scale revealed that the majority  still appears radio compact, with a flat spectrum in the GHz band \citep{baldi15,baldi19,cheng18,cheng21}. However, a small fraction of those exhibits kpc/pc-scale core-brightened jets, suggesting that FR~0s can actually produce collimated structures. The lack of substantially extended radio emission at kpc scale and the spectral flatness for the majority of these CRSs have led to the affirmation of the FR~0 nomenclature as a unique class of genuine compact RGs different from the other RLAGN classes.

In conclusion, in the last decade, different parallel studies have brought to light a revolutionary result, i.e. classical 3C FR~I/~IIs do not represent the ordinary picture of the RLAGN phenomenon in the local Universe, but FR~0-like LLAGN represent the bulk of the local RG population (Fig.~\ref{fr0panel}). The paucity of 
sources with weak extended radio structures in high
flux limited samples (such as in the 3C sample) is due to a selection bias, since the inclusion of such objects is highly disfavored.  In fact, in support to this interpretation, \citet{baldi09} showed that the lower flux threshold of B2 sample ($<$250 mJy at 408
MHz, \citealt{fanti78}) drastically reduces the selection bias and allows the inclusion of a larger fraction of core-dominated\footnote{The ratio of core to total extended emission (which in general includes the core emission for simplicity) is called the core-dominance parameter (generally the total and core emission measured, respectively,  at 1.4 GHz and $\gsim$5 GHz). Core-dominated galaxies have typically a core dominance $\gsim 1/3$.}  galaxies, consistent with being FR~0s.

\subsection{Selection of FR~0s}
\label{sec:selfr0}

Disentangling bona-fide FR~0s from the radio compact impostors (blazars, young RGs, RQAGN, compact star-forming galaxies) represents a multi-band selection process.  This can be harder at low luminosities (mJy-level at $z<0.3$). For example, \citet{best05a} used several optical photometric and spectroscopic diagnostics and radio properties to select RLAGN in the SDSS/NVSS sample, however a small fraction ($\sim$10\%) of a possible RQAGN contribution is still present after the selection. Because aligned and young RGs can be removed from the sample only on the basis of a spectral and temporal radio study which are often not available, the simplest method to select bona-fide FR~0 candidates is based on a shallow optical-radio (largely available) selection process which consists of a few steps to maximise the probabilities that the radio emission is associated with a compact RL active nucleus. Accordingly, \citet{baldi18} have compiled a catalogue of 104 FR~0 sources (namely, FR0CAT) from the SDSS/NVSS sample, by adopting the following criteria:

\begin{itemize}
    \item nearby (redshift $z \lsim 0.05$) galaxies.
    \item compact: the sources are unresolved in the NVSS maps at 45$\arcsec$ resolution.  More stringently, the source  must appear unresolved at FIRST resolution, 5$\arcsec$. The FR~0 candidates consist of unresolved  sources for which the deconvolved size is smaller than 4$\arcsec$. At $z = 0.05$ this corresponds to $\sim$5 kpc, that is, to a radius of 2.5 kpc. 
    \item FIRST 1.4-GHz flux density $>$ 5 mJy to increase the possibility of an accurate size and flux measurement.  This value corresponds to $\sim$30 times the noise level of the FIRST maps.
 \item LERGs. Selecting LINERs allows the exclusion of AGN with high-Eddington ratios (generally Seyferts/HERGs) and are more probably associated with RLAGN phenomena \citep{heckman14,panessa19}.
\end{itemize}

Follow-up observations at higher angular resolution than that of FIRST maps are needed to confirm whether the FR0CAT sources still remain unresolved at sub-kpc scale.

The resulting FR0CAT sample turns out to be  a population of RGs with a core dominance of a factor $\sim$30 higher than typical 3C/FR~Is \citep{baldi09,baldi19,whittam20}, where instead the core typically contributes to 1\% to the total radio emission \citep{morganti97b}. Their 1.4-GHz radio luminosities are in the range $10^{38}$\,--\,$10^{40}\, {\rm erg\, s}^{-1}$. These radio selections turned out to include mostly luminous ($-21 \gsim M_{r} \gsim -23$) red ETGs with BH masses $10^{7.5} \lsim M_{\rm BH} \lsim 10^{9}\,M_{\odot}$\footnote{All the BH masses reported in this work for FR0CAT, FRICAT, sFRICAT and FRIICAT objects are derived from SDSS stellar velocity dispersions $\sigma$ and considering the M$_{\rm BH}$-$\sigma$ relation of \citealt{tremaine02}.}. However, only a minor fraction of the selected FR~0s departs from this general behavior (galaxies with optical photometric and spectroscopic characteristics, typical of blue star-forming spirals and RQAGN, see Sect.~\ref{sec:host}), although a host (ETG) selection was not part of the selection criteria.

\begin{figure}
\includegraphics[width=0.5\linewidth,height=4.5cm]{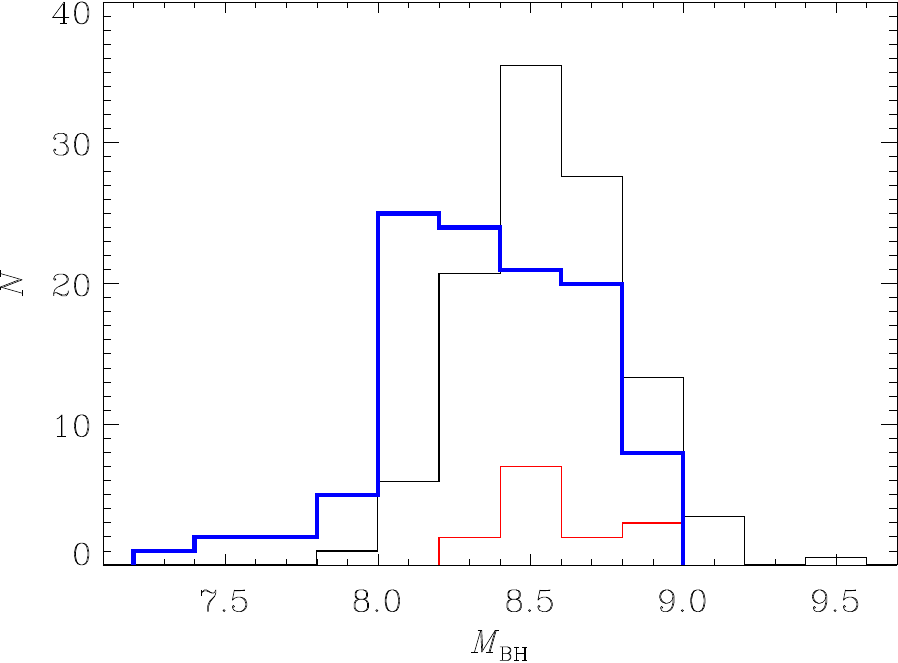}
\includegraphics[width=0.45\linewidth]{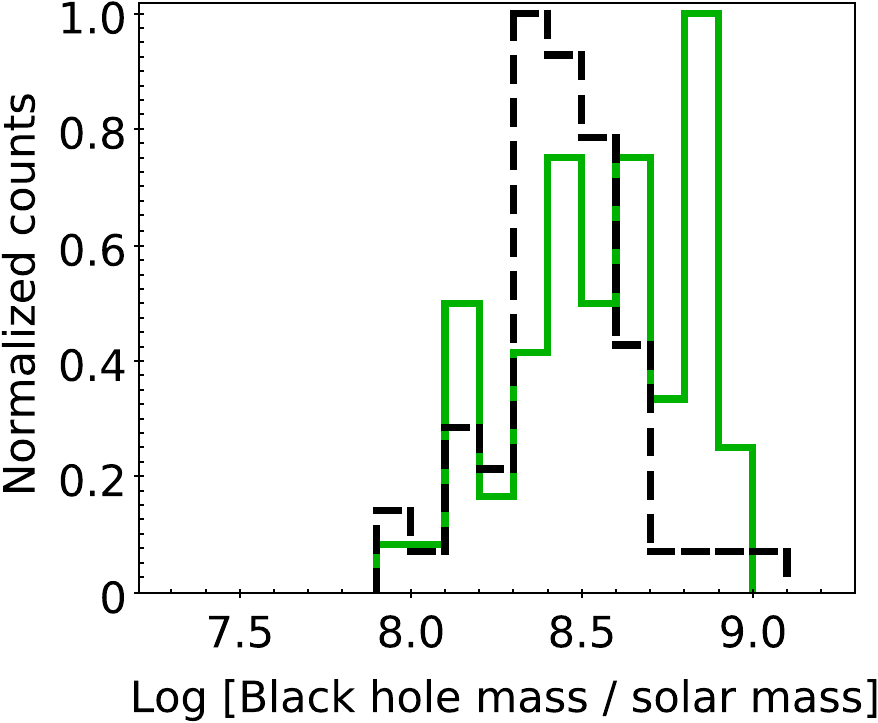}
\caption{Left panel: BH mass distribution (in M$_{\odot}$) of FR~0s (FR0CAT, blue line) with respect to FR~Is (FRICAT, radio size $>$ 30 kpc, black line) and small FR~Is (sFRICAT, 10 $<$ radio size $<$ 30 kpc, red line). Right panel: compact  (black) and extended RGs (green), when matched in radio core luminosities. 
Images reproduced with permission from [left] \citet{baldi18}, copyright by ESO; and from [right] \citet{miraghaei17}, copyright by the author(s).}
\label{fig_BHmass}
\end{figure}

As control samples with respect to the FR0CAT, other catalogues of low-luminosity FR~Is and FR~II have been selected from the SDSS/NVSS sample, a factor $\sim$10-100 weaker than 3C/RGs. \citet{capetti17a} selected  219 low-luminosity FR~Is, named FRICAT, with core-brightened radio morphology, redshift $\leq$ 0.15, and extending (at the sensitivity of the FIRST images) to a radius (r) larger than 30 kpc from the optical centre of the host. The authors also selected an additional sample (sFRICAT) of 14 smaller (10 $<$ r $<$ 30 kpc) FR~Is, limiting to $z < 0.05$.  The distribution of radio luminosity at 1.4 GHz of the FRICAT covers the range $10^{39}$\,--\,$10^{41.3}\, {\rm erg\, s}^{-1}$ and the sources are all LERGs. The hosts of the FRICAT sources are all luminous ($-21 \gsim Mr \gsim  -24$), red ETGs with BH masses in the range, $10^{8} \lsim M_{\rm BH} \lsim 10^{9.5}\,M_{\odot}$, slightly larger than FR0CAT BH masses (Fig.~\ref{fig_BHmass}). Similarly, \citet{capetti17b} selected 122 low-luminosity FR~IIs, named FRIICAT, with redshift $\leq$ 0.15, an edge-brightened radio morphology, and those with at least one of the radio emission peaks located at radius r $>$ 30 kpc from the optical galaxy center.  The radio luminosity at 1.4 GHz of the FRIICAT sources covers the range 10$^{39.5}$ -10$^{42.5}\, {\rm erg\, s}^{-1}$. The FRIICAT catalog mostly includes LERGs (90\%), which are luminous ($-20 \gsim Mr \gsim -24$), red ETGs with BH masses in the range $10^{8} \lsim M_{\rm BH} \lsim  10^{9}\,M_{\odot}$.

Other FR~0 samples were selected at lower and higher radio frequencies than the FIRST 1.4-GHz band (see Sect.~\ref{sec:radio} for details), which instead include a larger contamination from spurious sources than the FR0CAT. At low radio frequencies (hundreds of MHz) which is expected to be dominated by optically-thin emission, the vast majority ($\sim$70\%) of sources in the wide-area LOFAR \citep{hardcastle19,sabater19,mingo19,capetti20} and GMRT  Survey \citep{capetti19}, and in the deep well-studied field (e.g. ELAIS-N1 and BOOTES, \citealt{sirothia09,ishwara20}) appear compact with an angular resolution of a few arcsec and have $\alpha$ between 0 and 0.85, with the flat-spectrum sources more abundant than the steep-spectrum companions.

At higher radio frequencies (tens of GHz) which is expected to be dominated by the optically-thick emission, FR~0s have been selected by \citet{sadler14} from the AT20G-6dfGS sample and by \citet{whittam16} from the Cambridge 10C survey (mostly $z<3$) based on their radio morphological compactness (a few arcsec). Both the samples selected 70--80\% of CRSs, which include FR0-like LERGs and a large fraction of possible GPS/CSS sources.

 In conclusion, the selection of flat-spectrum weak CRSs in red massive hosts still remains the safest way to  select bona-fide FR~0s in relation with other compact and extended radio galaxies which can exhibit steeper radio spectra and bluer hosts (see next section).

\subsection{Host properties}
\label{sec:host}

The different radio-frequency selections of the FR~0s lead to a heterogeneous distribution of their host properties (e.g. galaxy type, colour, mass, $M_{\rm BH}$): selecting red massive ETGs represents the most secure criterion of identifying hosts of a FR~0. In fact, a prior host selection through several diagnostics can reduce the probability of inclusion of radio-compact impostors. The concentration index $C_{\rm r}$ is defined as the ratio of the galaxy radii including 90\% and 50\% of the light in the r band, respectively. ETG have higher values of concentration index than LTG, i.e. $C_{\rm r} > 2.6$ \citep{strateva01}. The Dn(4000) spectroscopic index is defined as the ratio between the flux density measured on two sides of the Ca~II break ($\sim$4000\,\AA) \citep{balogh99} and high values, Dn(4000) $>$ 1.7, are generally associated with old stellar populations ($\gsim$ 1 Gyr, \citealt{hernan13}) and, hence, with red passive galaxies \citep{best05b,capetti15}. Optical and infrared colour can also separate red ellipticals from blue spirals. The combination of these diagnostics with the FR0CAT criteria listed in Sect.~\ref{sec:selfr0} allows to identify the radio-compact red massive ETGs which have the highest probabilities of hosting a RLAGN.

\begin{figure}
\centering
\includegraphics[width=0.63\linewidth]{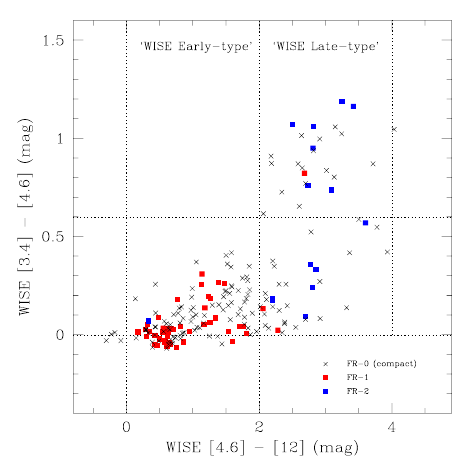}
\caption{WISE colour-colour plot (W2-W3 vs. W1-W2) for the host galaxies of FR~I (red squares), FR~II (blue squares) and compact (FR~0, black crosses) radio sources in the 20 GHz AT20G-6dFGS sample from \citet{sadler14}. The horizontal line at a $3.4-4.6\mu$m colour of 0.6 mag divides the AGN and normal galaxy populations. Objects where radiation from an AGN dominates the galaxy spectrum in the mid-infrared are expected to lie above this line, and objects where starlight dominates should lie below the line. The vertical line W2-W3 $>$2 identifies LTGs from ETGs. Image reproduced with permission from \citet{sadler16}, copyright by Wiley‐VCH.}
\label{fig_wise}
\end{figure}

The vast majority of  the FR0CAT, FRICAT and FRIICAT hosts are indistinguishable: red massive ETGs, based on the values of the $C_{r}$, spectroscopic Dn(4000) indices and broad-band colour. Their redness is confirmed by the photometric $u-r$ colour, measured over the whole galaxy. The WISE infrared colours  further support the general passive nature of the FRCAT  hosts (Fig.~\ref{fig_wise}, W1-W2 $<$ 0.2, \citealt{wright10}).  Nonetheless, a few galaxies of the FR0CAT extend to redder colours than those from the FRICAT and  there is a notable lack of blue host galaxies ($u-r > 2.5$) with respect to the general population of ETGs \citep{schawinski09}. In addition, the galaxy mass (and BH mass) of FR0CAT sources is on average smaller than those of FRICAT galaxies by a factor $\sim$1.4 (Fig.~\ref{fig_BHmass}), a possible effect of the selection of their lower radio luminosities since radio power and host mass are found to correlate in RL AGN (e.g. \citealt{best05b,capetti23}).

At high frequencies, \citet{sadler14} did not opt for a host selection and, in fact, found that the host galaxies of FR~0s display heterogeneous properties with a wide range in WISE colours, (33\%
in LTGs with some ongoing SF, see Fig~\ref{fig_wise}). This
implies that the selected FR~0 candidates, which make up the majority
of the AT20G-6dFGS sample, probably consists of a mixed bag of genuine FR~0s, young RGs and RQAGN. In fact, the bluer colour of the selected FR~0s is generally attributed to galaxies with a recent SF burst or to young RGs in gas-rich environments.

Since the radio core luminosity has been argued to be a better gauge of jet power than total radio luminosity\footnote{The radio core power is a measure of instantaneous power, rather than the total radio power, that is an averaged value over time and is also influenced by environment.}, 
\citet{miraghaei17} matched a sample of RL CRSs and extended RGs on the basis of the core luminosities. In terms of host properties, they found that CRSs and extended RGs differ only in the BH mass (Fig.~\ref{fig_BHmass}), similar to the result from the FR0CAT \citep{baldi18}.

The combination of the following criteria, i.e. the optical red colour, radio compactness and low radio powers (in mJy-level radio surveys), allows to 
increase the chances to exclude  radio-compact impostors and select mostly massive ETGs which harbour compact RL LLAGN, $<$10$^{23}\, {\rm W\, Hz}^{-1}$ \citep{best05a,sabater19,hardcastle19}, consistent with a FR~0 classification.

\begin{figure}
\centerline{\includegraphics[width=0.8\linewidth,angle=180]{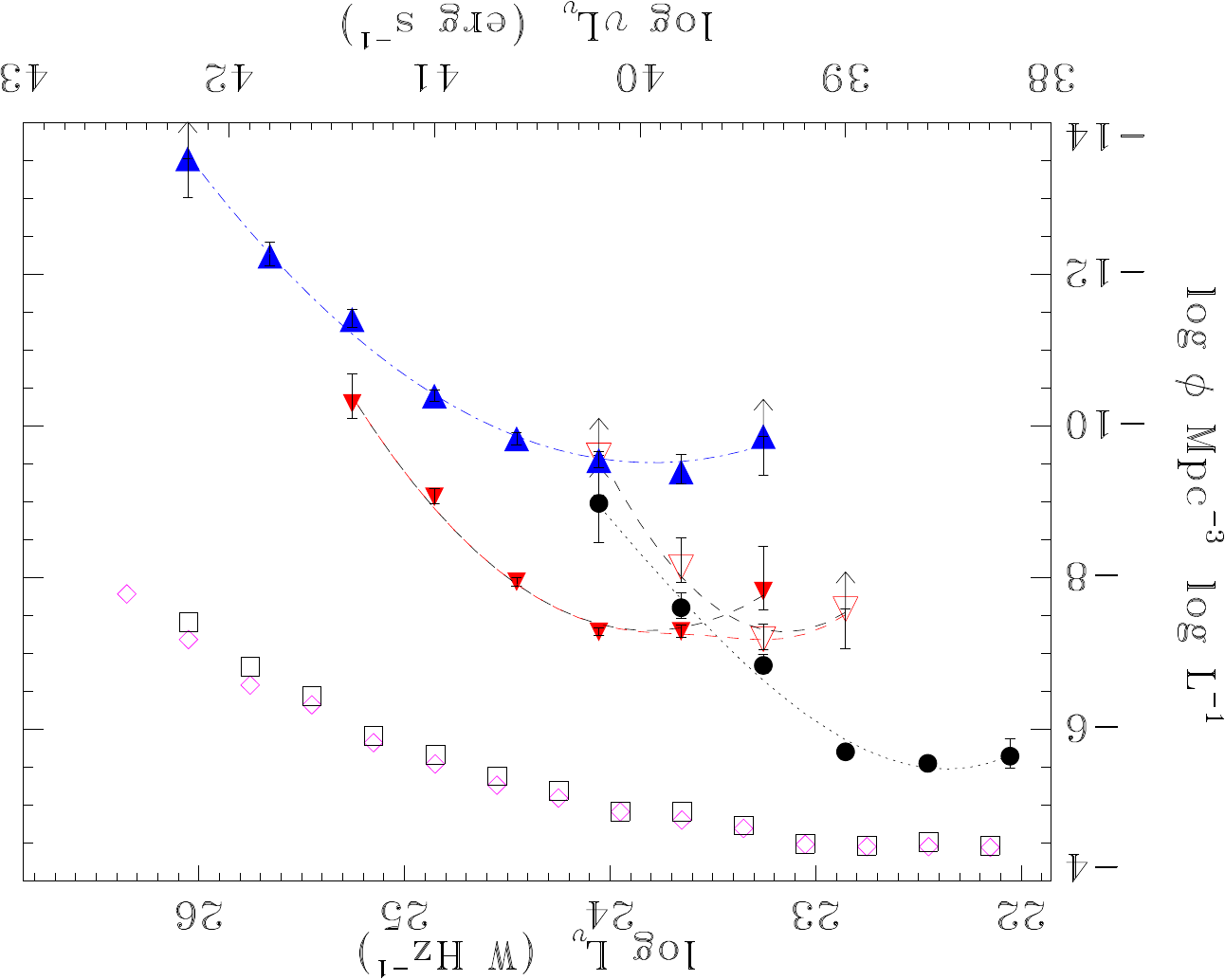}}
\caption{The local NVSS luminosity function at 1.4 GHz for RLAGN (pink diamonds) and LERGs (empty squares) from  the SDSS/NVSS sample \citep{best12}. The lower x-axis is expressed in erg s$^{-1}$ and the upper one in W Hz$^{-1}$. The other points represent the radio luminosity functions for FR~0s (filled circles), small FR~Is (empty red up-warded triangles), FR~Is (filled red triangles) and FR~IIs (blue filled down-warded triangles) from the FR0CAT \citep{baldi18}, sFRICAT/FRICAT \citep{capetti17a} and FRIICAT \citep{capetti17b}. The dot, dashed and dot-dashed lines are rough fits of the data-points, respectively, for FR~0s, FR~Is, and FR~IIs, to better visualize the luminosity functions.}
\label{RLF}
\end{figure}

\section{Radio luminosity function}
\label{sec:rlf}

We calculate the radio luminosity functions of the FRCAT sources as object density per unit logarithmic luminosity interval within the  maximum volume $V_{\max}$ in which the objects would be observed \citep{schmidt68,condon89}:

\begin{equation}
\Phi \left ( \log L_{\rm NVSS} \right ) = \frac{4 \pi}{\sigma} \sum_{i=1}^{N\left ( \log L_{*} \right )} \frac{1}{V_{\max\left ( i \right )}} \,, 
\end{equation}

\noindent where $\sigma$ is the area of the sky surveyed, $N\left ( \log L_{*} \right )$ is the number of objects in a given NVSS luminosity bin $L_{*}$, and $V_{\max\left ( i \right )}$ is given by the limiting magnitudes/fluxes in both the optical and radio properties of the sample, namely a radio cutoff of 5 mJy and SDSS optical cutoff of $r < 18$, as well as any imposed redshift limit for the analysis ($z < 0.05$ for FR0CAT and sFRICAT and $z < 0.15$ for FRICAT and FRIICAT). The sky area of the overlapping region between the SDSS DR7 spectroscopic
survey and the FIRST/NVSS radio survey is $\sigma$ = 2.17 steradians.  We place detected sources in bins of equal radio luminosities and estimate the uncertainties as in \citet{condon89}. We use Poisson statistics to
estimate uncertainties in luminosity bins with small numbers of sources ($N<7$). If $N=1$, we set 1$\sigma$ upper limit on the luminosity function in that bin.  The 1.4-GHz NVSS luminosities functions are derived for the FR~0s, FR~Is and FR~IIs from the FRCAT in Fig.~\ref{RLF} and tabulated in Table~\ref{tab_RLF}.

Figure~\ref{RLF} shows that, as expected, FR~0s dominate the radio
source population at relatively low radio luminosities $\lsim 10^{23.5}$ W\,Hz$^{-1}$, while the FR~Is and FR~IIs dominate at the highest luminosities.  Quantitatively, FR~0s represent the bulk of the RLAGN population of the local Universe ($z<0.05$) with a space density $\sim$4.5 times higher than that of FR~Is and $\sim$100 than that of FR~IIs. In relation to the luminosity function of ETGs, compact sources, consistent with a FR~0 morphology, are found in more than 60\% of the giant (K-band magnitude $\leq -25$) ETGs detected by LOFAR with 150-MHz luminosity $\geq$ 10$^{21}\, {\rm W\, Hz}^{-1}$ \citep{capetti22}.

\begin{table*}
\caption{\label{tab_RLF} The local NVSS radio luminosity functions at
  1.4\,GHz for FR0CAT, sFRICAT, FRICAT and FRIICAT (LERG) sources. The first column shows the range of 1.4\,GHz radio luminosities (erg s$^{-1}$) considered in each bin. The $N$ columns give the total number of radio sources and ${\log}_{10} \rho$ their space density (number per $\log_{10} L$ per Mpc$^3$, see Fig.~\ref{RLF}).}
\begin{tabular}{ccccccccc}
\hline
${\log} L_{\rm 1.4 GHz}$&
\multicolumn{2}{c}{FR0CAT} &
\multicolumn{2}{c}{sFRICAT} &
\multicolumn{2}{c}{FRICAT} &
\multicolumn{2}{c}{FRIICAT} 
\\
erg s$^{-1}$  &
N & ${\log}_{10} \rho$ &
N & ${\log}_{10} \rho$ &
N & ${\log}_{10} \rho$ &
N & ${\log}_{10} \rho$ \\
\hline
38.0--38.4 &  6 & $-5.64^{+0.15}_{-0.23}$ &  0  & --                       & 0  &  --                        & 0  & --                      \\
38.4--38.8 & 27 & $-5.55^{+0.08}_{-0.10}$ &  0  & --                       & 0  & --                         & 0  & --                      \\
38.8--39.2 & 48 & $-5.70^{+0.06}_{-0.07}$ &  1  & $<-7.58^{+0.51}$       & 0  & --                         & 0  & --                       \\
39.2--39.6 & 14 & $-6.84^{+0.10}_{-0.14}$ &  8  & $-7.19^{+0.13}_{-0.20}$  & 3  & $-7.82^{+0.24}_{-0.59}$ & 1  & $<-9.86^{+0.51}$       \\
39.6--40.0 &  8 & $-7.60^{+0.13}_{-0.19}$ &  4  & $-8.13^{+0.20}_{-0.39}$  & 32 & $-7.29^{+0.07}_{-0.09}$ & 7  & $-9.39^{+0.15}_{-0.23}$  \\
40.0--40.4 &  1 & $<-9.97^{+0.51}$     &  1  & $<-9.60^{+0.51}$        & 91 & $-7.28^{+0.05}_{-0.05}$ & 19 & $-9.54^{+0.10}_{-0.12}$ \\
40.4--40.8 &  0 & --                      &  0  & --                       & 70 & $-7.94^{+0.05}_{-0.06}$ & 36 & $-9.82^{+0.07}_{-0.09}$  \\
40.8--41.2 &  0 & --                      &  0  & --                       & 19 & $-9.06^{+0.09}_{-0.12}$ & 34 & $-10.39^{+0.07}_{-0.09}$  \\
41.2--41.6 &  0 & --                      &  0  & --                       & 4  & $-10.29^{+0.20}_{-0.39}$ & 16 & $-11.40^{+0.10}_{-0.13}$  \\ 
41.6--42.0 &  0 & --                      &  0  & --                       & 0  & --                         & 9  & $-12.23^{+0.13}_{-0.19}$  \\
42.0--42.4 &  0 & --                      &  0  & --                       & 0  & --                         & 1  & $<-13.52^{+0.51}$  \\
\hline  
\end{tabular}
\end{table*}

\section{Radio properties}
\label{sec:radio}

The radio band uniquely characterises the FR~0s as RL CRSs which lack of substantial extended radio emission. In this section, we focus on the radio properties of FR~0s  and RL CRSs in general. We provide an overview of the continuum observations  from different telescopes at different frequencies (from 150 MHz to mm-band) and resolutions (from arcsec to milli-arcsec) to probe the physical mechanism acting at various linear scales along the putative jet. Most studies of CRSs which are reported in the next sub-sections, are related to low-$z$ sources (unless explicited) and typically LERGs.

\subsection{Low resolution}
\subsubsection{GHz-band: sub/arcsec-scale with VLA}

For the large availability of shallow radio data in the band $\sim$1--5 GHz, the VLA has been the first telescope used to select and characterise the properties of FR~0s. In fact, for the large sky coverage, moderately high resolution and sensitivity, FIRST and NVSS 1.4-GHz surveys have been largely exploited to select CRSs and RGs in general in the local Universe, but other than these data the radio information was extremely limited. Later, follow-up VLA observations of 25 FR~0s at 1.4, 4.5, and 7.5 GHz revealed that two third still appear compact at the angular resolution of 0.3$\arcsec$ (a few hundreds of parsec)  and with a flat radio spectrum in the GHz band \citep{baldi15,baldi19}.  Only a third of
the sample exhibits twin or one-sided jets extended on a scale of $\sim$2--14 kpc  (see Fig.~\ref{fr0panel} as an example). The apparent radio compactness of most FR~0s at kpc scales could be caused by the fact that jet emission is below the surface brightness limit of most large-scale radio surveys. In fact, \citet{shabala17} demonstrated that VLBI-scale compact AGN could have lobes and plumes too faint to be detected by most surveys with the VLA and LOFAR. The absence of substantial extended jet emission, whether due to observational effects (no sufficient sensitivity to detect diffuse jets on larger scales) or to intrinsic reasons (intermittent jet activity, young radio activity, intrinsic jet inefficiency, see Sect.~\ref{sec:a&e} and \ref{sec:models}), represents the characteristic feature of the FR~0 class and their uniqueness with respect to the other classes of RGs.

Wide-area GHz-band surveys also revealed a large fraction of low-power CRSs, e.g., $\sim$93\% in the VLA-COSMOS Large Project at 3 GHz \citep{bondi18,vardoulaki21}. These FR~0s candidates are associated with less massive hosts $\sim10^{10.8}\,M_{\odot}$, with lower radio powers ($\lsim 10^{22}\, {\rm W\, Hz}^{-1}$) and at higher redshifts (median $z \sim 1.0$)  than the FR0CAT sources. In the Very
Large Array Sky Survey (VLASS; \citealt{lacy20}) at 3 GHz, \citet{nyland20} selected $\sim$2000 compact RGs, but the redshift information is not well characterised for the entire sample. The selected CRSs in these surveys consists of a heterogeneous population of AGN with red and blue colours, consistent with genuine FR~0s, star-forming galaxies, RQAGN and blazars. Furthermore, \citet{koziel20} found that $\sim$90\% of the optical SDSS galaxies  at $z<0.5$ with a FIRST counterpart appear compact with $L_{\rm 1.4\,GHz} \sim 10^{21}$\,--\,$10^{26}\, {\rm W\, Hz}^{-1}$, hosted typically by ellipticals, a similar result to the work by \citet{baldi10b}.

Other GHz-band studies on core-dominated LINERs with moderate radio-loudness  hosted in ETGs (e.g. \citealt{nagar00,filho00,filho02,verdoes02,filho04,kharb12,singh15b,dullo18,zajacek19,singh19}) strengthen the result that nearby elliptical galaxies tend to power RL LLAGN with galactic-scale jet structures,  in analogy to FR~0 galaxies.

\subsubsection{High-frequency up to mm-band}

Interferometric observations at $\nu \gsim 5$ GHz have the advantage of isolating better the compact optically-thick flat-spectrum core. In fact, at high frequencies the Australia Telescope Compact Array (ATCA)  played an important role in the early studies of FR~0s. \citet{sadler14} have cross-matched the Australia Telescope 20 GHz (AT20G) Survey with the optical spectroscopic 6dF Galaxy Survey (6dFGS; \citealt{jones09})
to produce a volume-limited sample of 202 high-frequency
CRSs associated with local galaxies (at a median $z \sim 0.06$) with 20-GHz flux density limit of 40 mJy. The angular resolution 10–-15$\arcsec$ corresponds to a projected linear size of 10–15 kpc. \citet{chhetri13} used data from the longest (6 km) ATCA baseline to
determine how much of the radio emission seen by the AT20G
survey arose in very compact components. They showed that generally almost all their 20 GHz radio emission comes from a central source $\lsim0.2\arcsec$  and almost half of the
AT20G sources have flat radio spectra  at 1–-20 GHz.  The selected FR~0s represent the dominant population ($\sim$70–-75\%) of the AT20G–6dFGS catalogue at radio powers between $\sim$10$^{22}$ and
$10^{26}\, {\rm W\, Hz}^{-1}$ in the local Universe. In addition, the high-frequency selected FR~0s consist of a heterogeneous population in terms of both optical AGN types (75\% LERGs, 25\% HERGs) and host galaxy types (67\%  ETGs, 33\% LTGs). Further studies of these 20-GHz CRSs confirmed that the flat-spectrum AT20G objects sources tend to preserve a similar spectral shape in polarisation and are hosted in bluer galaxies than standard ETGs
\citep{chhetri12,chhetri20,massardi12}.

\citet{whittam16} and \citet{whittam20} selected a complete sample of 96 faint ($>$ 0.5 mJy) RGs from the Tenth Cambridge (10C) survey at 15.7 GHz including LERGs and HERGs, mostly, within $z\sim3$. Sixty-five sources are unresolved in the 610-MHz GMRT radio observations, placing an upper limit on their angular size of $\sim2\arcsec$. The majority of these sources have flat spectra and are core dominated. The selected FR~0 population is the most abundant in the subset of sources with 15.7-GHz flux densities $<$1 mJy, extending the results of \citet{sadler14} at higher redshifts, $z \sim 1$.

Baldi et al. (in preparation) observed 25 FR0CAT sources at 15 GHz with the Arcminute Microkelvin Imager (AMI) telescope with an angular resolution of $\sim30\arcsec$, previously observed with VLA by \citet{baldi15} and \citet{baldi19}. The sources appear all unresolved and extend the spectral flatness of the FR0CAT SED at higher frequencies.

\citet{mikhailov21a,mikhailov21b} conducted  quasi-simultaneous radio observations of 34 FR~0s up to 22.3 GHz with the single-dish radio telescope RATAN-600 operating in transit mode with resolution varying from 11 to 80$\arcsec$. Quasi-simultaneous spectra in the range 2\,--\,8.2 GHz  are generally flat ($\alpha < 0.5$), but with a larger spread in the spectral index at higher frequencies. The key  result is that some FR~0s demonstrate a variability level of up to 25\% on a time scale of 1 year.

In the mm-band, a systematic study of FR~0s is still missing. Nevertheless, first studies on mm-band continuum observations of a sample of nearby ETGs and LLAGN found compact nuclear emission (on a scale 3--7$\arcsec$), showing flat or inverted spectra consistent with the scenario of small jets powered by RIAF discs (e.g. \citealt{doi11,marti17,chen23}). ALMA continuum observations of bright CRSs \citep{bonato18,bonato19,kawamuro22} reveal the presence of a minor population of flat-spectrum radio sources (possibly similar to FR~0s) in opposition to the abundant class of blazars.

\subsubsection{Low frequency}

Low-frequency observations ($<$ 1 GHz) have the advantage of probing the synchrotron-aged plasma and the optically-thin emission from an extended diffuse jet, crucial to test the duty cycles of FR~0s. Using the data release of the TIFR  (Tata Institute of Fundamental Research) GMRT Sky Survey (TGSS), \citet{capetti19} studied the low-frequency properties of 43 FR~0 galaxies  (FR0CAT, with 150-MHz flux densities $>$ 17.5 mJy) at 150 MHz at a resolution of $\sim25\arcsec$ (corresponding to 10 and 25 kpc).  No extended emission has been detected around the detected FR~0s, corresponding to a luminosity limit of $\lsim 4 \times 10^{23}\, {\rm W\, Hz}^{-1}$ over an area of 100 kpc $\times$ 100 kpc. The majority of the FR~0s have a flat or inverted SED (150 MHz\,--\,1.4 GHz, $\alpha < 0.5$): this spectral behavior confirms the general paucity of optically thin extended emission within the TGSS beam. By focusing on a sub-sample of FR~0s with 1.4-GHz flux densities $>$ 50 mJy and including 5-GHz data from the Green Bank survey  \citep{gregory96}, the authors found that $\sim$75\% of them have a slightly convex radio spectrum, with a smaller curvature than  powerful GPS sources. The typical FR~0 radio spectrum is better described by a gradual steepening toward high frequencies, rather than a transition from an optically-thick to an optically-thin regime as seen in young RGs.

Dedicated deep radio surveys on well-studied fields, such as ELIAS-N1,  have also detected large numbers of compact RGs: GMRT observations at 610 MHz \citep{ishwara20} and 325 MHz \citep{sirothia09} found CRSs with a median spectral index of $\sim$0.85 between 610 and 1400 MHz \citep{ishwara20}. The flat-spectrum sources, which are expected to be core-dominated, represent the FR~0 candidates.

The vast majority, $\sim$70\%, of the radio sources in the LOFAR Two-metre Sky Survey (LoTSS, \citealt{shimwell17,shimwell19,hardcastle19}) appear compact at 150 MHz with 6$\arcsec$ resolution, consistent with a FR~0 classification. \citet{capetti20} explored in details the LOFAR properties of the FR0CAT sources. Most of the objects still appear point-like structures with sizes of $\lsim$3--6 kpc.   However, $\sim$18\% of the FR~0s present resolved emission of low surface brightness, usually with a jetted morphology extending between 15 and 50 kpc. No extended emission is detected around the rest of FR~0s, with a typical luminosity limit of $\sim5 \times 10^{22}\, {\rm W\, Hz}^{-1}$ over an area of 100 kpc $\times$ 100 kpc. The spectral slopes of FR~0s between 150 MHz and 1.4 GHz span a broad range ($-0.7 \lsim \alpha \lsim 0.8$) with a median value of $\alpha \sim 0.1$; only 20\% of them have a steep spectrum ($\alpha \gsim 0.5$), which is an indication of the presence of diffuse emission confined within the spatial resolution limit. The fraction of FR~0s showing evidence for the presence of jets, by including both spectral and morphological information, is $\sim$40\%. 

In conclusion, the GMRT and LOFAR study of the FR~0s corroborates the result on the absence of  extended emission in most of the sources, even in the few hundred MHz regime, where optically-thin jet emission is expected to dominate over the core component, as seen in classical large-scale RLAGN.

\subsection{High resolution}

\begin{figure}
\centering
 \includegraphics[height=3.4cm]{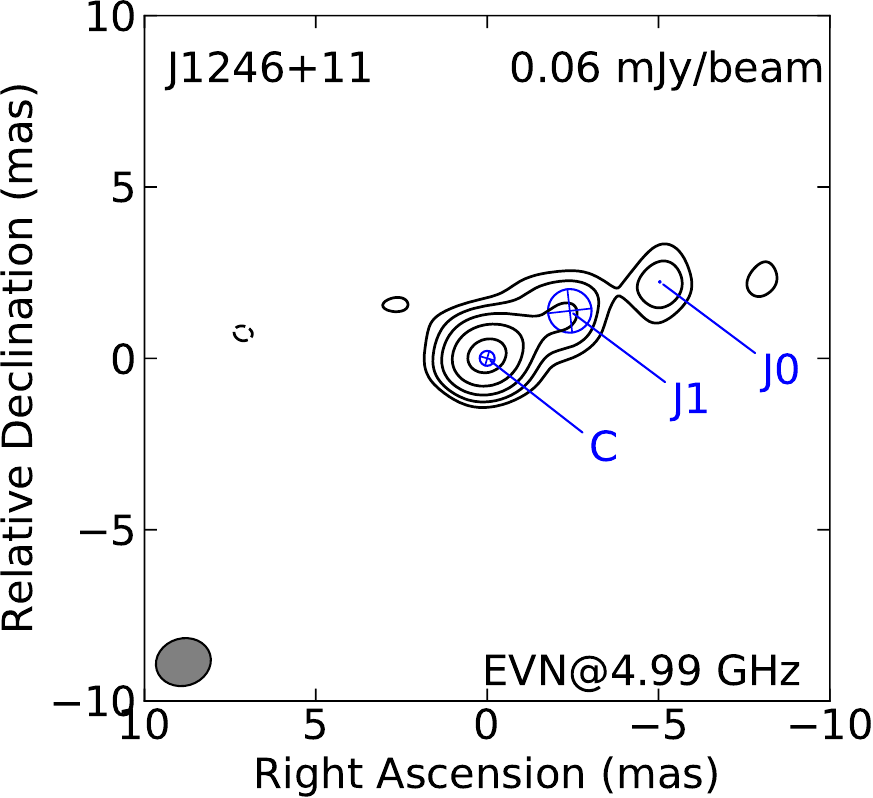}
 \includegraphics[height=3.4cm]{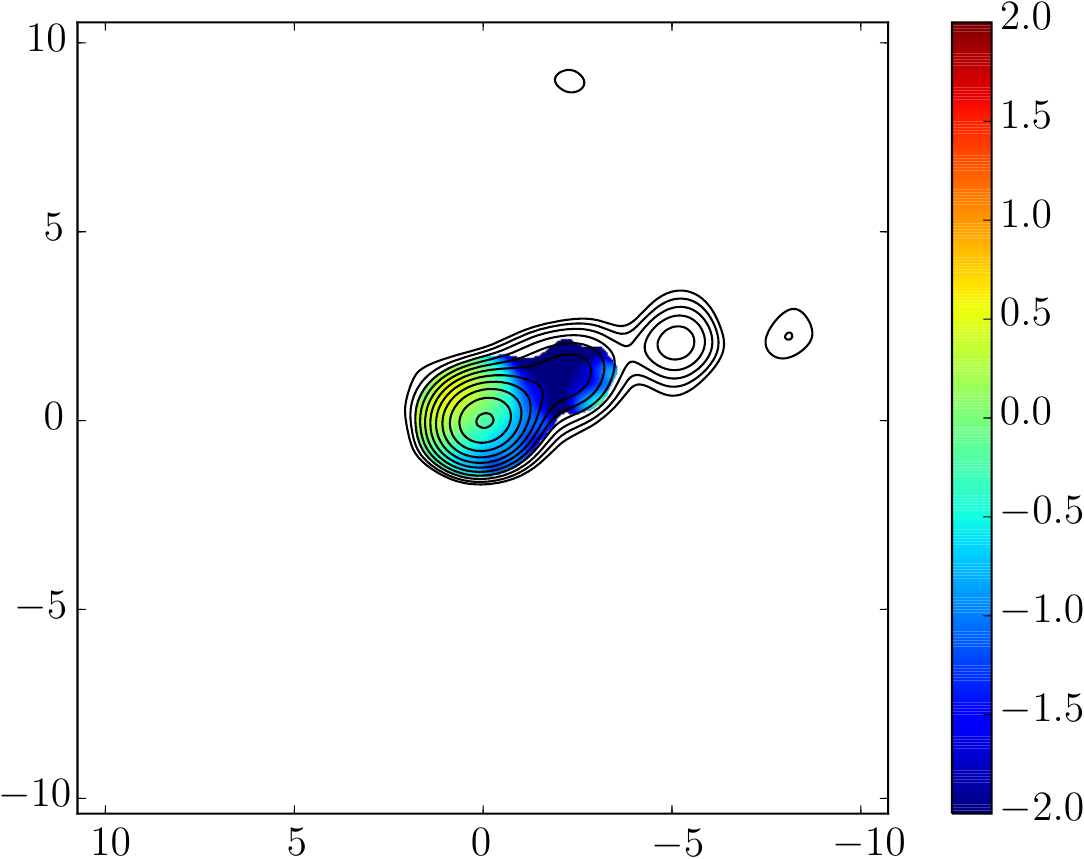}
 \includegraphics[height=3.4cm]{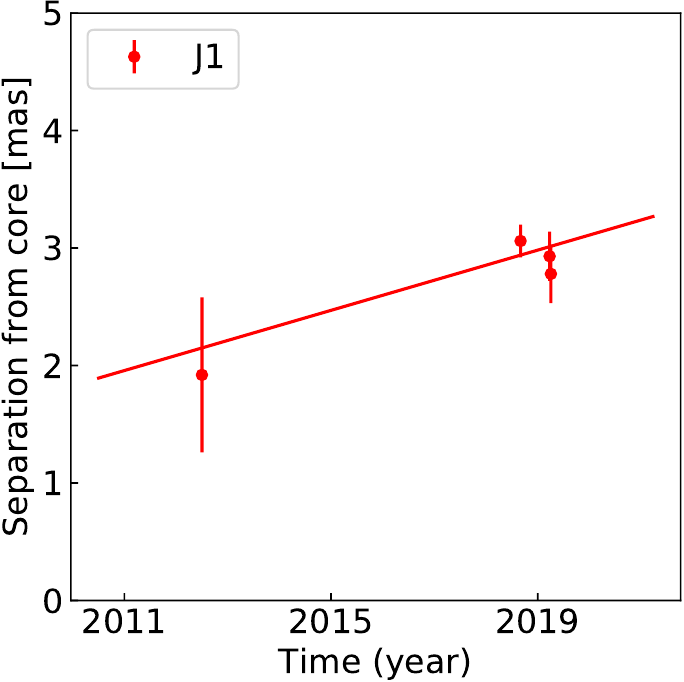}
\caption{EVN image at 5 GHz, spectral index map and jet proper motions of one FR~0 from \citet{cheng21} from left to right panel.  The grey-coloured ellipse in the bottom-left corner of each panel denotes the restoring beam. The spectral index maps (spectral index colour coded in the left palette) are obtained by using the EVN 5 and 8 GHz data. The proper motions of jet components are determined by the linear fit to the component positions as a function of time. Images reproduced with permission from \citet{cheng21}, copyright by the author(s).
}
\label{fig:VLBI}
\end{figure}

The VLBI technique enables to access to the pc-scale radio emission, a crucial region to study the jet properties of FR~0s closer to the launching site. \citet{cheng18} and \citet{cheng21} studied a sample of FR~0s with world-wide VLBI, the American Very Long Baseline Array (VLBA) and European VLBI Network (EVN)  and found resolved jets of a few pc for $\sim$80\% of the sample \citep{cheng18,cheng21} (see Fig.~\ref{fig:VLBI} as an example). The VLBI multi-epoch data and the symmetry of the radio structures indicate that the jet bulk speeds are mildly relativistic (between 0.08$c$ and 0.51$c$) with low bulk Lorentz factors (between 1.7 and 6) and large viewing angles. However, these VLBI-based studies focused on particularly bright FR~0s (flux densities $>$ 50 mJy, a factor 10 higher than the typical FR0CAT flux selection threshold, \citealt{baldi18}) with radio power $10^{23}$\,--\,$10^{24}\, {\rm W\, Hz}^{-1}$. Recent VLBI studies also target less luminous FR~0s. \citet{giovannini23}  studied pc-scale emission of 18 FR0CAT objects observed with the VLBA at 1.5 and 5 GHz and/or with the EVN at 1.7 GHz with flux densities a factor several lower than those of the FR~0s studied by \citet{cheng18} and \citet{cheng21}. All sources have been detected but one with radio core power down to $10^{21}\, {\rm W\, Hz}^{-1}$. Four sources remain unresolved at pc scale, while highly-symmetric jets have been detected in all other sources. High-resolution observations carried out with the eMERLIN UK-wide array  for a sample of 5 FR~0s at 5 GHz, reaching a resolution of $\sim$40 mas show sub-mJy core components \citep{baldi21c}. The pc-scale core emission contributes, on average, to 3--6\% of the total radio emission measured at kpc scale from NVSS maps, although an increasing core contribution for flat/inverted-spectrum sources is evident.  VLBI studies of FR~0s clearly demonstrate the jet-to-counter-jet flux ratios of FR~0s are significantly smaller that those of 3C/FR~Is \citep{baldi21c,giovannini23}, supporting the picture that jet bulk velocities in the FR~0s are lower (see Sect.~\ref{sec:a&e} for further discussion).

\begin{figure}
\centering
\includegraphics[width=5.5cm]{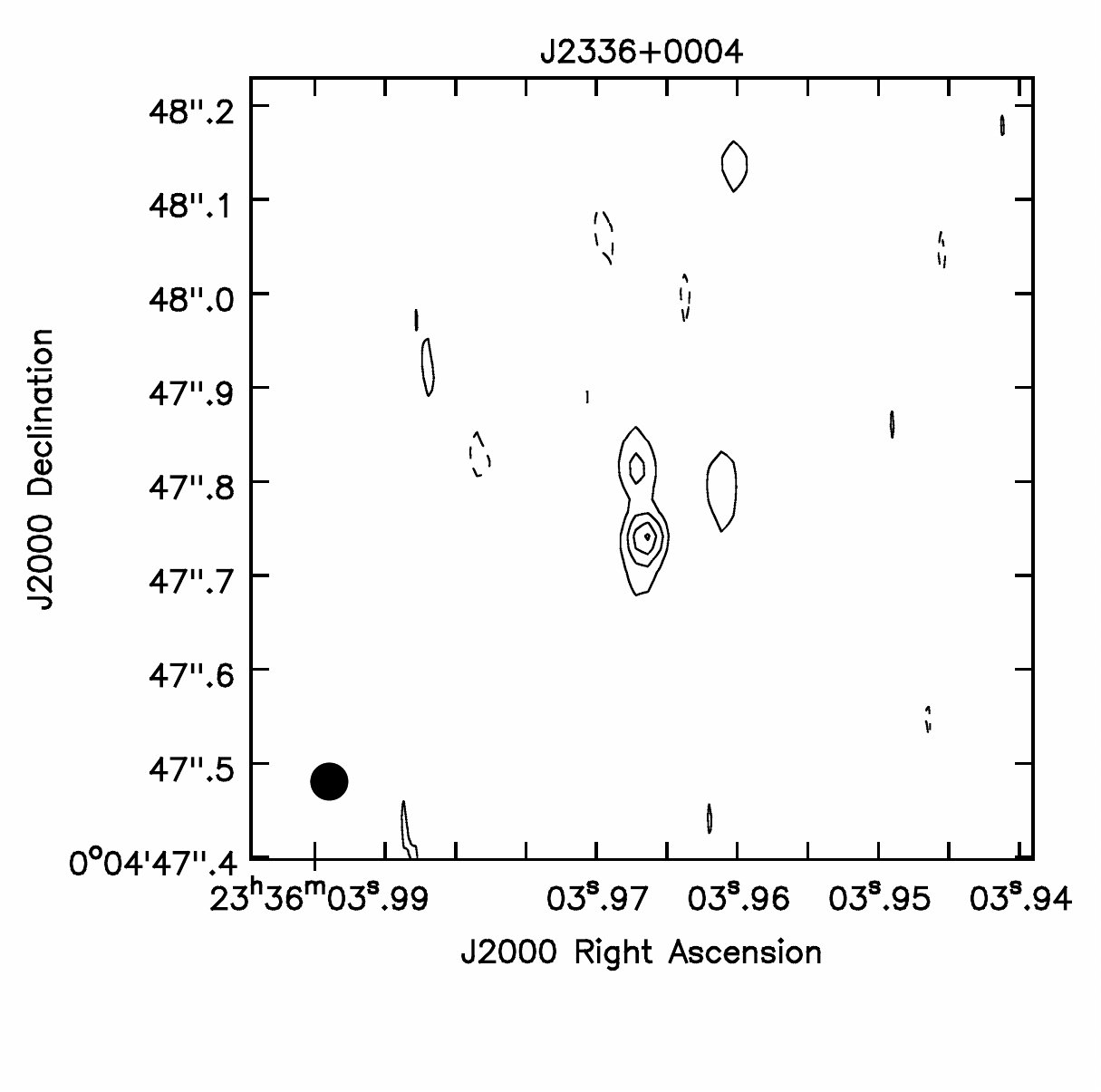}
 \includegraphics[width=5.5cm]{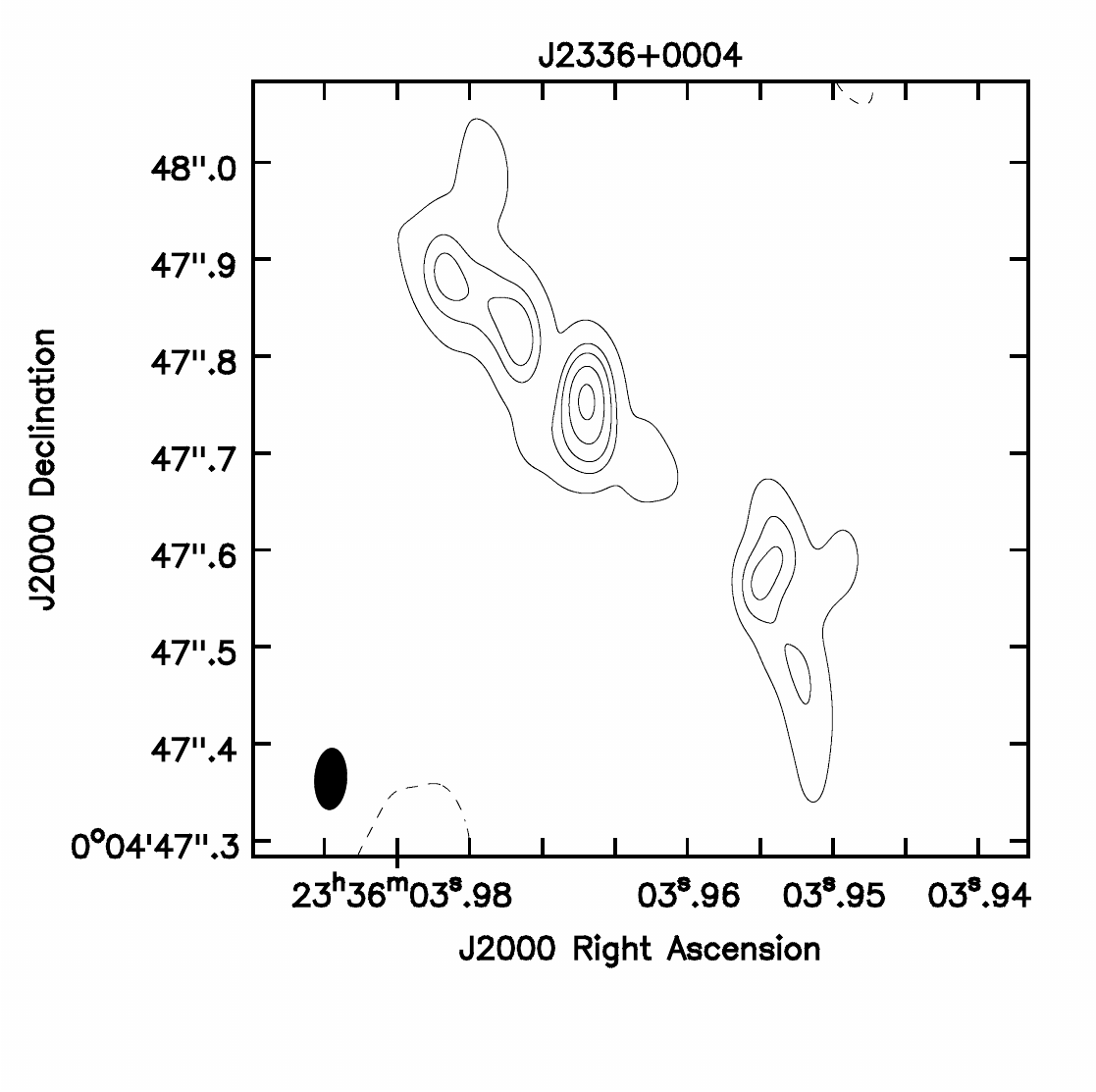}
\caption{The 5-GHz map of the one FR~0 (J2336+0004) observed with the eMERLIN array (resolution $\sim$40 mas) and its 4.9-GHz map (resolution $\sim$60 mas), obtained by combining eMERLIN and VLA visibilities. The filled area, shown at the bottom-left corner of the images, represents the restoring beam of the maps. Images reproduced from \citet{baldi21c}, copyright by the author(s).} 
\label{fig:comb}
\end{figure}

Apart from the cases ($\sim$30\%) where the VLBI core emission is higher than previous low-resolution data, possibly due to source variability and/or an inverted/peculiar radio spectrum, mas-scale radio emission is typically up to half of arcsec-scale core emission unresolved with VLA \citep{cheng18,cheng21, baldi21c,giovannini23}.  This suggests that a large fraction of emission is missed by moving from kpc to pc scale emission. \citet{baldi21c} combined, for the first time, the visibility datasets of the eMERLIN and VLA in the same band for five low-power FR~0s \citep{baldi15} in order to probe the intermediate scales of the jet length. This procedure turned out to be successful in detecting pc-scale jets for 4 objects, which were missing in the two original datasets (see Fig.~\ref{fig:comb} for an example) because unresolved in VLA maps and resolved out in the eMERLIN maps. We can thus conclude that FR 0s, although apparently lacking extended emissions, are effectively able to emanate pc-scale jets, whose both small size and low brightness  make them hard to isolate and detect. The combination of long and short baselines represents a powerful tool to study the jet properties of the FR~0 population.

In conclusion, VLBI studies of FR~0s reveal the presence of pc-scale jets, generally more symmetric than those of FR~Is, flowing with mildly relativistic jet bulk speeds. These results are in line with VLBI observations of nearby low-power LINERs (e.g. \citealt{ulvestad01b,falcke00,filho02b,nagar02b}).

\subsection{Radio SED}

\begin{figure}
\centerline{\includegraphics[angle=180,width=0.8\textwidth]{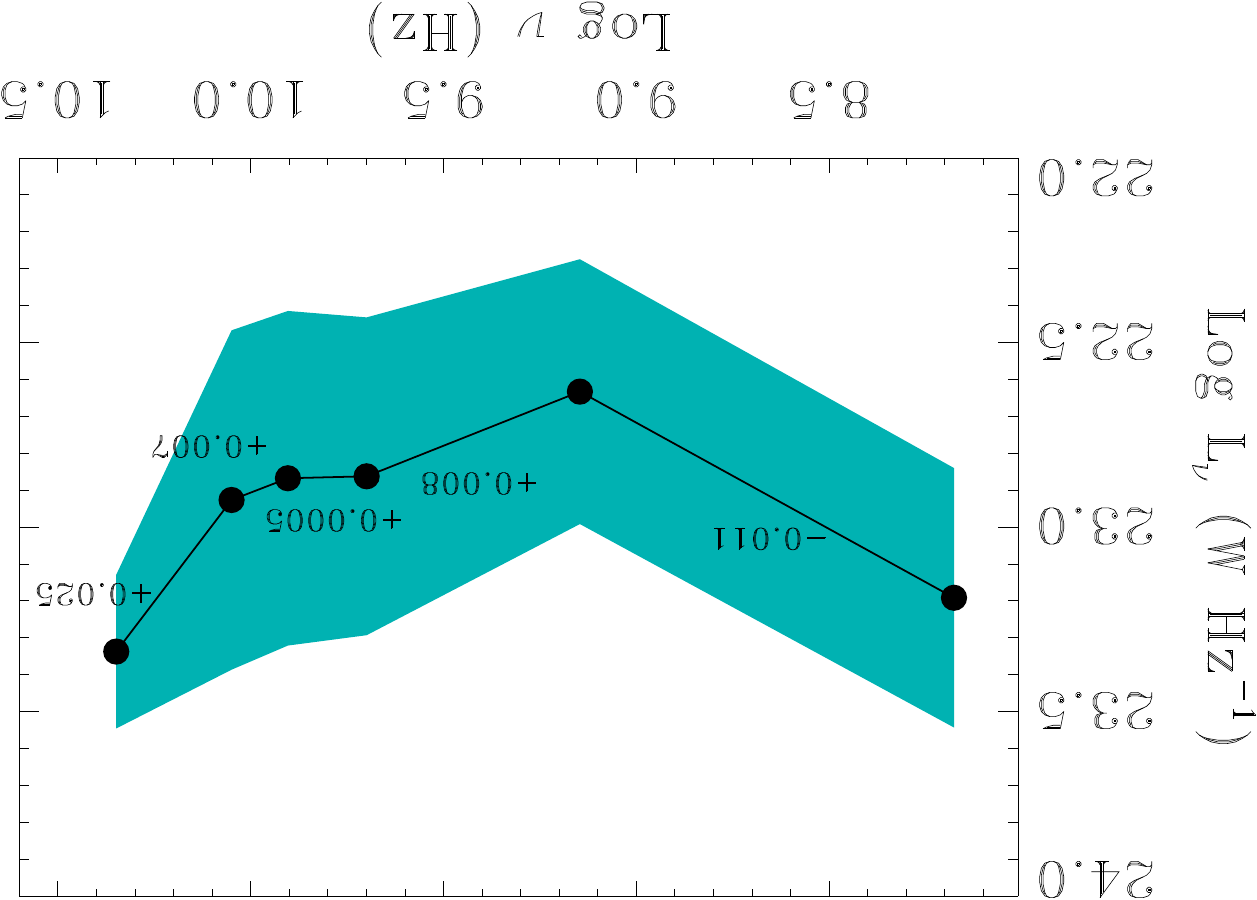}}
\caption{The mean radio spectra (L$_{\nu}$ vs $\nu$) of FR~0s from the FR0CAT \citep{baldi18} from 150 MHz to 22.3 GHz. The data are taken: 150 MHz from \citet{capetti19,capetti20}; 1.4 GHz from FIRST survey \citep{becker95}, 4.5--5 GHz from VLA data \citep{baldi19} and Green Bank 6-cm survey (GB6, \citealt{gregory96}); 7.5--8.2 GHz from VLA \citep{baldi19}  and RATAN-600 telescope \citep{mikhailov21a}; 11.2 and 22.3 GHz from RATAN-600 telescope \citep{mikhailov21a}. The colour filled area represents the 1 $\sigma$ distribution of the population. The numbers show the spectral indices in the 5 frequency segments (L$_{\nu} \sim \nu^{\alpha}$) and are all consistent with a flat spectrum.
\label{f:meansed}}
\end{figure}

To reconstruct the typical broad-band radio SED of a FR~0, we collect the multi-frequency radio data from MHz to GHz for the FR0CAT objects, available from low and high frequency surveys and single dish observations. Figure~\ref{f:meansed} depicts the mean radio SED (black solid line) from 150 MHz to 22 GHz with 1$\sigma$ dispersion (considering only detections). The main result is the overall flat spectral index ($-0.011 < \alpha < 0.025$), which confirms the general tendency of FR~0 population to be characterised by the lack of optically-thin component throughout the frequencies. The mean FR0 radio SED is flatter than the typical one derived for classical RLAGN, $\sim-0.6$\,--\,$-0.7$ \citep{elvis94},  even selecting the low-z sample of RLAGN \citep{shang11}. Non-thermal self-absorbed synchrotron emission from the basis of a core-dominated jet is most probably responsible to justify the observed spectral flatness.

At higher resolution, the (GHz-band) radio SED of the pc-scale cores is as flat as those derived from low-resolution radio observations (Fig.~\ref{fig:VLBI}, \citealt{cheng18,cheng21}). The jet components resolved with VLBI appear weak and have steeper spectra than those of cores, $\sim -1$\,--\,$-2$. This result confirms the small contribution of the extended optically-thin jetted emission to the total radio emission in FR~0s (i.e. high core dominance) and, indeed, sub-kpc scale jets can typically emerge from radio maps with hybrid angular resolution (e.g., combining short and long baselines) or with deep VLBI observations.

\section{Optical and infrared properties}
\label{sec:optIR}

In the optical band, the continuum and spectral information of genuine FR~0s is mostly limited to the SDSS data. For the FR0CAT host galaxies, the optical absolute magnitude distribution covers the range $-21 \lsim M_{\rm r} \lsim -23$, corresponding to masses $\sim10^{10-11}\, M_{\odot}$, consistent with massive ETGs, as also inferred from the infrared colours. Instead, from the nuclear point of view, a study of the optical and IR accretion-related emission of FR~0s, in analogy to what has been done with Hubble Space Telescope for nearby 3C/FR~Is \citep{chiaberge99,baldi10}, is still missing. The lack of a proper optical nuclear power estimate leads to the assumption of the optical galaxy emission as upper limit on the optical AGN.  Considering 5 mJy as radio flux cut from the FR0CAT sample, the radio-loudness parameter of the FR0CAT sources is at least $>$ 11.

\begin{figure}
\centerline{\includegraphics[width=0.65\textwidth]{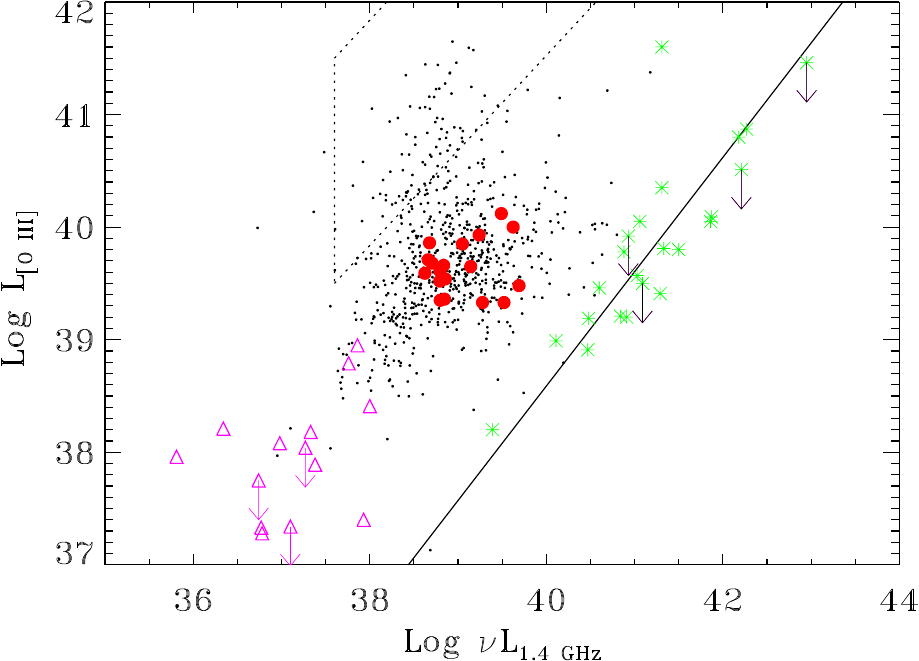}}
\centerline{\includegraphics[width=0.65\textwidth]{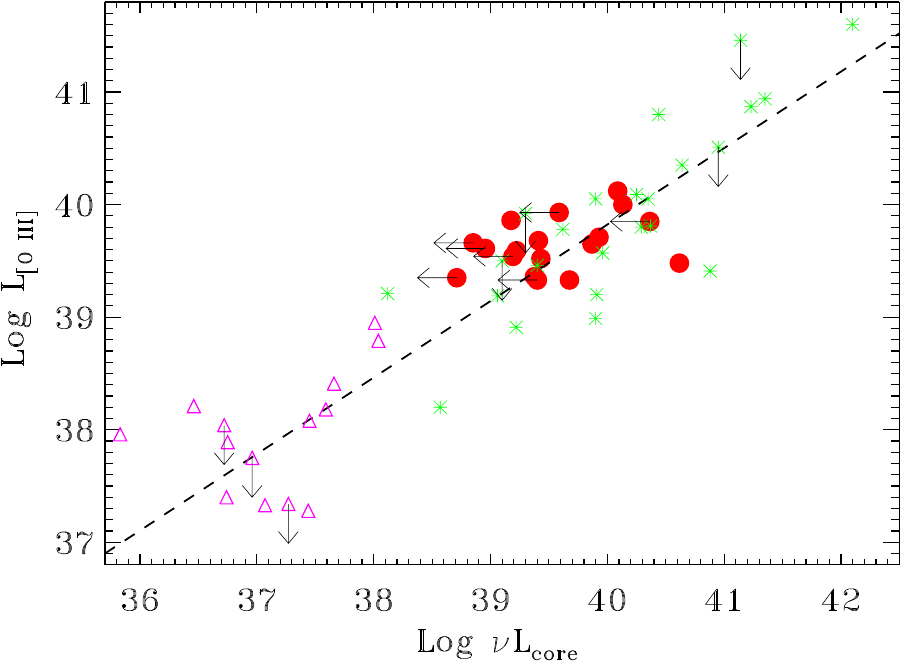}}
\caption{Upper panel: NVSS vs. [O~III] line luminosity (erg s$^{-1}$). The small points correspond to the SDSS/NVSS sample selected by \citet{best12}. The solid line represents the correlation between
line and radio luminosity derived for the 3C/FR~I sample (green stars) \citep{baldi19}. The dotted lines include the region where RQAGN (Seyferts) are found. The filled circles are FR~0s studied with the VLA by \citet{baldi19} and  the empty pink triangles are the CoreG.  Lower panel: VLA radio core (5 GHz) vs [O~III] line luminosity (erg s$^{-1}$) for 3C/FR~Is, FR~0s and CoreG and the dashed line represents their common radio-optical luminosity correlation.}
\label{fig:o3}
\end{figure}

An optical-band quantity which is widely used to characterise the AGN emission is the [O~III]$\lambda$5007 emission line, that is produced by continuum radiation from the accretion disc or jet which photoionises and heats the ambient gas. Since it is easily observed and largely available from SDSS spectra, its luminosity is usually used as a proxy of the bolometric AGN power \citep{heckman04} (see Sect.~\ref{sec:a&e} for details and caveats). While the line luminosities of FR~0s do not correlate with the total radio luminosities in analogy to CoreG, but in opposition to classical RLAGN  (upper panel, Fig.~\ref{fig:o3}), they do with the radio core luminosities, once the sub-arcsec core emission is resolved. In fact, FR~0s lie on the radio-line correlation valid for FR~Is and CoreG \citep{baldi15,baldi19} (lower panel, Fig.~\ref{fig:o3}). This common core-[O~III] relation valid for LERG-type RGs (FR~0s, FR~Is, FR~IIs) is generally interpreted as measurement of non-thermal radiation from the jet base at different bands (see Sect.~\ref{sec:a&e}, e.g. \citealt{hardcastle00,baldi19}). Since [O~III] line is mostly isotropic, this shared correlation implies that the radio compactness of FR~0-like RGs is not due to geometric effects and also sets an universal accretion-ejection coupling at the nuclear level for all LERG-type RGs. Similarly, \citet{miraghaei17} found that compact RGs have $L_{\rm [O~III]}$ distribution analogous to that of extended RGs, $10^{39}$--$10^{40}\, {\rm erg\, s}^{-1}$, when matched in radio core luminosities. Conversely, the total radio luminosity of the FR~0s and CoreG does not scale with AGN bolometric luminosity, as, instead, it is valid for LERG FR~I and FR~IIs \citep{buttiglione10}, but a strong deficit of total radio emission with respect to the 3C/FR~Is (not due to orientation), by a factor 100-1000 lower at the same AGN power, is notable. This shortage of total jet power suggests a lower jet efficiency of FR~0s than that of the other RLAGN classes (see Sect.~\ref{sec:a&e} for a deeper discussion).

The high detection rates  of optical and IR nuclei and the lack of evidence for thermal emission at IR wavelengths have been interpreted as the absence of a dusty torus in 3C/FR~Is and generally for LERGs (e.g \citealt{chiaberge99,leipski09,baldi10b,vanderwolk10,antonucci12,dicken14,tadhunter16a}). This scenario has also been applied to LINER-like LLAGN in general (FR~0s included), which find similar optical and IR characteristics of FR~Is (e.g. \citealt{ho08,muller13}), consistent with a luminosity-dependent model of a torus that disappears at very low accretion rates \citep{elitzur06,balmaverde15,gonzalez15}.

\section{High-energy properties}
\label{sec:HE}

The study of high-energy (HE, $>$0.1 keV) properties of jetted AGN can help to investigate the accretion and ejection mechanisms in action. The current and upcoming generations of HE detectors are revolutionising our picture of how the engines at the center of the RLAGN are able to launch plasma at relativistic speeds and extend their spectra to very-high energies (up to TeV, \citealt{rani19,rulten22}). In addition, the detection of HE emission and neutrinos associated with low-luminosity, 
 misaligned AGN and BL~Lacs (e.g., \citealt{abdo10b,icecube18b,icecube22,torresi20}) has opened a new window on the physics of particle accelerations and jets even in AGN with less extreme conditions than that expected in powerful blazars.

FR~0s, $\sim$4.5 times more numerous than FR~Is in the local Universe ($z<0.05$), represent potentially interesting targets at high and very-high energies (from X-ray to TeV) and could make a non-negligible contribution to the extragalactic HE background \citep{stecker19}.  Here we discuss the HE properties (from keV to TeV) of FR~0s, in analogy with the review by \citet{baldi19rev}.

\subsection{X-ray}

The X-ray emission represents an optimal proxy to study the accretion properties of active BHs, because the keV band can probe the HE photons  produced by the corona and disc. \cite{torresi18} performed the first systematic study in the X-ray (2--10 keV) band of a sample of 19 nearby FR~0s selected from \cite{best12}, for which X-ray data were available in the public archives of the \textit{XMM-Newton}, \textit{Chandra} and \textit{Swift} satellites.  Their FIRST 1.4-GHz flux densities ($>$30~mJy) are higher than those of the FR0{\sl{CAT}} sources. \citet{torresi18} found that the X-ray spectra of these FR~0s are generally well represented by a power-law $\Gamma \sim$ 1.9 absorbed by Galactic column density and do not require an additional intrinsic absorber, confirming the optical-IR results on the absence of a dusty torus, similar to 3C/FR~Is (e.g. \citealt{donato04,balmaverde06a}). In some cases, the addition of a thermal component is required by the data: this soft X-ray emission could be related to the extended intergalactic medium or to the hot corona typical of nearby ETGs \citep{fabbiano92}. The X-ray luminosities of FR~0s, $L_{\rm X}$, range between $10^{40}$ and $10^{43}\, {\rm erg\, s}^{-1}$, similar to those of 3C/FR~Is \citep{balmaverde06a,hardcastle00}.

\begin{figure}
\centerline{\includegraphics[width=0.63\textwidth]{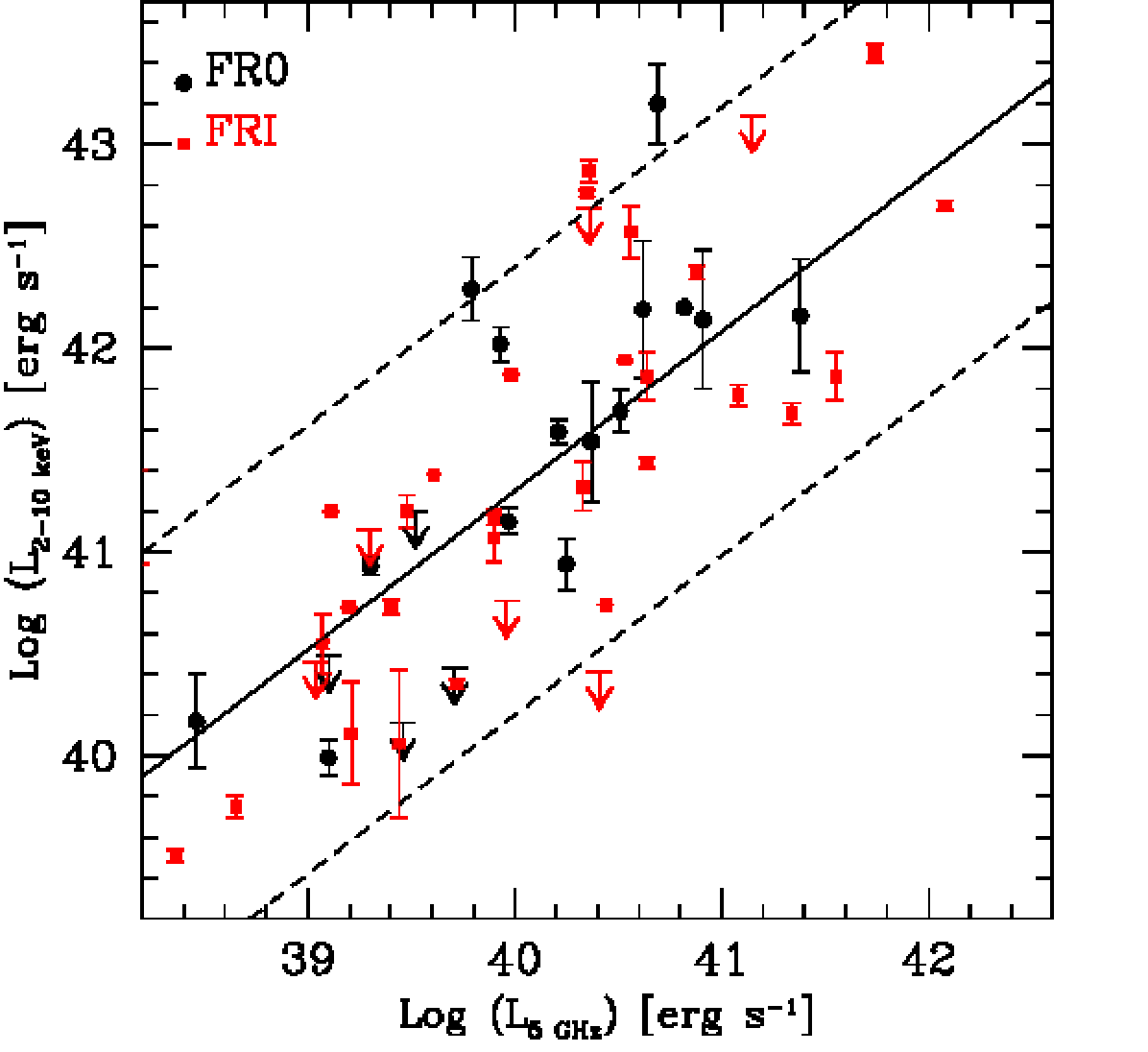}}
\caption{X-ray (2--10 keV) luminosity versus 5-GHz radio core luminosity for
FR~0s (black circles) from SDSS/NVSS sample and 3C/FR~Is (red squares). Arrows indicate upper limits. The black solid line is the linear regression for the overall sample of FR~0s and FR~Is, excluding the upper limits. The black dashed lines represent the 1$\sigma$ uncertainties on the slope. Image reproduced with permission from \citet{torresi18}, copyright by the author(s).}
\label{fig:lxlcore}
\end{figure}

When the X-ray luminosity is compared to that of the radio core, a statistically significant  correlation is established (Fig.~\ref{fig:lxlcore}), valid for FR~Is and FR~0s. This result corroborates the common interpretation that the X-ray emission in low-power RGs, FR~0s, FR~Is and LERGs in general, has a non-thermal origin from the jet (e.g. \citealt{balmaverde06b,hardcastle00,hardcastle09}). The X-ray luminosities of FR~0s also support the idea that the central engine of FR~0s is powered by a sub-Eddington RIAF-type disc, $\dot{L}_{E} \sim10^{-3}$\,--\,$10^{-5}$, analogous to 3C/FR~Is and different from powerful 3C/FR~IIs (HERGs) \citep{baum95,evans06,hardcastle09}. 
Since the study from \citet{torresi18} is slightly biased towards high-luminous FR~0s, a dedicated study of the accretion properties with deep Chandra data would be required for a statistical confirmation.

\subsection{Gamma-ray}

Gamma rays ($>$ 100 keV) are generally produced under extreme relativistic conditions and offer a unique view of the physical mechanisms in jet launching and propagation \citep{blandford19,hada19}. In such a band, blazars are known to be the most luminous class of $\gamma$-ray emitters and have been thoroughly studied \citep{fermi22}. Conversely, the HE properties of low-luminosity and misaligned AGN are generally less explored than their luminous counterparts, because of their lower flux densities \citep{abdo10b,angioni17,rieger18,demenezes20}. In fact, there are only a few cases of $\gamma$-ray detection of FR~0s in literature.

\begin{figure}
\centerline{\includegraphics[width=0.55\textwidth]{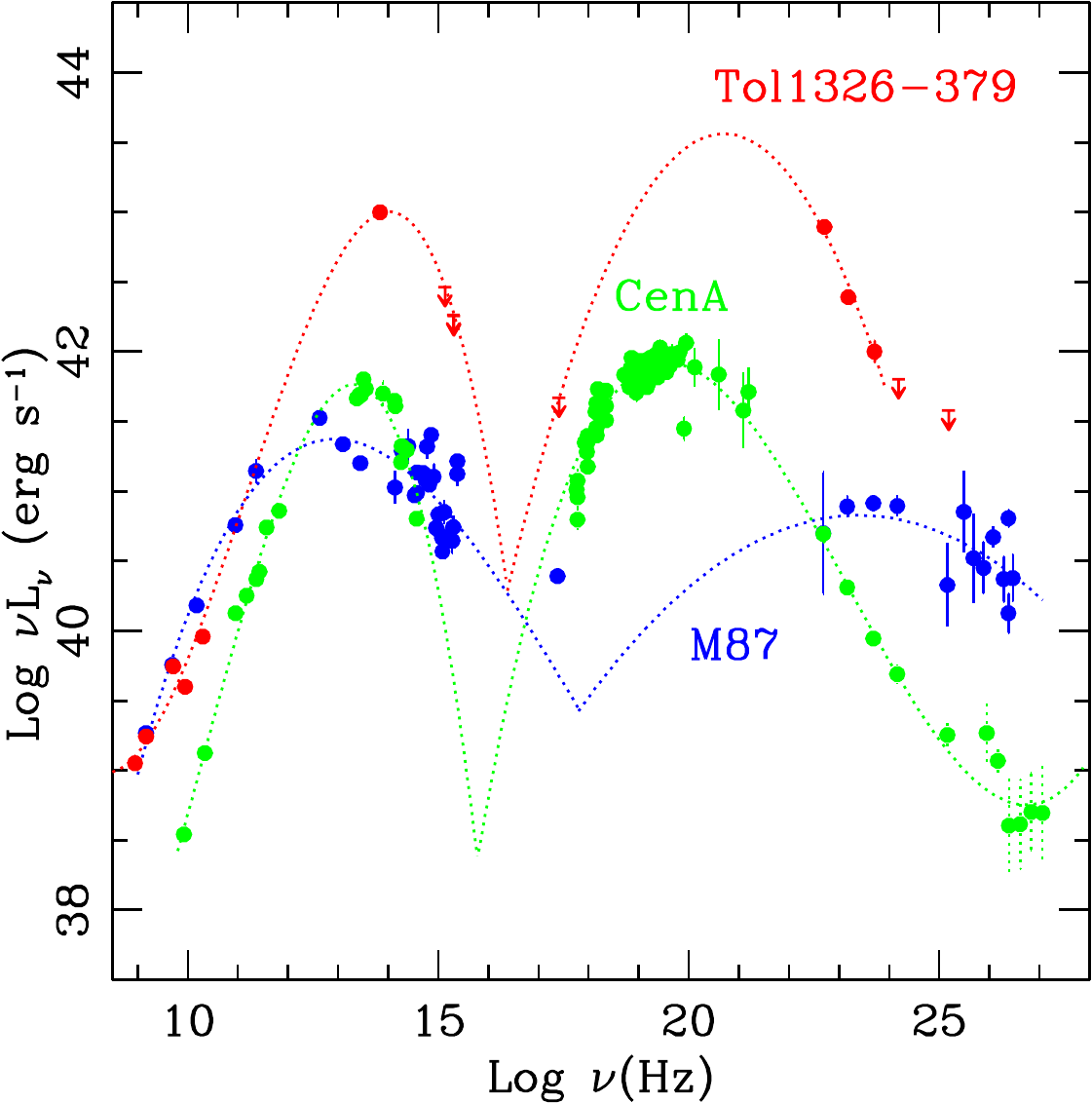}}
\caption{Multi-band SED (from radio to $\gamma$-ray) of the Fermi-detected FR~0, Tol~1326-379 (red symbols) compared to those of two nearby prototype FR~Is, HE emitters: Cen~A (green) and M~87 (blue). The dotted lines are polynomial functions connecting the data-points and do not represent model fits to data. Image reproduced with permission from \citet{grandi16}, copyright by the author(s).}
\label{fig:sedtol}
\end{figure}

\citet{grandi16} claimed the first Fermi $\gamma$-ray detection of a FR~0, Tol~1326-379, with a GeV luminosity of $2\times10^{42}\, {\rm erg\, s}^{-1}$, similar to FR~Is. Its radio-GeV SED is double-peaked \citep{maraschi92}, similar to other jet-dominated RLAGN (Fig.~\ref{fig:sedtol}, see the SEDs of M87, \citealt{abdo09}, and Cen~A, \citealt{abdalla20}), where non-thermal synchrotron and inverse-Compton emission dominate in any band over the disc and host emission. While the GeV luminosity reconciles with the detection of local FR~Is, the prominent Compton peak, brighter than the synchrotron one, makes this source similar to flat-spectrum radio quasars, while the steep $\gamma$-ray spectrum  makes it conversely more similar to low-luminosity BL~Lacs.  Nevertheless, the best scenario which can reproduce the whole SED is a misaligned RG which emits synchrotron and synchrotron self-Compton radiation with a total energy flux of the order of a few $10^{44}\, {\rm erg\, s}^{-1}$ \citep{grandi16}. Later, \citet{palyia21} reports the $\gamma$-ray identification of the other three FR~0s from the FR0CAT above 1 GeV using more than a decade of the Fermi Large Area Telescope (LAT) observations. By stacking present large datasets, other FR~0 candidates and compact core-dominated RGs have been recently claimed to be detected 
\citep{best19,demenezes20}. In addition, based on the sensitivities of upcoming MeV–TeV telescopes,  a significant population of low luminosity-RGs emitting at HE will be unearthed  in the near future \citep{baldi19rev,balmaverde20}. In fact, it has been estimated that nearby core-dominated RGs (FR0~s and CoreG) can account for $\sim$4\%--18\% of the unresolved $\gamma$-ray background below 50~GeV observed by the LAT instrument on-board \textit{Fermi} \citep{stecker19,harvey20}. Unfortunately, no evident FR~0s have been listed among the  non-blazar AGN list in the recently released Fourth LAT AGN Catalog (4LAC, \citealt{fermi20,fermi4}) and the $\gamma$-ray identification of Tol~1326-379 has also been questioned \citep{fu22}.

In addition, \citet{tavecchio18} proposed that FR~0s can accelerate HE protons in the jet and be powerful enough to sustain the neutrino production detectable by the IceCube experiment, above several tens of TeV \citep{jacobsen15}. \citet{merten21,merten22} argued that FR0 jets can generate ultra-high-energy cosmic rays through stochastic shear acceleration up to $\sim10^{18}$\,--\,$10^{19}$ eV \citep{lundquist22}. In opposition, \citet{mbarek21} argued that
the lower bulk Lorentz factors of FR0 jets than those of FR~I/IIs could disfavour their HE emission in general.

In conclusions, although HE studies on FR~0s are still sparse, the main result is that FR~0s and FR~Is share common X-ray and $\gamma$-ray properties, suggesting similar generic accretion and ejection phenomena in the vicinity of the BH (e.g. accretion disc properties and relativistic acceleration of particles at GeV energies in the jet).

\section{Accretion and ejection}
\label{sec:a&e}

Current magneto-hydrodynamic simulations have produced a wide range of accretion discs coupled with jets (e.g., \citealt{meier01b,ohsuga09,yuan14}). In the low-accretion regime (where $\dot{L}_{\rm E}$ is typically less than 2\% of the Eddington limit, \citealt{heckman14}), ADAF  discs are akin to launch jets \citep{narayan95}. An ADAF system can evolve under standard and normal evolution (SANE, e.g. \citealt{narayan12b}) and magnetically arrested disc (MAD, e.g. \citealt{bisnovatyi74,narayan03,tchekhovskoy11}) configurations: in the former the disc is not significantly threaded with  poloidal magnetic flux, while in the latter the magnetic flux threading the  BH horizon becomes so large that the magnetic pressure of the jet can temporarily stop the flow of matter into the BH. Current interest in MAD accretion is driven by the discovery that it leads to low and powerful relativistic jets. In fact, for M87, only strongly magnetized (MAD) disc models remain the most favourable solutions to reproduce the EHT results (e.g. \citealt{EHT21}).  This result strengthens the common interpretation that low-power RGs (generally FR~Is, such as M~87)  are probably powered by ADAF (MAD-type?) discs with low $\dot{m}$ and low radiative efficiencies, which channel a small fraction  of the disc plasma into the  relativistic jet (e.g., \citealt{nagar00,falcke00,ho02,hardcastle00,balmaverde06core,zanni07,ho08,balmaverde08,hardcastle09}). To study the accretion and ejection characteristics of low-power RGs, broad-band empirical relations have been used to gauge the disc and jet energetics.

For the accretion-related argument,  we must rely on various proxies for the bolometric AGN luminosity based on the radiation that is not fully obscured by the torus and escapes or is reprocessed.  For its large availability, the radiative bolometric luminosity or accretion power can be estimated from the optical [O~III] emission line, L$_{\rm Bol}$ = 3500 L$_{\rm [O~III]}$ (for LLAGN, \citealt{heckman04}),  as the AGN emission excites the gas clouds in the narrow line region, which re-emit [O~III] line almost isotropically. This quantity is a good, but not optimal, proxy since internal obscuration and stellar contamination can affect the measurement. L$_{\rm [O~III]}$ represents an upper limit on the accretion power for jet-dominated AGN, LERGs (generally not affected by nuclear dust obscuration), where jet shocks can cause [O~III] emission, instead of the underluminous RIAF disc \citep{capetti:cccriga}.

The AGN jets are observable through their synchrotron emission. The mechanical (kinetic) power of the jets, $L_{\rm Mech}$ has been estimated by using different assumptions. Monochromatic radio luminosity represents only a small fraction of the energy carried by the jets, about 2 orders of magnitude smaller than total $L_{\rm mech}$ \citep{scheuer74}. However, recalibrating this relationship with physical constraints (e.g. synchrotron spectral ageing,  radiative loss, content of particles and magnetic fields)  has yielded to 
\begin{equation}
 L_{\rm mech} = 7 \times 10^{36} f ( L_{\rm 1.4 \,GHz} /10^{25}\, W \,Hz^{-1})^{0.68} \, W  
 \label{eq4}
\end{equation}
 
 \noindent estimated by \citet{heckman14}. This relation was obtained by studying the  jet mechanical energy as  $pV$  work done by the jet to inflate cavities found in hot X-ray emitting halos \citep{rafferty06,birzan08,cavagnolo10}.  The jet  energy can also be estimated from synchrotron emission using the minimum energy condition in the radio lobes in an equipartition regime (i.e. the internal energy is almost equally divided
between magnetic field and relativistic particles) \citep{willott99,odea09,daly12}.  The $f$
factor includes all the uncertainties on the physical state of the lobes, such as for example 
the particle composition, SED, volume filling factor, possible deviation from the equipartition and adiabatic condition, turbulence, additional heating from shocks. \citet{heckman14} adopted $f = 4$ based on the best linear relation of the data. We note that these empirical assumptions, set on samples of FR~Is and FR~IIs, may not be entirely applicable to FR~0s \citep{grandi21}. However we choose to use this value to be consistent with previous works on low-power RGs (e.g. \citealt{heckman14}).

\begin{figure}
\centering{
\includegraphics[width=0.495\textwidth,height=5cm,angle=180]{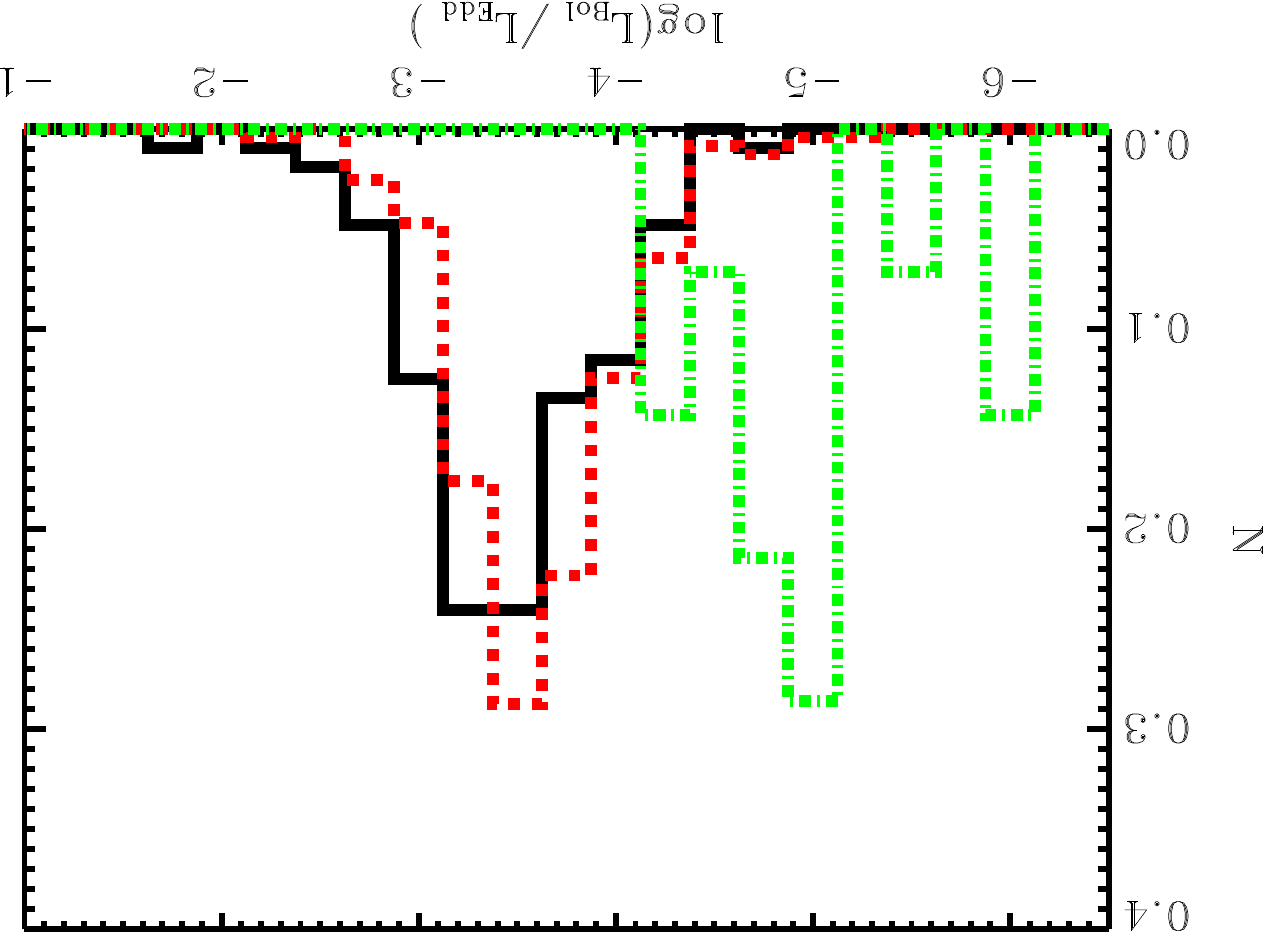}
\includegraphics[width=0.495\textwidth,height=5cm,angle=180]{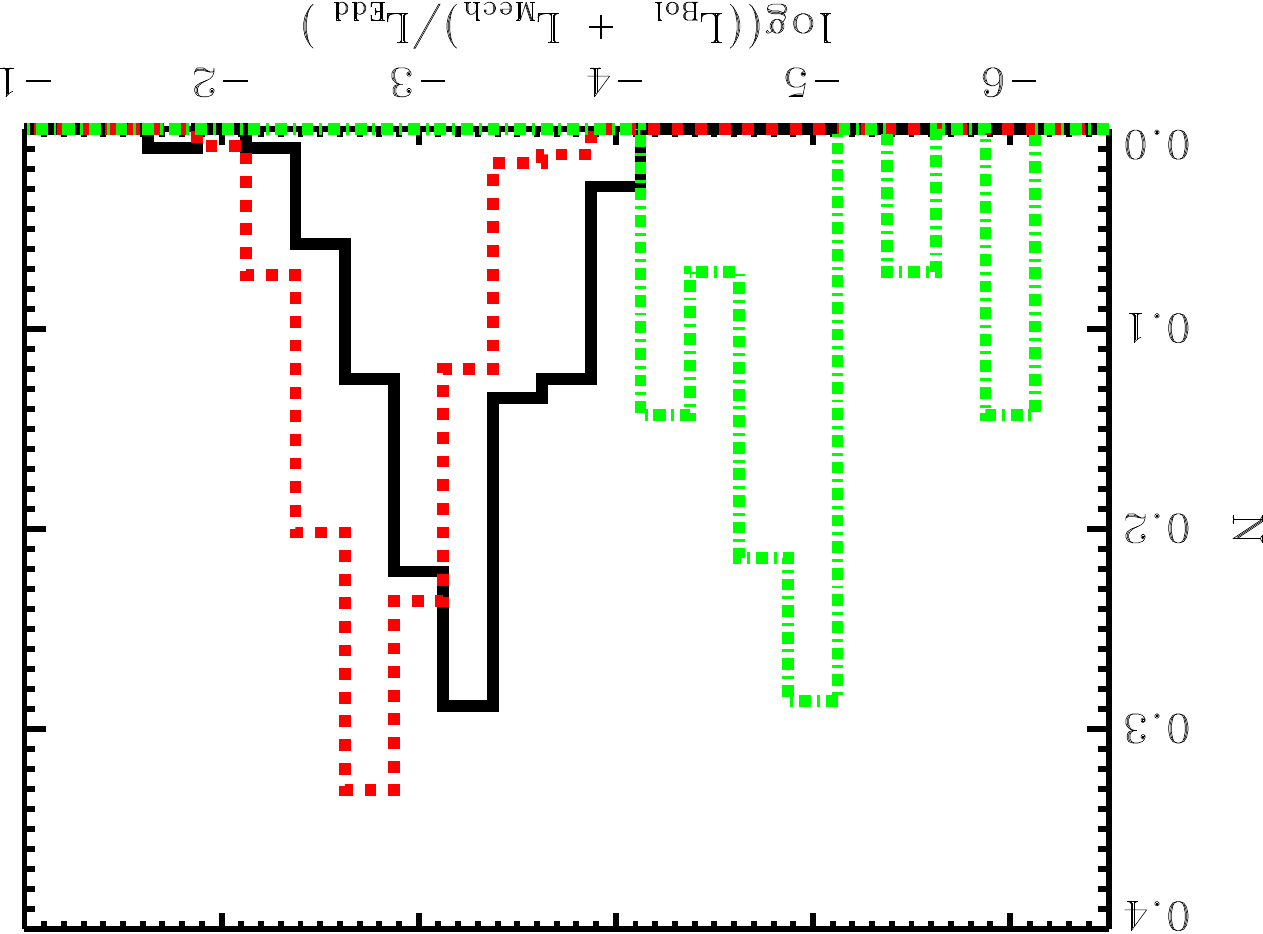}}
\caption{Histograms of BH accretion rates estimated as Eddington ratio, L$_{\rm Bol}$/L$_{\rm Edd}$ (left panel), and as total accretion rate, (L$_{\rm Bol}$ + L$_{\rm Mech}$)/L$_{\rm Edd}$ (right panel) for FR0CAT objects (black solid line), FRICAT objects (red dashed line) and CoreG (green dot-dashed line).}
\label{histo}
\end{figure}

Left panel of Fig.~\ref{histo} depicts the  Eddington ratio ($L_{\rm Bol}/L_{\rm Edd}$) distributions for FR0CAT, FRICAT and CoreG galaxies. FR~0s and FR~Is have similar rates, $10^{-5}$\,--\,$10^{-2}$. A Kolmogorov–-Smirnov (KS) statistic test confirms that two distributions are not drawn from different populations with a probability $P=0.0059$. Conversely, CoreG have significantly lower accretion rates $<10^{-4}$.

Since a large amount of the falling gas is launched into the jet without feeding the BH \citep{zanni07}, another method to estimate the total accretion is by adding the jet kinetic power to the radiative power as follows $\dot{L}_{\rm E, tot} = (L_{\rm Bol} + L_{\rm Mech})/L_{\rm Edd}$. The right panel of Fig.~\ref{histo} shows the distribution of this total accretion rate estimator for the different groups of
sources. The CoreG generally have lower total accretion rates than
the FR~0s and FR~Is. However, there is a considerable overlap
between the populations of FR~0s and FR~Is, $\dot{L}_{\rm E, tot} \sim 10^{-4}$\,--\,$10^{-2}$. A KS test confirms that the cumulative distribution function of FR0s is not significantly different from that of FR~Is ($P=5.0 \times 10^{-17}$). These results confirm the X-ray study from \citet{torresi18} that FR~0 BHs are fed at low rates, consistent with a jet-mode AGN and RIAF-type accretion states \citep{heckman14}. CoreG, being low-power FR~0s, also have lower accretion rates than FR0CAT objects.

\begin{figure}
\centering{
\includegraphics[width=0.75\textwidth,angle=180]{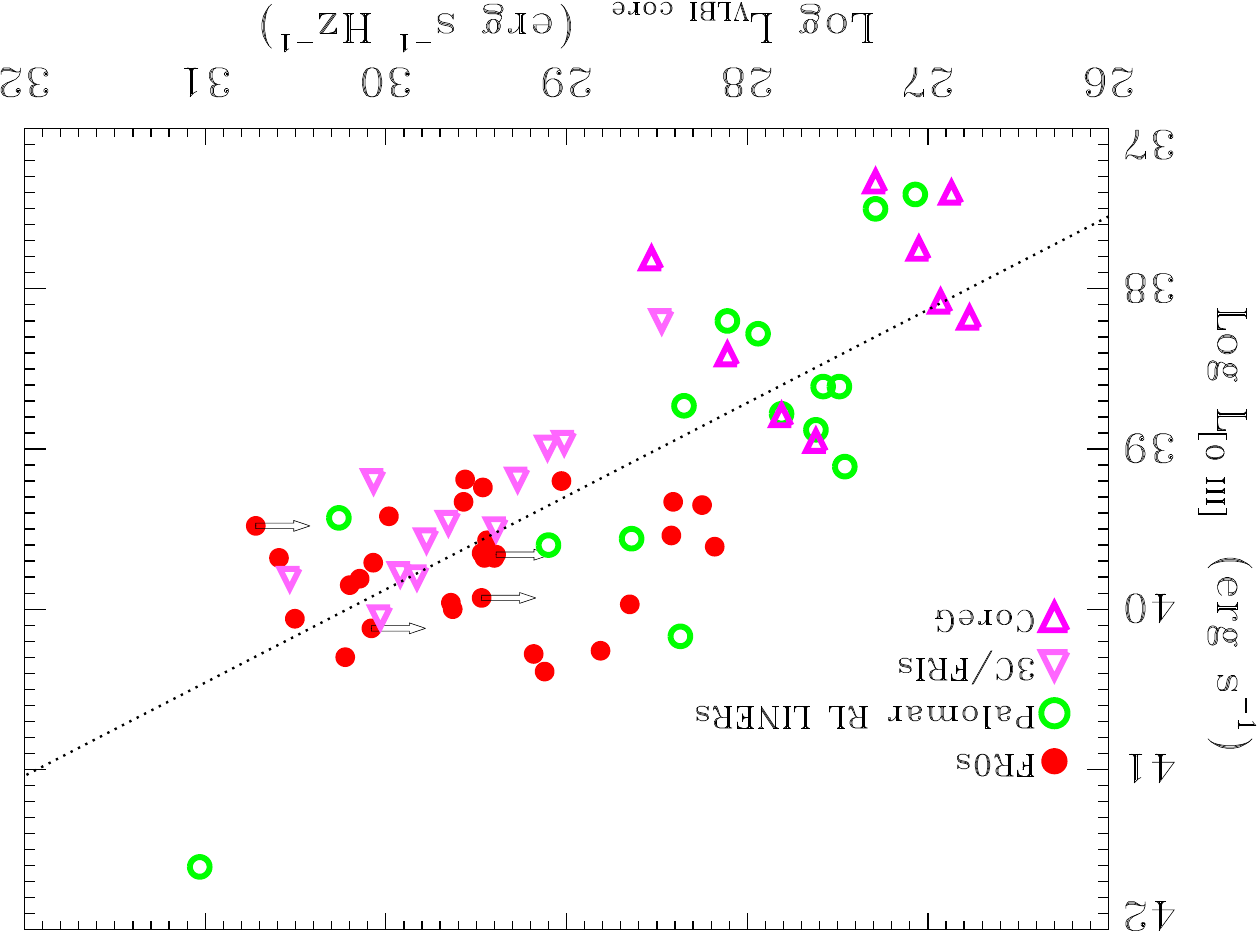}}
\caption{Parsec-scale core radio power (erg s$^{-1}$ Hz$^{-1}$) vs. [O~III] line luminosity (erg s$^{-1}$) for different samples of LINER-type RLAGN hosted in ETGs with core-brightened morphologies (see the legend): FR~0s (red filled dots) from \citet{cheng18,cheng21,baldi21c,giovannini23}, RL LINERs from the Palomar sample from \citet{ho95} (green empty dots),  
3C/FR~Is (upwards orange triangles), Core Galaxies (downward pink triangles). The dotted line indicates the best linear correlation.}
\label{lo3lvlbi}
\end{figure}

Broad-band proxies for accretion and kinetic jet powers are expected to broadly correlate in RLAGN, corresponding to two parallel empirical relations valid for the two accretion states (e.g. \citealt{rawlings91,willott99,buttiglione10,capetti23b}). For AGN-dominated RLAGN (HERGs),  the correlation between radio and optical (continuum) or X-ray emission  probably results from a combination of thermal and non-thermal emission from disc and jet (e.g., \citealt{chiaberge:fr2,hardcastle00,baldi19b}). For jet-dominated RLAGN (LERGs), the correlation between two luminosity proxies is best explained as the
result of a single emission process in the two bands\footnote{The caveat is that the optical and X-ray emission represent only an upper limit on the actual accretion power for LERGs.}, i.e. non-thermal
synchrotron emission from the relativistic jet (e.g., \citealt{chiaberge99,balmaverde06a,mingo14}), launched by a RIAF disc as supported by multiple theoretical and analytical studies (e.g. \citealt{meier01,begelman12,mckinney12}). This result has also been found valid for low-luminosity AGN, where compact jet dominates the broad-band continuum emission (e.g. \citealt{nagar02,ho08,ontiveros22}). \citet{balmaverde08} found that for 3C/FR~Is and CoreG the accretion power correlates linearly with the jet power, with an efficiency of conversion from rest mass into jet power of $\sim$0.012.  An [O~III]-radio correlation found for FR~Is, FR~0s, CoreG, and RL low-power LINERs (e.g. \citealt{verdoes02,nagar05,balmaverde06core,baldi15,baldi19,baldi21b}, see also Fig.~\ref{fig:o3}) suggests a similar ionising central source, where a scaled-down accretion rate for the core-dominated sources explains a likewise scaled-down jet power with respect to the more powerful 3C/FR~Is \citep{balmaverde08}. By focusing on the parsec-scale radio emission, the higher resolution of VLBI observation probes a section of the jet base `closer' to the launching site, which is thus more sensitive to the BH-accretion properties, than that detected with the VLA at arcsec resolution. In fact, an analogous $L_{\rm [O~III]}$--$L_{\rm VLBI\,\ core}$ correlation has been reported  by \citet{baldi21c} over $\sim$4 orders of magnitudes (Fig.~\ref{lo3lvlbi}) for RGs with comparable properties, e.g. hosted in massive ETGs and characterised by a LINER spectrum (FR~I, FR~0s, CoreG, RL LLAGN).   By also including the new VLBI data for FR~0s from \citet{giovannini23}, we fit the data points present in this sequence with a power-law relation. We find a robust correlation in the form $L_{\rm [O~III]} \propto L_{\rm VLBI \,\ core}^{0.58\pm0.06}$  with a Pearson correlation coefficient of 0.767 which indicates that the two quantities do not correlate with a probability smaller than $8\times 10^{-14}$. This statistically-robust relationship corroborates the idea that  the model of  RIAF disc with core-brightened jets of FR~Is is also applicable to  FR~0s and LINER-like RLAGN in general. The large scatter of the correlation, $\sim$0.28 dex, could be caused by Doppler boosting, nuclear variability and non-flat spectral index (1.4\,--\,8 GHz). However, there is no clear evidence for strong Doppler-boosted effect in FR~0s  (higher radio luminosities than implied by the linear correlation) for the one-sided jets or highly variable sources, suggesting that the jet spine is not highly relativistic and/or prominent (see Sect.~\ref{sec:models} for more discussion on jet structure).

\begin{figure}
\centering
\includegraphics[width=0.7\textwidth,angle=180]{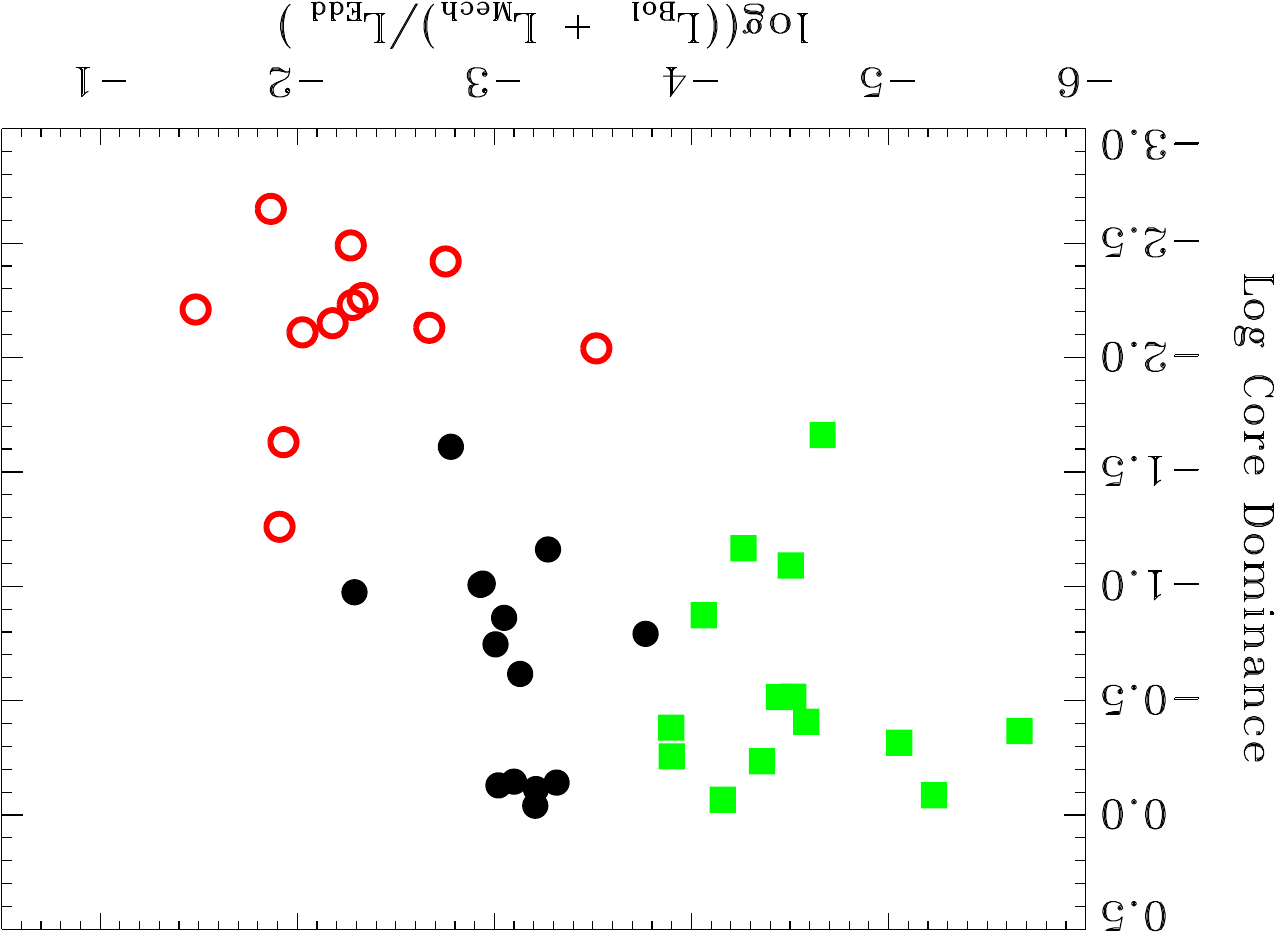}
\caption{The core dominance measured as ratio between VLA 5-GHz core and NVSS 1.4-GHz flux densities for FR0CAT sources (black filled dots), FRICAT sources (red circles) and CoreG (green squares) as function of total accretion rate (L$_{\rm Bol}$ + L$_{\rm Mech}$)/L$_{\rm Edd}$. We exclude sources with core dominance $>$1 because probably affected by variability or systematic errors.}
\label{cd_accretion}
\end{figure}

The accretion-ejection coupling can also be explored by comparing the core dominance, a proxy of the jet  brightness structure (i.e. how much the core shines over the extended jet emission), with the total accretion rate, $\dot{L}_{\rm E, tot}$. Figure~\ref{cd_accretion} presents the distribution of these two quantities for FR0CAT, FRICAT and CoreG galaxies (excluding the few sources with core dominance $>$1 possibly due to variability or systematic errors).  Although the core dominance naturally saturates at 1 as the source becomes weaker (and radio spectrum flatter, \citealt{debhade23}), there is a general tendency for RGs to increase their core dominance with decreasing accretion rate. These results suggest that the capability of a RG to develop kpc-scale structures is related to accretion properties: more core brightened structures are associated with lower-$\dot{m}$ sources.

\begin{figure}
\centering{
\includegraphics[width=0.65\textwidth]{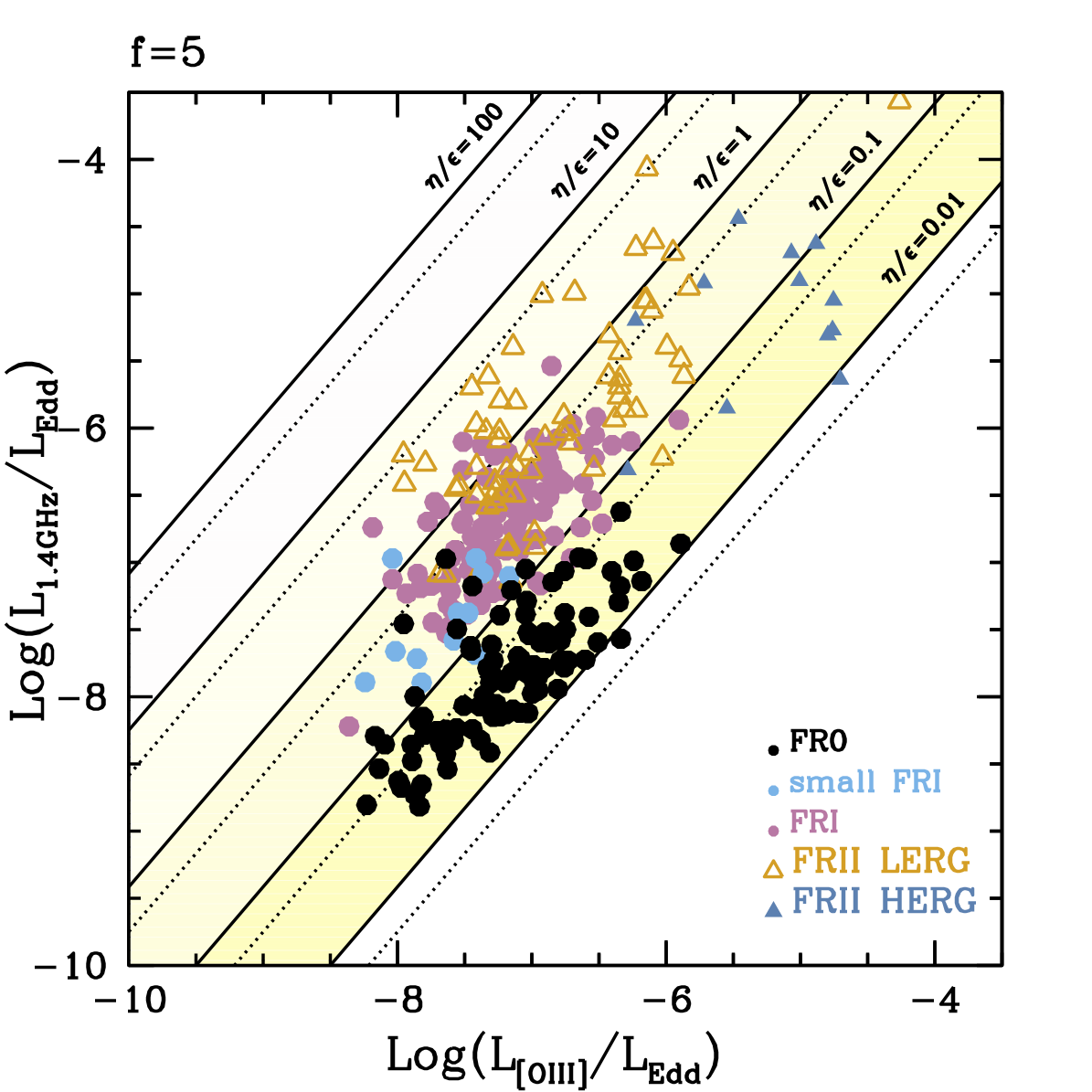}}
\caption{$L_{\rm 1.4 \,GHz}/L_{\rm Edd}$ versus $L_{\rm [O~III]}/L_{\rm Edd}$ of FRCAT sources (FR~0, FR~Is and FR~IIs) compared to the predicted values of kinetic jet power estimated by Equation~\ref{eq4}  assuming $f=5$. Each line in the plots corresponds to a different value of $\eta/\epsilon$. Since a change of the BH mass can have a minor impact on the predicted $\eta/\epsilon$ curves, we plot
(solid and dotted) lines corresponds to $M_{\rm BH}=10^{7.5}$ and $M_{\rm BH}=10^{9.5}\,M_{\odot}$. Image reproduced with permission from \citet{grandi21}, copyright by the author(s).}
\label{jeteff}
\end{figure}

The jet  efficiency, i.e. the fraction of the kinetic jet power produced with respect to the AGN  accretion power, offers a good diagnostic to investigate the nature of the nuclei of RGs.  The $L_{\rm Mech}/L_{\rm Bol} \sim \eta/\epsilon$ ratio ($\eta$ and $\epsilon$ are the fraction of gravitational energy converted into jet power
and thermal radiation, respectively) directly measures the ability of the system to channel gravitational energy into the jet rather than to dissipate it in thermal radiation. Figure~\ref{jeteff} depicts  $\eta$/$\epsilon$ for FR0CAT, FRICAT and FRIICAT objects \citep{grandi21}.  Neglecting the $f$ and $M_{\rm BH}$ effect on the jet efficiency, whereas HERGs favour
a thermal dissipation of the gravitational power, different LERG types, powered by similar inefficient accretion flows, launch jets with different luminosities and different jet efficiencies: FR~0s appear less efficient in extracting energy from the BHs into the jets than FR~Is.

\begin{figure}
\centering{
\includegraphics[width=0.7\textwidth]{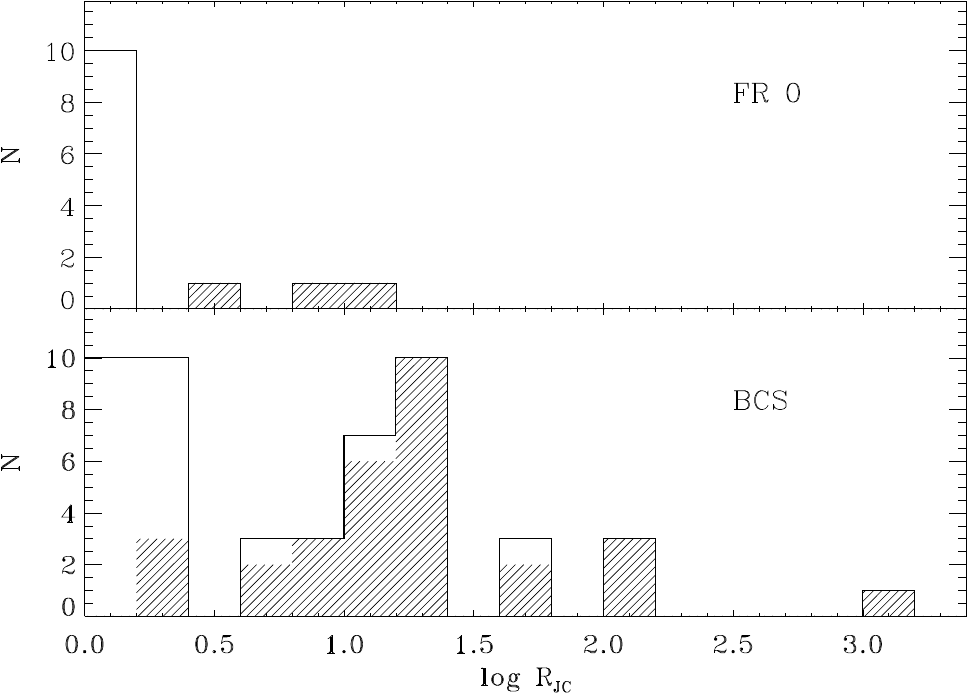}}
\caption{Distributions of the logarithm of the $R_{\rm JC}$, the jet-to-counter-jet flux ratio  for the FR~0  (top panel) from \citet{giovannini23} and FR~I/FR~II RGs from the Bologna Complete Sample (BCS, bottom panel, from \citealt{liuzzo09}). The dashed histograms correspond to lower limit on the jet sidedness. Image reproduced with permission from \citet{giovannini23}, copyright by the authors.}
\label{jetsided}
\end{figure}

At parsec scale, a comparison between FR~0 and 
 classical 3C/FR~I jets can help us understand the reason why FR~0s do not develop large structures. 3C/FR~Is generally exhibit core-brightened radio morphologies with VLBI observations \citep{fanti87,venturi95,giovannini05} and FR~0s show occasionally similar morphologies when resolved. However, the degree of jet asymmetry and the ratio between one-sided and two-sided jets appears different between the two classes.  In FR~Is, the effect of Doppler boosting on the jet sidedness, i.e. the jet-to-counter-jet flux ratio,  decreases from VLBI to VLA observations and is typically larger than 3  at parsec scale \citep{bridle84,parma87,giovannini90,venturi95,xu00,giovannini01}. On the basis of the presence of a link between jet speed and asymmetry, this result is interpreted as a change of the FR~I jet bulk speed from relativistic, $\Gamma >$3, to sub-relativistic speeds on kpc scales by decelerating, possibly due to entrainment of external material \citep{bicknell84,bicknell95,bownman96,laing14,perucho14}. For FR~0s the jet sidedness is less prominent:  only one
third of FR 0s has jet sidedness larger than 2 at parsec scale (Fig.~\ref{jetsided}). This is a clear observational evidence that the jet bulk speed of FR~0s is significantly smaller than that of FR~Is.  Following the procedure discussed by \cite{bassi18}, we can roughly estimate the bulk Lorentz $\Gamma$ factor of the jet, but with a strong assumption on the unknown orientation: the $\Gamma_{\rm bulk}$ for the angle of the jet to the line of sight $\theta_m$ that maximizes $\beta = v/c$. With these assumptions, considering the range of jet sidedness observed $<$ 10, $\Gamma_{\rm bulk}$ for FR~0s is typically $<$2.5.  This result concurs with the low jet proper motions studied by \citet{cheng18} and \citet{cheng21}. In conclusion, although FR~0 and FR~Is share comparable accretion properties, the jets of the former appear less efficient and slower, mildly relativistic at parsec scales with a bulk velocity which does not exceed 0.5$c$. However, a proper systematic analysis on larger samples of FR~0s is needed to draw a final conclusion on their accretion-ejection state.

\section{Environment}
\label{sec:environment}

The kpc- and Mpc-scale environmental properties (e.g. clustering, ICM, location within the cluster/group, relative galaxy velocity) can regulate the accretion and ejection states of an active BH:  e.g. bright cluster galaxies at the centre of dense environments typically host a RG and have different merger histories and fueling properties than galaxies at the cluster outskirts moving away from the centre (e.g. \citealt{lin10,vattakunnel10,shlosman13,kormendy13,conselice14}). The understanding of the relationship between RLAGN activity and their environment is essential for a comprehension of BH-host evolution, AGN triggering and life cycles, and for calibrating feedback processes in cosmological models (e.g., \citealt{husko23}). However, the role of the environment in shaping RLAGN is still not clear (e.g. \citealt{best04,ineson13,ineson15,ching17,macconi20}). The FR~I/II  dichotomy is  believed to depend on jet interaction with the environment (e.g. \citealt{laing94,kaiser97}), or due to host properties \citep{ledlow96}, apart from  mechanisms associated with jet production itself
(e.g. \citealt{meier01}).

At small scales, the similarity between the host types of FR~0s and FR~Is suggests that the galactic gas conditions between the two classes are rather comparable. Precisely, the smaller optical host masses of the FR~0s than those of FR~Is argue against the idea of a dense galaxy-scale environment which could cause the jet deceleration and disruption through the interaction with ISM \citep{kaiser07}. No evidence of a denser hot-gas halo with respect to that of FR~Is hosts, which typically permeates the atmosphere of elliptical galaxies, can be inferred from the sparse X-ray studies of FR~0s.

The large-scale environment is typically invoked to explain the deceleration and confinement of FR~I jets with respect to FR~IIs, since FR~Is typically reside in denser environment (denser coronae and richer groups/cluster, e.g. \citealt{prestage88,hill91,zirbel97,gendre13,laing14,massaro19,massaro20b}). Several studies on the Mpc-scale environment of RL CRSs have confirmed that they inhabit dense environment, but the presence of  environmental differences with respect to FR~Is have been questioned. \citet{torresi18} found that at least 50\% of the FR0s live in a dense X-ray environment,  which reflects massive dark matter halos in which these objects are embedded.  \citet{vardoulaki21}, studying the VLA-COSMOS Large Project, found that FR~I/IIs and compact AGN are found in all types and density environments (group or
cluster, filaments, field), regardless of their radio structures. \citet{miraghaei17} only found a marginal trend of RL CRSs in denser environments. In this direction, \citet{capetti20b} found that FR0CAT sources do indeed live in rich environment but with lower density by a factor of 2 on average, than FR~Is, and that about two thirds of FR~0s are located in groups containing $<$15 members. A similar result was found by \citet{prestage88} who argued that RL CRSs lie in regions of lower galactic density than extended sources. In addition, \citet{massaro20} concluded that nearby BL~Lacs share similar clustering properties with FR~0s, suggesting a common parental population.  In conclusion, there is growing evidence of an environmental difference (at least at large scales) between FR~0s and FR~Is (and extended RGs in general), which would imply a different cosmological evolution between the two classes.

\section{Feedback}
\label{sec:feedback}

AGN feedback comes in two flavours: quasar and radio (or maintenance) mode (e.g. see \citealt{croton06,best07,fabian12,bower12,heckman14,harrison17}). While the former mode is associated with powerful radiatively dominated AGN, i.e. quasars (and HERGs), associated with high Eddington ratios ($\dot{L}_{E}>$ 0.01), the radio mode is attributed to BHs with low accretion rates ($\dot{L}_{E}<$ 0.01, mainly LERGs). The latter releases most of their energy in the form of jets, preventing strong cooling flows in galaxy clusters (e.g. \citealt{fabian03}), and regulating the level of SF in their host galaxies (e.g. \citealt{best06}). It is only with the advent of deep multi-band radio surveys, with their combination of high sensitivity to both compact and extended emission \citep{shimwell17}
that we are now able to systematically study the effects of galactic-scale feedback from RL CRSs (e.g. \citealt{bicknell18}). In opposition, powerful
quasars have jets that rapidly ``drill''
through the ISM, depositing most of the energy in the
intergalactic medium. Observational evidence continues to mount that
lower-power ($L_{\rm 1.4\,GHz} \lsim 10^{24}\, {\rm W\, Hz}^{-1}$) jetted
AGN may have a significant impact on their hosts through jet-ISM
interactions on small ($\sim$1--10 kpc) scales, where the jets heat,
expel, or shock the ambient ISM, thereby altering the SF efficiency
(e.g., \citealt{nyland13,jarvis19,jarvis21,webster21a,webster21b,grandi21,venturi21}). State-of-the-art jet simulations (e.g. \citealt{sutherland07,wagner11,mukherjee16,mukherjee18a,bicknell18,rossi20,talbot22,tanner22}) provide further support to this scenario, demonstrating that lower-power jets are susceptible to disruption and entrainment, which increases the volume and timescale of the feedback, as well as the amount of energy transferred to the ISM ($>$ 5--10\% of bolometric power).

FR~0s, showing galaxy-scale jetted emission, could play a critical role in the radio-mode feedback. In fact, they are the best candidates to offer continuous energy injection into the ISM, although at low regimes ($\dot{L}_{E}< 0.02$), but fully inserted in the host and  with most of the energy deposited in the ISM, on smaller physical (galactic medium) scales than those (inter-galactic medium) affected by the full-fledged jets of FR~I/IIs. Nevertheless, the role of FR0s and the jetted population of LLAGN in general in the context of feedback has just started to be explored (e.g., \citealt{kharb23,krause23,gold23}).

\citet{vardoulaki21} showed a comparable radio-mode quenching of SF in the hosts of RL CRSs and of FR~I/IIs. In fact, while compact RGs can also be  found in  less massive hosts ($10^{9.5}$\,--\,$10^{11.5}\,M_{\odot}$) than FR~I/IIs, the former also have low specific SF rates and large time from the last burst of SF derived from SED fitting \citep{delvecchio17} similar to those of the latter. RL CRS hosts lie in cooler X-ray groups than extended RGs with average inter-galactic medium temperatures of $\sim$1 keV. Additionally, the older the episode of SF, the cooler the X-ray group in which RL CRSs lie, suggesting a SF shutdown by kinetic feedback.

A dense cold- or hot-phase in the ISM can increase the chances of detecting signatures of an active radio-mode feedback.  \citet{best00} showed that compact radio sources smaller than 90 kpc have emission line nebulae with lower ionization, higher luminosity, and broader line widths than in larger radio sources, consistent with shocks driven by the jets or outflows, typically observed in dust-shrouded young RGs . Low-luminosity jet can also carry enough power to shock and remove the cold/hot gas (e.g. \citealt{morganti19,morganti22,murthy22}), as demonstrated by some cases with observed  disturbed gas kinematics, absorption features and LINER-like line emission in compact sources (e.g. \citealt{holt08,glowacki17,baldi19b,tadhunter21}). The detection of X-ray cavities in low-power RLAGN ($<10^{23}\, {\rm W\, Hz}^{-1}$) demonstrates the ability of their jets to inflate bubbles in the hot-gas atmosphere \citep{birzan04,allen06}.  Nevertheless, ordinary FR~0s are not expected to drive strong outflows in dense ISM.

\begin{figure}
\centering{
\includegraphics[width=1.0\textwidth]{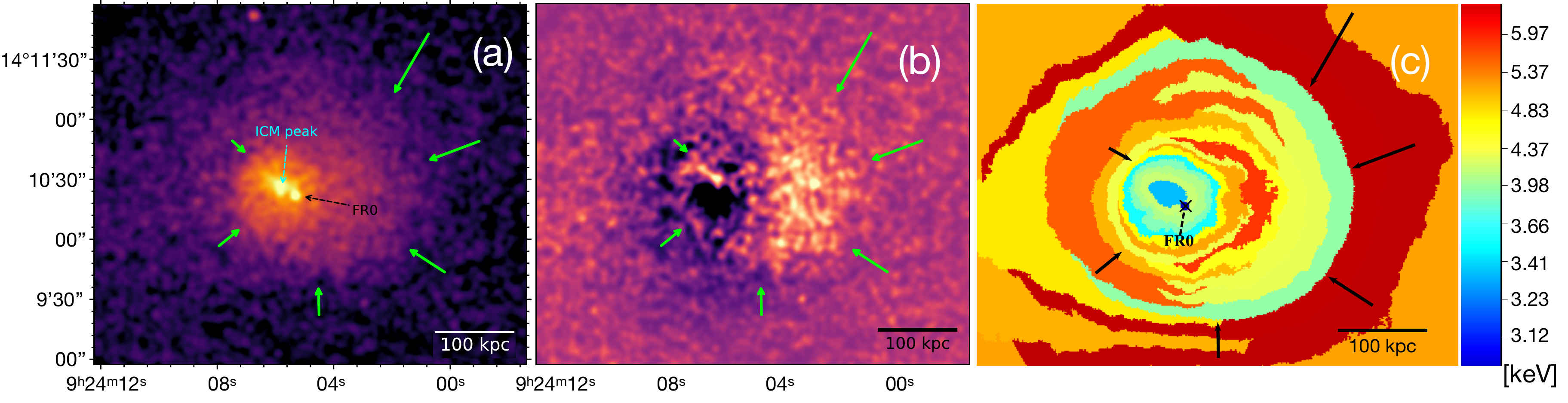}}
\caption{Panel a: Chandra image (0.5--2 keV) of the cluster A795; the ICM peak and the position of the FR~0 are indicated. Panel b: ICM brightness profile model subtracted to the  Chandra image over the same region of panel a. Panel c: Temperature map of A795. In each panel, the arrows highlight the ICM spiral geometry. Image reproduced with permission from \citet{ubertosi21b}, copyright by the author(s).}
\label{fig:A795}
\end{figure}

The first dedicated study which has observationally  addressed the radio-mode feedback for FR~0s is by \citet{ubertosi21}, who found two putative X-ray cavities and two prominent cold fronts possibly associated with the jet activity of a FR~0 (with a bolometric luminosity of the order of 10$^{40}\, {\rm erg\, s}^{-1}$), associated with a brightest cluster galaxy (cluster Abell~795) (Fig.~\ref{fig:A795}). The estimated cavity power and the cooling luminosity of the ICM follow the well-known scaling relations (e.g. \citealt{mcnamara07,mcnamara12}), providing a strong evidence for the self-regulated feedback in this source. Being fuelled by the inflow of a cold spiral-shape ICM, the central AGN inflate radio cocoons that excavate X-ray depressions and drive shocks in the ICM which slosh and heat gas, establishing a feedback loop. However, a systematic study of the feedback for a large sample of FR~0s is still missing, mainly because of the great difficulty to detect low-brightness X-ray cavities related to small jets.

To roughly estimate the impact of FR~0 jets on galaxies,  assuming that the X-ray atmosphere is regulated by the jet activity, we compare the internal energy within the jets with the energy of the hot X-ray emitting gas in the host, similar to the analysis performed by \citet{webster21a} for galaxy-scale jets. First, to calculate the jet energetics, we assume that the radio emission comes from a cylindrical region of 5$\arcsec$ long with a radius of 0.3$\arcsec$ (radio observation based, \citealt{baldi19}). By using a Python code (pysynch\footnote{\url{https://github.com/mhardcastle/pysynch}}, \citealt{hardcastle98c}) we derive the minimum energy density and the minimum total energy, which is of the order of $\sim5\times10^{48}$\,--\,$10^{50}$ J, by considering radio flux densities between 5 and 500 mJy, consistent with FR0CAT sources. Second, to estimate energy within the hot ISM, several assumptions are needed. Since the small jets of FR~0s should have a larger impact on the bulge, we estimate the bulge mass from the BH mass distribution of FR0CAT, $10^{7.5}$\,--\,$10^{9}\,M_{\odot}$, using the \citet{mcconnell13} scaling relation for ETGs. Then we fix the hot gas mass fraction to 5\% (e.g., \citealt{dai10,trinchieri12}).  Then assuming an average particle mass of $0.62\, m_{\rm proton}$ and a typical gas temperature of 0.5 keV \citep{goulding16},  we are able to estimate the internal energy of the hot phase in the bulge, which is of the order of $\sim10^{50}$\,--\,$3\times 10^{51}$ J. Finally, the  total jet energy of FR~0s turns out to be $\sim$3--5\% of the total binding energy of the bulge. However, bear in mind that  minimum jet energy estimates represent only a lower limit, since the jets must also displace the ISM and produce shocks and its enthalpy for a relativistic gas undergoing adiabatic expansion could be $> 4 pV$ \citep{birzan04,croston07,hardcastle13}. In addition, the internal estimated jet energy could be lower than the kinetic jet energy, which can be calculated by using the  method by \citet{willott99} by using the 151-MHz luminosity. In fact, by considering the LOFAR 150-MHz luminosity of the FR0CAT, $10^{38}$\,--\,$10^{40}\, {\rm erg\, s}^{-1}$, the jet output is of the order $1 \times 10^{43}$\,--\,$4 \times 10^{44}\, {\rm erg\, s}^{-1}$. This evaluation also considers the uncertainties on the factor $f$ ($<$20, \citealt{hardcastle07} for FR~Is),  which includes the effect from the jet structure and its environment \citep{willott99}. Assuming a lifetime of the jet activity of 10$^{7}$ yr, the kinetic jet energy would range $\sim3 \times 10^{50}$\,--\,$6\times 10^{51}$ J. These can be considered as upper limits of the jet energetics. In this case, the ISM energy would balance jet energetics. We conclude  that the FR0 jets are potentially capable of
affecting the ISM properties, at least in  the bulge.

Current hydrodynamical simulations (Horizon-AGN, \citealt{dubois14b}; Illustris, \citealt{vogelsberger14}; EAGLE, \citealt{schaye15}; MUFASA, \citealt{dave16}; SIMBA, \citealt{dave19}; SWIFT, \citealt{schaller23}) implement quasar- and radio-mode feedback with a typical efficiency of 5-10\%, assuming that the energy deposited back into the ISM, scales directly with accretion rate. The ratio  (L$_{\rm Mech}$)/(L$_{\rm Bol}$+L$_{\rm Mech}$)  provides a measure
of the fraction of the total accreted energy released back into the ISM
in mechanical form in jets. We measured this ratio for FR~0 and FR~Is from FRCAT and  found
that all deposit more than 10\% (on average 30\% for FR~0s) of their accreted energy back into the galaxy. This calculation confirms the result from \citet{whittam18,whittam22} that LERGs in general have higher feedback efficiencies and thus thought to be more
responsible for the maintenance mode of mechanical feedback  than HERGs, which, as more powerful FR~IIs on average than LERGs, generally deposit their energy at larger distances in the ICM.

\section{Comparison with FR~II LERGs}
\label{sec:friilerg}

Deep optical-radio surveys have unearthed a large population of low-luminosity FR~II LERGs \citep{capetti17b,jimenez19,webster21a}, which show kpc-scale edge-brightened radio morphologies, smaller ($>$30 kpc) and  less luminous ($\sim10^{41}\, {\rm erg\, s}^{-1}$) than the Mpc-scale powerful 3C/FR~II LERGs. Their nuclear properties (luminosity, accretion rates) can still be reproduced by a RIAF disc, consistent with the general jet-mode LERG population \citep{heckman14}. \citet{macconi20} suggested that FR~II LERGs are characterised by intermediate properties between  FR~Is and FR~II HERGs, since they populate an intermediate region of a correlation between accretion rates and environmental richness. Conversely, \citet{capetti23b} found FR~II LERGs are  among the most luminous radio sources in the Universe (up to radio power 10$^{35}\, {\rm erg\, s}^{-1}$ Hz$^{-1}$). \citet{tadhunter16a} argued that FR~II LERGs represent a phase of  the RG evolution, when the accretion has recently switched off or leveled down from an FR~II HERG high state, after exhausting the cold gas.  However, preliminary studies on the properties of the warm ionised and  cold molecular gas in RGs \citep{balmaverde19,torresi22} possibly rule out the presence of a  statistical difference between FR~II LERGs and HERGs, weakening the evolution scenario and, instead,  suggesting that jet properties in powerful FR~IIs do not depend on the accretion mode or the disc structure \citep{capetti23b}.

The connection between FR~0 and FR~II LERGs is established by their common affinity with FR~Is, since they all share a LERG optical spectrum and are generally interpreted to be jet-dominated RGs powered by a RIAF disc. In fact, \citet{baldi18} envisaged that FR~0s, FR~Is and FR~II LERGs belong to a single  continuous population,  with similar BH mass, galaxy and accretion properties, regardless of their different jet morphologies. Differences related to intrinsic intimate BH properties (spin and magnetic field at its horizon, and marginally different BH mass) shape the whole LERG population \citep{miraghaei17,grandi21}: when these parameters are maximized,  highly relativistic jets are launched and form full-fledged FR~I/FR~II LERGs, while FR~0s would originate from less
extreme values of these parameters (see \citealt{baldi18} and Sect.~\ref{sec:models} for discussion).

\section{Models for FR~0s}
\label{sec:models}

Here we will discuss two possible scenarios to account for the multi-band results on FR~0s  where the jet and nuclear properties of FR~0s 1) are intrinsically different from those of the other FR classes and do not evolve; 2) evolve within a context of RLAGN population where FR~0 represents a particular phase of this evolution.

\subsection{Static scenarios}

In a non-evolutionary scenario, where the intrinsic properties of the FR~0 class remain unchanged across their lifetime, we will review the main features which can determine the accretion and ejection in FR~0s in relation to FR~Is.

Magneto-hydrodynamic simulations of jet launching (e.g. \citealt{mckinney04,hawley06,mckinney06,tchekhovskoy11}) predict the formation of a light, relativistic outflow powered by the rotational energy of the BH, as described in the work of \citet{blandford77} (BZ),
as well as of a heavier and mildly relativistic outflow powered by
the accretion disc, as originally proposed by \citet{blandford82} (BP). LERGs, which are jet-dominated sources, are generally interpreted as BZ powered, while HERGs, which have quasar-type discs, are generally interpreted as powered by BZ and BP for the presence of both relativistic jets and strong outflows \citep{heckman14}.  FR~0s as well as FR~Is  are expected to launch BZ jets (with a possible contribution from a BP process for the outer jet layer in case of a stratified jet, see below).

For RLAGN jets  generated by BZ-type process in RIAF discs  \citep{tchekhovskoy11,liska22}, the ratio of jet and accretion powers (jet efficiency) is maximum when the BH is
both rapidly spinning and has accumulated a substantial amount of
large-scale poloidal magnetic flux by accretion (see e.g. \citealt{komissarov01,tchekhovskoy10}). The BZ jet power does not directly depend on the accretion rate, but the outflowing plasma is surely a fraction of the accreting flow. As discussed in Sect.~\ref{sec:a&e}, although they can share similar accretion rates, core-dominated RGs and FR~0s show a less jet efficiency than more powerful FR~Is. The small fraction of plasma within the  disc that is actually channeled into the jet could justify the paucity of matter to accelerate to relativistic speeds in the FR~0 jets.

The BZ jet power depends on $M_{\rm BH}$, the magnetic field strength $ B$ threading the BH and the magnitude of its spin  $\bar{a}$ \citep{chen21}. In Newtonian physics, in a ballistic model, the jet height is proportional to the ratio of initial speed to gravity. Since the gravity is proportional to the BH mass and the initial jet speed is set by BZ process as the $\sim E_{\rm kin}^{1/2} \bar{a}^{2} M_{\rm BH}^{2} B^{2}$, the maximum jet length is $\sim \bar{a} B$. This mathematical approximation suggests that the limited length of FR~0 jets could, in fact, depend on spin and magnetic field.

{\bf $M_{\rm BH}$}, the mass of the central compact object  is often used as an indicator of BH activity
as AGN are preferentially associated with massive systems
(e.g., \citealt{chiaberge11}). The jet power roughly establishes the likelihood of the source being radio-jet dominated \citep{cattaneo09}. Kinetic jet power and BH masses are connected in  radio active nuclei, as AGN tend to become more radio powerful (i.e. more
radio loud) at larger BH masses (e.g. \citealt{best05b}). Furthermore, the $L_{\rm mech}$--$M_{\rm BH}$ relation mirrors the mass dependence on the accretion rate estimated  with the Bondi accretion flow expected from the hot
hydrostatic gas halos surrounding the galaxies (e.g. \citealt{allen06,balmaverde08}). The slightly smaller BH masses of FR~0s can constitute a limit on the jet power, but cannot simply justify the substantial lack of extended jet emission.

{\bf Magnetic field $B$}, plays a primary role in the processes of
jet formation, acceleration, and collimation (e.g. \citealt{blandford77,blandford82,nakamura01,lovelace02}). Its azimuthal and poloidal components,  originated by rotation of the accretion disc and BH, are required to form and then hold the jet, which extracts angular momentum from the disc surface by torque. The magnetic field integrated on BH horizon sets the jet power. The magnetic flux paradigm by \citet{sikora13} suggests that the radio loudness is determined by the deposition of magnetic flux close to the BH,  which occurs more efficiently during the hot  RIAF-type (ADAF) phase and facilitates the jet launching. In fact, as counter-example to stress the important role of $B$ in jet production, the low magnetic field strength measured with VLBI in the the radio-intermediate quasar  III Zw~2 has possibly determined its failure to develop a powerful jet \citep{chamani21}. The amount of magnetic flux accumulation and the geometry  of the external field can differentiate between powerful and weak RGs, including FR~0s  \citep{osullivan15,grandi21}. Moderate jet activity as in FR~0s can also be triggered by the dissipation of turbulent fields in accretion disc coronae \citep{balbus91,brandeburg95}. In conclusion, a low intensity of the magnetic field structure of FR~0s represents a plausible scenario to describe their limited jet capabilities, although there is not still clear evidence.

{\bf BH spin $\bar{a}$}, is the primary ingredient in separating the formation of different jets: the spin paradigm for AGN \citep{sikora07,garofalo10} is a phenomenological scale-invariant framework based on BH-disc parameters for understanding BH feeding, feedback and jet launching mechanisms across the BH mass scale. This model, also named gap paradigm, involves the physics of energy extraction from BH via the BZ effect, the extraction of accretion disc rotational energy via BP jets and 
disc winds \citep{pringle81,kuncic04,kuncic07}. The total outflow
power (BZ jet, BP jet, disc wind) is based on the size of the gap region between the BH event horizon and the disc.  The BH spin still mediates launching the jet and determines the upper bound on the radio loudness \citep{sikora07}. Retrograde and prograde BH spin configuration with  the accreting material rotating opposite or parallel  to the direction of the BH can determine the gap region and so jet power: high retrograde BH spin for greater jet power and low spinning prograde BHs for weak jets \citep{garofalo09}. The latter scenario would fit with the FR~0 class.

Recently, in the framework of BZ jet model,  it has been found that the measured poloidal jet magnetic field $\phi_{\rm jet}$ threading a BH \citep{narayan03,tchekhovskoy11,mckinney12,yuan14} correlates over seven orders of magnitudes with the disc luminosity for a sample of aligned and misaligned RLAGN, in the form $\phi_{\rm jet} \sim L_{\rm Bol}^{1/2} M_{\rm BH}$ \citep{zamanisab14}, as predicted by a MAD model. This relation suggests that the magnetic field twisted by the rotation of the BHs which powers the BZ jets, dominates the plasma dynamics of the MAD disc, prevents the gas infall, and slows down the rotation by removing angular momentum into collimated relativistic outflow. Although we cannot directly measure field strength at the BH horizon $\phi_{\rm BH}$, this quantity is the same as $\phi_{\rm jet}$  by the flux freezing approximation for BZ jets.  Assuming that $\phi_{\rm jet}$ is set by the BZ mechanism, $\phi_{\rm jet} \sim  L_{\rm jet}^{1/2} \bar{a}^{-1} M_{\rm BH}^{-1}$, and the empirical relation from \citet{zamanisab14}, we can derive a rough estimate of the BH spin as $\bar{a} \sim L_{\rm Mech}^{1/2}  M_{\rm BH}^{-2} L_{\rm Bol}^{-1/2}$ by deriving accretion and jet power, respectively, from [O~III] and radio luminosities (Eq.~\ref{eq4}). This approximate calculation performed for the FR0CAT and FRICAT leads to the conclusion that FR~0s have on average a smaller BH spin than those of FR~Is by a factor 0.7$\pm$0.3.  A similar result is obtained if the BH spin is estimated by using the empirical correlation with jet power \citep{narayan12}.

The smaller BH spin of FR~0s would reflect to a lower bulk Lorentz factor $\Gamma$ than those of FR~Is, as suggested by \citet{baldi15,baldi19}. The maximisation of the BH parameters (M$_{\rm BH}$, B, $\bar{a}$)  would lead to high-$\Gamma$ jets with a FR~I/II morphology. This is in line with theoretical works which suggest a link between BH spin and jet speeds (e.g. \citealt{thorne86,meier99,maraschi12,chai12}). While  an initial disc-jet magnetization is needed, high spins are possibly required to launch the most relativistic jets, but observational evidence for the connection between BH spin and the jet is controversial and RQAGN with high
spins have been observed \citep{reynolds14},  breaking the one-to-one correspondence between high BH spins and presence of jets. However, the lower BH spin of FR~0s would certainly contribute to the lower jet bulk speeds, observed in the form of lower jet sidedness than that of FR~Is.

To reconcile the common pc-scale $L_{\rm core}$--$L_{\rm Bol}$ luminosity correlations valid for FR~0s and FR~Is, the lower jet sidedness of FR~0s, their lack of kpc-scale emission, their putative $\gamma$-ray emission, the invoked (static, but valid also for a dynamic scenario) jet model for FR~0s comes from the well-known  (stratified jet) ``two-flow model'' \citep{sol89}: an outer jet layer with a mildly relativistic velocity ($v \sim 0.5c$) surrounds an inner electron-positron jet spine, which
moves at much higher relativistic speeds (bulk
Lorentz factor $\sim$ 10). The existence of two flows at different velocities provided
 a good agreement with both theoretical and observational constraints of RGs in general \citep{ghisellini05}. This model can provide a simple way to solve the discrepancy between the required high Lorentz factors to produce the observed $\gamma$-ray emission and the slower observed motion in jets at pc scales \citep{cheng18,chen21}. Based on this model, the inner beam of FR~0 jets is slower than that of FR~I'. Similarly to what happens in FR~I jets \citep{bicknell84,bicknell95,bownman96,laing14,perucho14}, the jet spine of FR~0s could decelerate on kpc scale to sub-relativistic speeds for entrainment of external material. As suggested by the similar Mpc-scale environment, the unification of the FR~0s and weak BL~Lacs in a single class of RGs characterised by a fainter slower spine than that of FR~Is, finds supports from recent results  which identify a large number of
BL~Lacs showing `non classical' blazar-like properties and analogies with FR~0s (e.g. \citealt{liuzzo13,massaro17,dammando18}).
 
 Another parameter which can play a role in two-model flow jets for FR~0s  is the prominence of one of the two components over the other. In fact, the picture of FR~I jets as decelerating flows with
transverse velocity gradients and with an intrinsic emissivity (prominence) differences between the spine and the sheath (a slow-moving
boundary layer being more prominent than faster material near
the centre, \citealt{komissarov90}) finds observational support in resolved jet structures of individual sources (e.g.,  3C~84, \citealt{giovannini18}; Cen~A, \citealt{janssen21}). Similarly, in FR~0s, the large loss of radio emission from pc-scale structure with respect to the arcsec-scale cores indicates that the jet emissivity does not remain constant and the sheath emission dominates over the spine emission, which is hardly seen, even if boosted, with VLBI observations. In addition, an intrinsically weak spine and a brighter slower shear, supported by BZ and BP processes respectively, could account for the possible loss of jet stability for galaxy medium entertainment. On kpc scale the spine dies out, dragging the layer to disruption.   Another advantage of a  sheath-dominated jet is the formation of relativistic shocks between the jet layer of the two flows moving at different speeds, which can accelerate particles along the shock front and produce $\gamma$-ray emission by Inverse Compton \citep{wang23}. This would justify the $\gamma$-ray detection of FR~0 candidates \citep{baldi19rev}.

Another parameter which takes part in shaping the jet structure is the composition, which is one of the major uncertainties in AGN physics. In powerful RGs,
protons (or huge Poynting flux with a very low particle content) are needed in the spine to support the jet kinetic energy \citep{deyoung16}. Conversely, pure leptonic pair (electrons/positrons) jets are excluded, because the jet would be slowed down by Compton interactions.  \citet{croston18} suggested that FR~Is are likely dominated by hadrons  (mostly protons) and FR~IIs are dominated by leptons,  FR~0 jets could be lighter than FR~I/II jets, with a smaller hadronic component. This scenario would reduce the necessity for very high bulk $\Gamma$ factors for FR~0 jets and consequently would probably favour their jet instability by crossing the host galaxy.

\subsection{Dynamic scenarios}

\begin{figure}
\centerline{\includegraphics[width=0.82\textwidth]{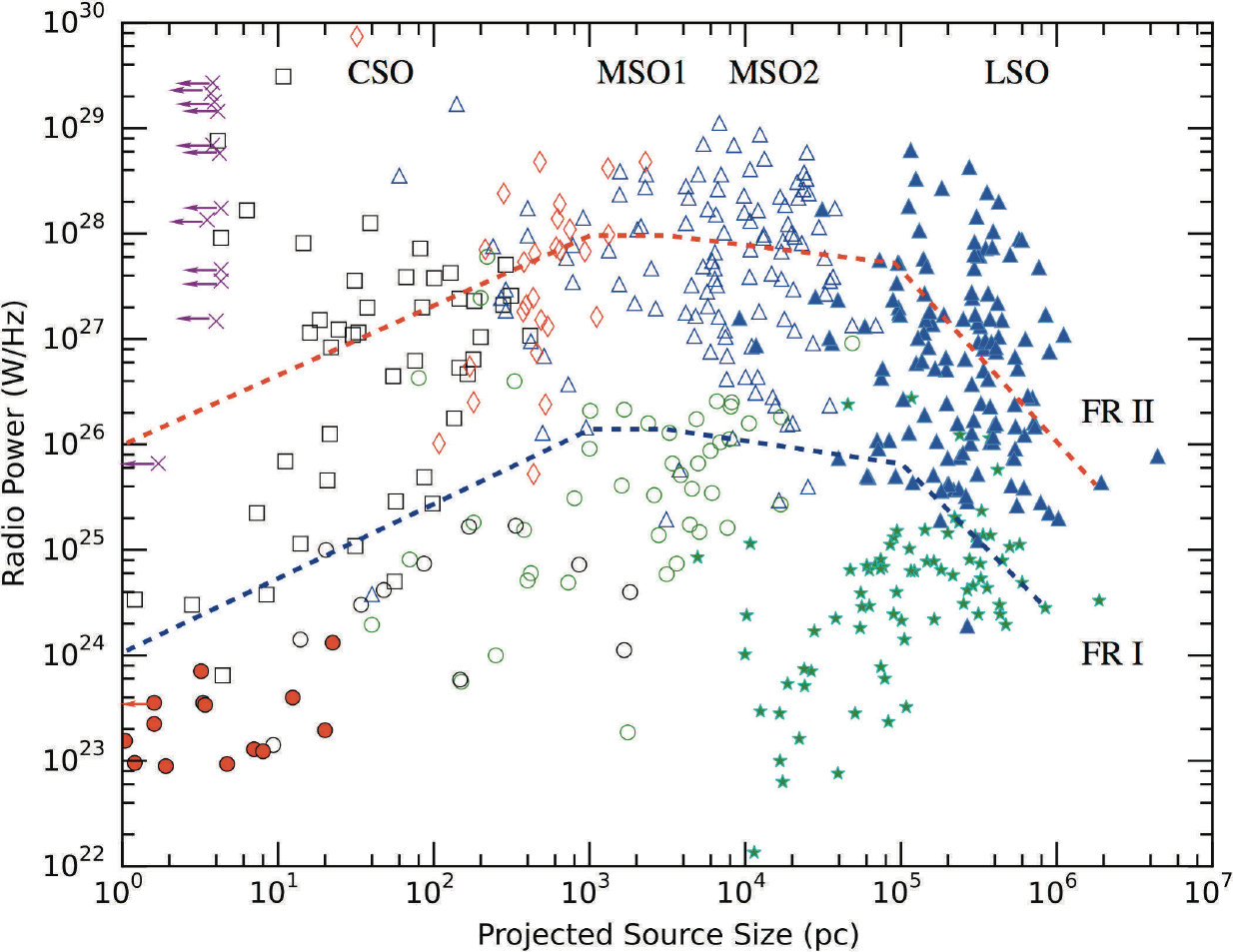}}
\caption{Radio power vs.\ source size ($P$-–$D$ diagram) of RGs adopted from \citet{cheng18}  with data take from \citet{an12}. Black squares are CSO, black circles are low-power GPSs, red diamonds are high-power GPSs, purple crosses are HFPs, green circles are low-power CSSs, blue open triangles are high-power CSSs, blue filled triangles are FR~IIs, and green filled stars are FR Is. A further morphological sub-classification is also considered  that distinguishes among CSO ($<$ 1 kpc),  MSO (1--15 kpc), and large symmetric objects (LSO; $>$ 15 kpc, FR~I/IIs) \citep{readhead95}. Red and blue dashed lines are illustrative of the evolutionary tracks based on parametric modeling for the high-power and low-power sources, respectively. The pc-scale FR~0s (red filled circles) studied by \citet{cheng18} are situated in the bottom-left corner, occupied by low-power CSOs and some compact low-power MSOs.}
\label{PDdiagram}
\end{figure}

The inclusion of a temporal variation of the accretion-ejection parameters across the RG lifetime span can better reproduce the different observed classes of RLAGN. The tracks in Fig.~\ref{PDdiagram} based on parametric modeling present the expected
evolutionary routes of a radio source that begins as a CSO and successfully evolves to FR~I or FR~IIs under conditions of long-duration AGN activity. If this standard evolutionary scenario is also applicable to the FR~0s and we consider FR~0s as progenitors of FR~Is, FR~0s would correspond to the population of the low-power ($P_{\rm 1.4\,GHz}<10^{24}\, {\rm W\, Hz}^{-1}$) CSOs in their earliest evolutionary phase. Instead, the VLBI-resolved FR~0s with pc-scale jets may shift horizontally their position in the $P$-–$D$ diagram into the region of MS0s. According to the evolutionary model of FR~0s into low-power FR~Is proposed by \citet{an12}, it is necessary for the jet structure to remain preserved before breaking out of the host galaxy and for the AGN activity to last longer than $10^{4}$ yr. However, due to their low radio power and susceptibility to jet fragmentation, only a small fraction of FR~0s would be capable of evolving beyond a few tens of kpc and becoming low-power FR~Is. Furthermore, the much larger space density of FR~0s with respect to FR~Is clearly clashes against the picture of all FR~0s as young FR~Is and necessarily,  not all RL CRSs may be destined to evolve into double RGs \citep{fanti90,fanti95}.  In fact, a uniform distribution of  total lifetimes of RLAGN in the range 0–1000 Myr, estimated from low radio frequencies data, reproduces well the distributions of projected linear sizes of the powerful sources, $10^{25} \lsim L_{\rm 150\,MHz} \lsim 10^{27}\, {\rm W\, Hz}^{-1}$, but diverges from the expectations for the large number of compact/small sources at lower luminosities,  even when surface-brightness selection effects are taken into account \citep{hardcastle19}. To break this tension, the presence of RLAGN populations with distinct lifetime distributions or accretion-ejection mechanisms (e.g., FR~0 vs FR~I/II) needs to be considered.

A possible scenario to resolve the problem of the large abundance of compact RGs concerns an intermittent AGN activity. \citet{baldi18} stated that a radio activity recurrence, with the duration of the active phase covering a wide range of values and with short active periods of a few thousand years strongly favored with respect to longer ones, might account for the  large density number of FR~0s.  This would explain why their jets do not develop at large scales  \citep{sadler14}. An occasional fueling of the central BH can  significantly reduce the accretion rate and cause a discontinuous plasma injection in the jet and its possible rapid deceleration and instability within the galaxy. Particular conditions of magnetic field loop, which trap gas and grow magnetic instabilities, could lead to a strangulated BH \citep{czerny09,yuan14,inayoshi20}.  An `aborted' jet scenario was invoked by \citet{ghisellini04} to account for the jetted RQAGN  where the BH fails to eject an extended relativistic particle jet, if the central engine works intermittently. According to this model, a small difference in BH masses, as seen between FR~Is and FR~0s, could play a role in aborting the nascent extended jets. \citet{gopal08} suggested that a dependence of the jet phenomenon on the BH mass probably could drive a large amount  of gas tidally stripped from stars by the central BH, which could truncate the jets in the BH vicinity due to mass loading from the stellar debris.  In addition, there is recent evidence that some compact sources, possibly a fraction of the FR~0 population,  are turning-off/fading \citep[e.g.,][]{kunert05,kunert06,giroletti05,orienti10b} and short-lived due to accretion-related criticality (e.g. 
\citealt{czerny09,kunert10,an12,kiehlmann23}). However, there is not still observational proof of different nuclear gas distribution between FR~0s and FR~Is, which might lead to an intermittent BH feeding or a jet frustration of the former with respect to the long-lasting secular accretion and ejection of the latter \citep{balmaverde06a}.

A temporal evolution of the BH spin within the gap paradigm predicts a FR~0 as a specific phase of a continuous activity in the family of RLAGN \citep{garofalo10}.  As the gap region reduces in size with BH spin, the BZ/BP jet decreases in power. Instead, continuous mass accretion spins the BH up towards the angular momentum value of the accretion flow. An evolution of the  BH spin configuration with the disc angular momentum can reduce or increase the gap region and change the BH spin magnitude. This dynamic process can accommodate the formation of a FR~0 population within two different scenarios: an accretion-driven or a merger-driven one.

\begin{figure}
\centerline{\includegraphics[width=0.87\textwidth]{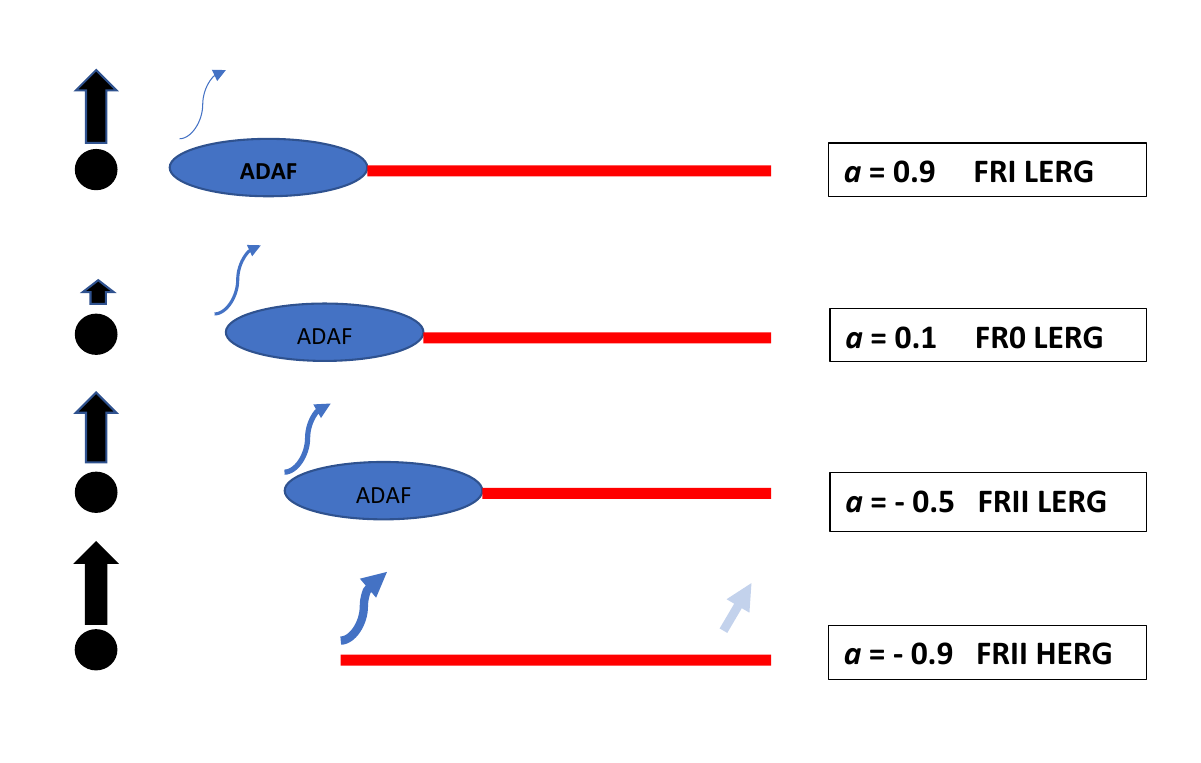}}
\caption{Focus on the temporal evolution of RGs according to the gap paradigm from high-powered FR~II HERGs  to FR~I LERGs in a accretion-driven scenario. FR~0s represent a stage of this evolution, as prograde low-spinning BHs. $a$ is the BH spin, the black thick and the blue curved arrows represent the BZ and BP jets, the blue oval and the red line represent the ADAF and the REAF discs,  while the gray arrow in the REAF disc represents the disc wind, which is absent in ADAF states.
Image reproduced with permission from \citet{garofalo19}, copyright by AAS.
}
\label{fig:garofalo}
\end{figure}

In a scenario where the BH spin depends on the accretion history of the system, the gap paradigm has been applied to FR~0s as low, prograde, spinning BHs whose
progenitors are powerful FR~II quasars \citep{garofalo19} (Fig.~\ref{fig:garofalo}). In gas rich mergers, powerful (FR~II) HERGs emerge from a BH accreting in a cold mode, surrounded by a thin REAF disc  with a retrograde accretion. Due to the powerful jet feedback, the disc moves into a RIAF disc on a timescale of about a few million years. The continuous accretion across the duty cycles  will spin down the BHs, moving the system to lower luminosities with a FR~II jet, as the retrograde BH approaches to zero (LERGs). As the BH spin moves to a prograde regime, the BZ-jet power increases as the spin increases. In this low BH spinning regime, jet is weaker than in the FR~II stage and tends to level off in a stable state. In this region of BH-jet parameter space, FR~0s find their location, where weak, compact jets are found. As the system keeps on feeding the low-spinning prograde BH, the FR~0 moves to a full-fledged FR~I,  when the spin is sufficiently higher than 0.2 and the BH must accumulate 30\% of its original mass.

The large abundance of FR~0s with respect to the other FR classes can also be interpreted as a result of the limited gas availability in nearby ETGs. The paucity of gas in the FR~0 (small- and large-scale) environments slows down the transition from FR~0s to FR~Is because of the low accretion rates. Therefore  FR~0s are not young sources, but they are the result of a prolonged slow accretion of prograde low-spinning massive BHs  over timescales of hundreds of millions -- billions of years. One-sixth of this population succeeds to funnel sufficient fuel to the BH and ultimately turns into a FR~I. The poorer Mpc-scale environment of FR~0s and the slightly smaller BH masses (galaxy masses) than those of FR~Is are the two main evidences of a different cosmological evolution of FR~0s with respect to that of FR~Is. Therefore, FR~0s are predicted to grow both in small groups (primarily) and in  rich clusters.

In a merger-driven scenario, major mergers are known to be 
the main mechanism for spinning-up BHs \citep{martinez11,bustamante19}. 
Since such objects are the result of BH-BH coalescence event, galaxies with higher masses are more likely to
have undergone more mergers  and therefore own high-spinning BHs. The simulations performed by \citet{dubois14} indicate that indeed the most massive BHs (M$_{\rm BH} \gsim$ 10$^{8}\,M_{\odot}$), in particular those associated with gas-poor galaxies, acquire most
of their mass through BH coalescence. In a poor environment, major mergers of galaxies with similar masses are rare, causing a limit on the formation of highly-spinning BHs. Although large-scale environment seems to generate a sort of difference in the BH spin distribution, the nature of the connection between environment and BH spin is still under debate, because opposite results have also been found (e.g.. \citealt{smethurst19,beckmann22}). However, in the standard picture, this merger-driven scenario would agree with the observational result that FR~0s and FR~Is live in different environment. Consequently, in this scenario, a positive link between local galaxy density, BH parameters (mass and spin), and accretion rate is set. The poorer neighborhood of FR~0s, on a statistical basis, determines a longer phase of their lower BH spin than those of their companions FR~Is, which live in richer environment. FR~0s in clusters of galaxies are likely formed recently and have not yet accreted a sufficient amount
of mass onto their central BH to turn into a FR~I. Conversely, rare FR~Is in poor groups have likely undergone particular conditions (magnetic field, gas availability, reciprocal galaxy velocity, position in the cluster/group),  which have led to an acceleration on their evolution from a FR~0 stage or a different duty cycle. However, the physical process which controls the connection between large-scale environment (Mpc scale) and the BH accretion (Bondi radius, tens and hundred pc) still remains to be understood and there is
currently limited observational evidence to support the two proposed scenarios.

In the nearby Universe, $\sim$70--80\% of the RLAGN phase is spent in a compact-jet configuration. Given that $\sim$30\% of the most massive galaxies are active and the activity must be constantly re–triggered so that the galaxy spends over a
quarter of its time in an active state \citep{best05b}, the FR~0 phase is an important stage of the evolution of an ETG where their galactic-scale jets are  continuously operating in maintenance mode. The large excess of RL CRSs over what would be expected from models in which all sources live to the same age (i.e. constant age models),  particularly evident at lower
radio luminosities \citep{shabala08,hardcastle19,shabala20}, 
suggests that the actual process of FR~0 evolution is longer than the phase spent as FR~I and FR~II. Assuming a monotonic jet expansion,
the limited size of FR~0s would point to irregular duty cycles, where  shorter active phases occur more often than the longer ones \citep{baldi18,baldi19}. However, this would conflict with the LOFAR result that the most massive galaxies are always switched on at some level at $L_{\rm 150\,MHz} \gsim 10^{21}\, {\rm W\, Hz}^{-1}$ \citep{sabater19}. Therefore, the  large uncertainties on the origin and nature of FR~0 jets, the role of environmental and internal conditions on the duration of the compact phase, complicate the estimate of the duty cycle of FR~0s.

\section{Conclusions and future perspective}
\label{sec:fututre}

The BH accretion-ejection mechanism provides a major power source in the Universe and is believed to regulate the evolution of galaxies, by injecting energy and momentum. However, the details of how and when this occurs
  remain uncertain, particularly at low luminosities, where the
  majority of active BHs are expected. There is compelling
  evidence, supported by numerical simulations, that low-luminosity RGs
  channel the bulk of their accretion power into compact and galactic-scale jets ($\sim$1--10 kpc) which  may have a significant impact on their hosts,
  regulating the SF, because they plough energy in the ISM more efficiently than powerful jets. Yet, a poor characterization of the jet physics and the AGN-host connection at low luminosities
 hampers our comprehension of the accretion-ejection paradigm,  feedback and hence the galaxy
  evolution.   The cross-correlation of high-sensitivity radio and optical surveys showed that the vast majority of
local RGs ($\sim$80\%) appear unresolved on arcsecond scales and shed light on a `new' class of low-luminosity RGs, \emph{FR~0s}, which lack of kpc-scale extended radio
emission.  This review about recent results on the multi-band properties of FR~0s collected enough evidence to conclude that FR~0s constitute a unique class of CRSs, which {\it can} launch pc-scale jets with mildly relativistic bulk speeds, probably due to small (prograde) BH spins or lower magnetic fields in the BH vicinity.

To solve the long-lasting question about the large abundance of RL CRSs with respect to what expected by standard RG evolution models, the puzzling nature of FR~0s and their impact on BH-galaxy evolution, an accurate census of the accretion-jet properties  is needed with the following characteristics: i) a statistically complete sample to
include all galaxy and AGN diversity to explore the role of each
physical parameter that controls the accretion-ejection and feedback processes;
 ii) in the
radio band, because long-baseline radio arrays can isolate the
low-brightness nuclear emission far better than any other instruments
at higher energies; iii) at luminosities as low as possible to probe the very end of
luminosity functions, ideally down to Sgr~A* luminosity ($\sim$10$^{15.5}$ W Hz); iv) in the local Universe to enable pc-scale spatial
resolution to disentangle the relative AGN-SF contribution  and
probe small jet structures.

The current and upcoming generation of radio arrays and surveys done with these facilities,  LOFAR \citep{best08,shimwell19,hardcastle20} and the International LOFAR Telescope \citep{morabito22,morabito22b}, ASKAP \citep{norris11,riggi21}, MeerKAT \citep{jarvis16,heywood22}, SKA \citep{falcke04a,kapinska15}, ngVLA \citep{nyland18b,nyland18}, uGMRT \citep{gupta17,lal21},  DSA-2000 \citep{hallinan21} and other radio antennae (e.g. ALMA, WSRT),  will provide the cornerstone of our understanding of BH activity in the local Universe at
low luminosities, across a wide range of galaxy types and environments. Because of their sub-arcsecond resolution and $\mu$Jy-level sensitivity, they will uncover the bulk population of CRSs,  opening a new window onto the physical properties of FR~0s. For example, within the wide sky coverage of the LOFAR observations, the census of nearby
active BHs at 150 GHz will count $\sim$3000 LLAGN with luminosities
$< 10^{40}\, {\rm erg\, s}^{-1}$ at $z<0.03$ \citep{sabater19}. The next
step will be with the advent of SKA and ngVLA, which will survey vast numbers of
nearby galaxies  with unprecedented sensitivities at
sub-arcsecond resolutions on a large range of radio frequencies (reaching
$\sim$1 $\mu$Jy at $<$1 GHz over 30 deg$^{2}$ will detect $\sim$300,000
LLAGN, \citealt{prandoni15,padovani16}). A multi-band cross-match with other surveys at higher frequencies (optical, X-ray) will trace a demography of local low-power jetted BHs
and their interplay with galaxies, providing firmer constraints on models of accretion-ejection coupling in ordinary AGN (not quasar type) \citep{prandoni14,prandoni15}.

Low-frequency ($<$1 GHz) radio surveys with SKA precursors (e.g. ASKAP, MWA), LOFAR and GMRT are already extremely valuable for studying the putative extended emissions of FR~0s, because it remains crudely true that the observed duty cycle of AGN increases with decreasing frequency: this is because of the
longer synchrotron lifetimes of the lower-energy relativistic
particles at lower frequencies. Deep sub-arcsecond international LOFAR telescope observations could reveal the true extent of the penetration of
FR~0 jet into the galaxy,  by discovering
synchrotron-aged plasma from past injection events. This would lead to
a better characterization of the physical properties, duty cycles and kinetic power of FR~0 jets

Combining hundreds-MHz information with GHz-observations can help to characterise the spectral shape of FR~0s to infer the fraction of optically thin, hence extended, emission
present in FR~0 jets and eventually isolate the fraction of genuine young
radio sources  erroneously included in this class. High-resolution radio
observations with  long baseline arrays (e.g. eMERLIN, EVN, VLBA) are crucial to establish the fraction of jetted FR~0s on pc scale and derive the jet asymmetry and velocity distribution.

With those ideas in mind,  the future research on FR~0s will address the following key topics:

\begin{itemize}
\item \textbf{Pc-scale accretion-ejection.} The origin of the inability of such a large population to grow kpc-scale jets is still a mystery. The separation of the genuine population of FR~0s with respect to other compact impostors (star forming galaxies, RQAGN, young RGs, blazars)  is fundamental to identify the crucial aspects which can diagnose their jet limitations. Accretion and ejection studied with non-radio high-resolution data (e.g. Chandra, eROSITA, JWST, VLT, ELT) can help to disentangle the different contribution in RL CRS population and  constraining models of disc and jets.

\item \textbf{High energy.} Several FR~0 as $\gamma$-ray emitters have been detected at the present time and are expected to be multi-messenger sources. It is important to continue the search for $\gamma$-ray emission from RL CRSs and LLAGN in general to study particle acceleration mechanisms at low powers.

\item \textbf{AGN feedback.} Several studies point to the result that RL CRSs can have a more efficient feedback on galaxy than powerful extended RGs. A single studied case of FR~0 driving turbulence and creating cavities in the X-ray atmosphere of a cluster is not sufficient to derive robust results on the effect of low-power jets of FR~0s in the surrounding medium. Systematic studies with deep multi-band data,  combined with VLBI observations will provide a unique data set for advancing our comprehension of the interaction of the FR~0s with their environments.

\item \textbf{High redshifts.} There is evidence that the local FR~0 population has an important counterpart also at higher redshifts ($z>1$). A systematic study of the genuine FR~0s at the cosmic noon from deep fields would help to understand the formation and cosmic evolution of low-power RLAGN with respect to the other classes of RGs. 

\item \textbf{Numerical simulations.} High resolution, 3D numerical simulations of low-power jets (total jet power $< 10^{44}\, {\rm erg\, s}^{-1}$) can help to clarify the formation, propagation and impact of FR~0 jets in the galactic medium. 
\end{itemize}

\begin{acknowledgements}
I am very grateful to friends and colleagues who provided thoughtful comments on the manuscript, especially A.~Capetti, who helped me through fruitful discussions and inspired this review, and M.~Brienza, G.~Giovannini, P.~Grandi, G.~Migliori, and E.~Torresi,  who triggered a productive discussion on FR~0s. I thank L.~Ferretti for offering me this wonderful opportunity to write this review and the anonymous referees for constructive comments and suggestions that greatly helped improve the manuscript. R.D.B. acknowledges financial support from INAF mini-grant ``FR0 radio galaxies" (Bando Ricerca Fondamentale INAF 2022). This research has made use of NASA's Astrophysics Data System Bibliographic Services. This research has made use of the NASA/IPAC Extragalactic Database (NED), which is funded by the National Aeronautics and Space Administration and operated by the California Institute of Technology.
The author has no conflicts of interest to declare. I certify that the submission is original work and is not under review at any other publication.
\end{acknowledgements}

\phantomsection
\addcontentsline{toc}{section}{References}
\bibliographystyle{spbasic-FS}      
\bibliography{fr0-review2}   

%
%

\end{document}